\documentclass[review]{elsarticle}
\usepackage{setspace}
\usepackage[margin=1in]{geometry}
\usepackage{graphics}
\usepackage{graphicx}
\usepackage{subcaption}
\usepackage{amsmath}
\usepackage{amssymb}

\def\pmb#1{\setbox0=\hbox{#1}
	\kern-.025em\copy0\kern-\wd0
	\kern.05em\copy0\kern-\wd0
	\kern-.0125em\raise.0433em\box0 }
\def\bxi {\pmb {$\xi$}}
\def\beta {\pmb {$\eta$}}
\def\bzeta {\pmb {$\zeta$}}
\def\bpsi {\pmb {$\psi$}}
\def\bc {\pmb {$c$}}

\journal{Journal Name}

\begin{document}

\begin{frontmatter}

\title{Machine Learning for Physics-Informed Generation of Dispersed Multiphase Flow Using Generative Adversarial Networks}


\author[ccmt]{B. Siddani}

\author[ccmt]{S. Balachandar\corref{correspondingauthor}}
\cortext[correspondingauthor]{Corresponding author}
\ead{bala1s@ufl.edu}

\author[ccmt]{W. C. Moore}

\author[ccmt]{Y. Yang}

\author[biomed]{R. Fang}

\address[ccmt]{Center for Compressible Multiphase Turbulence, University of Florida, Gainesville, FL 32611, USA}
\address[biomed]{J. Crayton Pruitt Family Department of Biomedical Engineering, University of Florida, Gainesville, FL 32611, USA}

\begin{abstract}
Fluid flow around a random distribution of stationary spherical particles is a problem of substantial importance in the study of dispersed multiphase flows. In this paper we present a machine learning methodology using Generative Adversarial Network framework and Convolutional Neural Network architecture to recreate particle-resolved fluid flow around a random distribution of monodispersed particles. The model was applied to various Reynolds number and particle volume fraction combinations spanning over a range of [2.69, 172.96] and [0.11, 0.45] respectively. Test performance of the model for the studied cases is very promising.   
\end{abstract}


\end{frontmatter}

\section{Introduction}
Dispersed multiphase flows are flow systems that contain dispersed elements, such as particles, droplets or bubbles that are distributed in a continuous phase \cite{Turb_disp_mltphs}. These flows are prevalent both in nature and industry: Sediment transport, fluidized bed and pneumatic transport of gas-particle mixture are some widespread applications. It is intuitive that the existence of a dispersed phase affects the continuous phase, but the extent of their interaction is highly dependent on spatial distribution, size, density, volume fraction and other dispersed phase properties.

Fluid flow around a random distribution of stationary spherical particles within a periodic box can be considered as a geometrically-simplified canonical problem of substantial importance in the study of dispersed multiphase flows. This problem has been studied extensively both experimentally and numerically with the focus on better understanding pseudo turbulence generated by the arrangement of particles and the back effect of pseudo turbulence on the hydrodynamic forces on the particles. This understanding enables model development of pseudo turbulent Reynolds stress and parameterization of particle forces, which can be leveraged in more complex scenarios like particle-laden flows and flow through packed-beds.
Due to its fundamental significance, the problem of flow through a random distribution of stationary particles has been simulated over a range of Reynolds numbers and volume fractions using a variety of numerical methods: {Finite Volume Methods} \cite{MAGNICO20035005}, {Immersed Boundary Methods}, \cite{UHLMANN2005448, tenneti2011drag,zaidi2014new, AKIKI201634} and {Lattice Boltzmann Methods} \cite{Beetstra2007, RONG201344, bogner2015drag}.   

The rapid advancement of Machine Learning ({ML}) algorithms and computer technology in recent years has brought about great interest in its many applications. The impacted applications of interest are not only from the field of Computer Science but also from other streams of sciences like physics, chemistry, various branches of engineering, and data analytics. This interest is mainly due to the success of ML in solving the complex data-rich problems of different domains. In particular, in the field of fluid mechanics ML has found many emerging applications, which have been discussed in recent reviews \cite{Brunton_ML_FM}. For example, \citet{turb_ML} addressed the use of ML in turbulence modeling, ML applications in flow visualization using convolutional neural networks (CNN) was discussed in \citet{autodesk_cnn}, and prediction of flow over an airfoil using CNN was presented in \citet{Bhatnagar2019}. \citet{raissi2017physics} introduced \textit{Physics Informed Neural Networks} (PINN) based on the idea that neural networks designed for physical systems should obey the governing equations of these systems. This approach was also successful in reproducing the underlying partial differential equations of systems \cite{raissi2017physics2}. Recently, PINNs were also used to approximate Euler equations in one-dimensional and two-dimensional domains \cite{pinn_euler}. In the domain of multiphase flows, \citet{Qi_2019} used ML for computing curvature for Volume of Fluids methods. Of particular relevance is the recent work of \citet{farimani2017deep} who used a conditional variant of generative adversarial network (GAN) \cite{goodfellow2014generative, goodfellow2016nips} to generate synthetic steady-state velocity and pressure fields in 2D incompressible cavity flows. Similarly, \citet{tempoGAN} proposed their tempoGAN ML architecture, which is a temporally coherent generative adversarial network model, for obtaining super-resolution of fluid flows from an input of coarse-grained flow fields. 

Here we will present a ML methodology that will recreate particle-resolved fluid flow within a periodic box of random distribution of monodispersed particles. This work will make use of multiple CNNs in combination with the GAN algorithm to achieve its objective. In doing so the following three questions must first be addressed: (i) what information regarding the random distribution of particles and the external mechanism driving the flow is needed as {\em input} to the ML algorithm? (ii) what precise information regarding the flow needs to be predicted as {\em output} by the ML algorithm? and (iii) what prior information connecting the input to the output will be provided as training data in order to train the ML algorithm? These three aspects of ML methodology will be introduced below and will be elaborated in this paper.

The flow prediction question that we plan to address with the ML methodology is as follows: Given the location and motion of a {\em reference} particle within a random distribution of particles in a multiphase flow, along with the location and motion of its few nearest neighbors (say for example 10 nearest neighbors), we desire to accurately predict the flow around the reference particle. Important distinctions must be made between the above ML quest and the prediction goal of conventional computational fluid dynamics (CFD) approach. In a particle-resolved direct numerical simulation (PR-DNS) of flow over an array of randomly placed particles, a large periodic domain of size much larger than the correlation length is chosen containing $O(10^3)$ or more particles \cite{Beetstra2007, zhou-rot2, bogner2015drag, tenneti2011drag, zaidi2014new, tang2015new, akiki2016force, akiki2017pairwise}. The domain is then discretized with millions of grid points and the flow around all the particles over the entire computational domain is simultaneously solved. Such a {\em global approach}, where the flow around all the particles is solved simultaneously, is neither feasible with the current ML algorithms, nor necessary due to the manner in which ML predicts the flow. Therefore, in contrast to conventional CFD, a {\em local approach} will be pursued with ML, where the flow around each particle is predicted with a deterministic knowledge of the precise state of a few nearby particles, and a statistical knowledge of distribution of all other distant particles. Once the flow associated with each particle is determined, the overall flow can be obtained through appropriate superposition. 

In the present study, following the PR-DNS simulations mentioned above, we will restrict attention to steady flow over a random distribution of stationary particles. In this limit, when each particle within the system is chosen as the reference particle, the {\em input} to ML is the list of relative position of its nearest neighbors that are located within a chosen neighborhood (i.e., location of neighbors within a sub-volume centered around the reference particle). In addition, the {\em {input}} will include the Reynolds number of the flow within the sub-volume calculated based on the average fluid velocity within the sub-volume and the particle diameter. The desired {\em output} of ML is the accurate prediction of the velocity and pressure fields within the sub-volume, which can be further refined to focus only on the region in the immediate vicinity of the reference particle, without extending into the neighboring particles.

We now address the last question pertaining to the training data. The results of PR-DNS simulations of flow over a random distribution of particles for a range of Reynolds number and volume fraction \cite{akiki2017pairwiseJCP, akiki2017pairwise} will be used for both training and testing purposes. The raw simulation results will be curated to obtain the input and output data for each particle of the random array for training the ML algorithm. This curation is an important step, since it allows the number of training samples to scale as the number of particles within a CFD computational box. For each case (i.e., for each Reynolds number and volume fraction combination) we consider multiple CFD realizations to ensure sufficient amount of training data for fully converged training. Most of the CFD realizations will be used for training, but a few that are not used for training will be reserved for testing.

A simpler version of this flow prediction problem has recently been achieved within the framework of pairwise interaction extended point-particle (PIEP) approach \cite{moore2019hybrid, moore-bala2019}. The perturbation flow induced by each particle was defined as its superposable wake, which was taken to depend only on the average statistical presence of all its neighbors, and thus was parameterized as a function of the local particle Reynolds number and volume fraction. The key aspect of the PIEP approach was that the flow around any particle can be calculated as a summation of its perturbation flow along with those of all its neighbors.  As a result, in this approach, the {\em input} to prediction of flow around a reference particle is still the relative position of its neighbors. However, the important simplification comes from the assumption that the influence of each neighbor can be taken pairwise, i.e., one at a time. Due to this assumption, the predicted flow was observed to range in accuracy from 56\% to 83\%, for varying Reynolds number and volume fraction. 

Here we pursue a more advanced ML algorithm, which allows us to account for the perturbation flow of all the neighbors taken together without the pairwise approximation. By addressing this multi-particle interaction problem directly without the pairwise interaction assumption, we attempt to predict the flow around a random distribution of particles far more accurately than the PIEP model.

The paper describes a successful implementation of the generative adversarial network (GAN) methodology and architecture that models the Navier--Stokes equation and generates accurate dispersed multiphase flow. Once developed, the GAN can generate velocity and pressure fields around a random distribution of particles at a computational cost that is orders of magnitude cheaper than conventional CFD. However, we must emphasize two important facts: (i) CFD results are needed in the first place to train GAN, and (ii) the GAN results, though quite accurate, will not perfectly reproduce the CFD results. Nevertheless, the ability to very quickly generate synthetic flows over large arrays of particles will be of great value, since this capability can provide  us with new opportunities. For example, by considering very large multiphase flow systems or by considering many repeated realizations, more converged higher-order statistical information of pseudo turbulence can be obtained. Similarly, the large amount of synthetic data can be used to develop better particle force models that depend not only on average Reynolds number and volume fraction, but also on the relative location of nearby neighbors.

Summary of succeeding sections of this paper is as follows. In section 2, Direct Numerical Simulation (DNS) methodology and analysis of the data used in this work is presented. In section 3, the GAN methodology is explained. Section 4 deals with performance of the proposed GAN model and discussion of the results. The presented neural network and GAN methodology can also act as a well-defined starting point to design more compact networks for similar applications.               

\section{Simulation Data and Its Curation}
Direct Numerical Simulation (DNS) results are required for the training of a neural network to produce synthetic flows, and the current work makes use of simulation results presented in \cite{moore2019hybrid,AKIKI201634}. The basic geometric structure of these simulations is a random distribution of stationary, monodispersed spherical particles in a cubic domain. The non-dimensional diameter of the particles is unity. They are randomly distributed within the domain with uniform probability, and the size of the cubic box is $3\pi$. The domain is subjected to periodic boundary conditions along \textit{x, y} directions. Symmetric, no-stress boundary conditions are applied along the \textit{z} direction. A mean macroscale pressure gradient is imposed along the \textit{x} direction making it the streamwise direction. A grid resolution of $376 \times 376 \times 401$ points along $x$, $y$, and $z$, respectively, ensures that the fluid is fully resolved around each particle in every simulation. Incompressible Navier-Stokes equations in non-dimensional form are solved to steady state with no-penetration and no-slip boundary conditions on each particle's surface using immersed boundary method. 
The different cases simulated can be categorized using two macroscale parameters: particle Reynolds number ($ Re$) and particle volume fraction ($\phi$). A list of all the cases considered is presented in Table~\ref{tab:cases}. \citet{akiki2017pairwise} considered only the inner 64\% along \textit{z} direction to avoid the effects of no-stress boundary condition. This inner portion will be referred to as the {\em inner flow region} in succeeding sections of this paper.

The variables have been non-dimensionalized using particle diameter $d_{*}$ as length scale, the mean streamwise fluid velocity within the inner flow region $U_{*}$ as velocity scale, and $\rho_{*}U_{*}^{2}$ as the pressure scale. Here, $\rho_{*}$ denotes fluid density. Here and henceforth in this study, all dimensional quantities will be indicated with $*$ as subscript and all non-dimensional quantities will be represented without the subscript.     

\begin{table}
	\begin{center}
		\resizebox{\textwidth}{!}{%
		\begin{tabular}{|cccccccccc|}
			\hline
			Case&$Re$ & $\phi$ & Realizations & $\sigma_u$ & $\sigma_v$ & $\sigma_w$ & $\sigma_p$ & $N_{tr}$ & $N_{te}$\\
			\hline
			1&39.47 & 0.11 & 10 & 0.5026 & 0.1810 & 0.1844 & 0.2749 & 418 & 46 \\
			2&69.88 & 0.11 & 10 & 0.5031 & 0.1781 & 0.1806 & 0.2357 & 418 & 46 \\
			3&172.96 & 0.11 & 10 & 0.4968 & 0.1947 & 0.1960 & 0.1973 & 418 & 46 \\
			4&16.49 & 0.21 & 8 & 0.6131 & 0.2821 & 0.2877 & 0.7473 &  636 & 91 \\
			5&86.22 & 0.21 & 7 & 0.5960 & 0.2580 & 0.2609 & 0.3807 & 635 & 90 \\
			6&2.69 & 0.45 & 5 & 0.8204 & 0.4740 & 0.4732 & 17.3688 & 772 & 193 \\
			7&20.66 & 0.45 & 5 & 0.8227 & 0.4584 & 0.4631 & 2.9072 & 772 & 193 \\
			8&114.60 & 0.45 & 5 & 0.8006 & 0.4340 & 0.4421 & 1.0467 & 772 & 193 \\
			\hline
		\end{tabular}
	   }
		\caption{The Reynolds number and volume fraction of the different cases considered, the corresponding RMS values of $u, v, w, p$ and the average size of training and testing datasets}
		\label{tab:cases}
	\end{center}
\end{table}

\subsection{Sub-domain Selection}
As discussed in the introduction, though the raw DNS results of each realization includes flow information over all the particles within the entire computational domain, each training data set will be centered around an individual {\em reference } particle and will cover only a portion of the computational domain centered around the particle. To decrease the effects of the top and bottom no-stress boundaries in the training date set, only the particles that are inside the inner 27\% along \textit{z} direction will be considered as reference particles. This inner 27\% portion along the \textit{z} direction will be referred to as \textit{inner particle region}. Thus, each realization of each case listed in Table~\ref{tab:cases} will yield $N_i$ training data sets, where $N_i$ is the number of particles within the inner particle region of the $i^{th}$ realization.  The first step of this curation process is to determine the size of the sub-domain that will surround the reference particle, so that the flow information within this sub-domain will form the training data of the reference particle. 

The size of the sub-domain will be chosen to satisfy the following properties: (i) It must be large enough to contain sufficient number of neighbors, since the relative location of these neighbors will be the key input to the ML algorithm. In this regard, we require the sub-domain to be larger along the streamwise direction than along the transverse directions, since the neighbor's influence on the reference particle is expected to be dominant along the flow direction. (ii) The sub-domain must not be too large, for otherwise the implementation of the three-dimensional (3D) ML algorithm will be computationally impossible (this issue will be elaborated later); (iii) The average properties of the flow within the sub-domain must be reasonably close to the ensemble average. This condition has been imposed so that variation in the average flow properties around the different particles when chosen as reference is mainly due to the distribution of its neighbors and not due to differences in the average macroscale flow within their sub-domain. 

In order to identify the sub-domain size that satisfies the above requirements, we first calculate the mean and rms values of velocity components and pressure of all the cases. These statistics were first computed for the inner flow region of the DNS taking into account all the realizations. These global mean and rms values define the reference against which corresponding values computed over the sub-domains can be compared to determine the appropriate sub-domain size. Since the mean pressure gradient was applied only along the \textit{x} direction, mean velocities along \textit{y} and $z$ are zero and the mean non-dimensional streamwise velocity is unity. Standard deviations of $u, v, w$, and $p$ are also tabulated in Table~\ref{tab:cases}. 

\begin{figure}
	\centering
	\includegraphics[width=\linewidth]{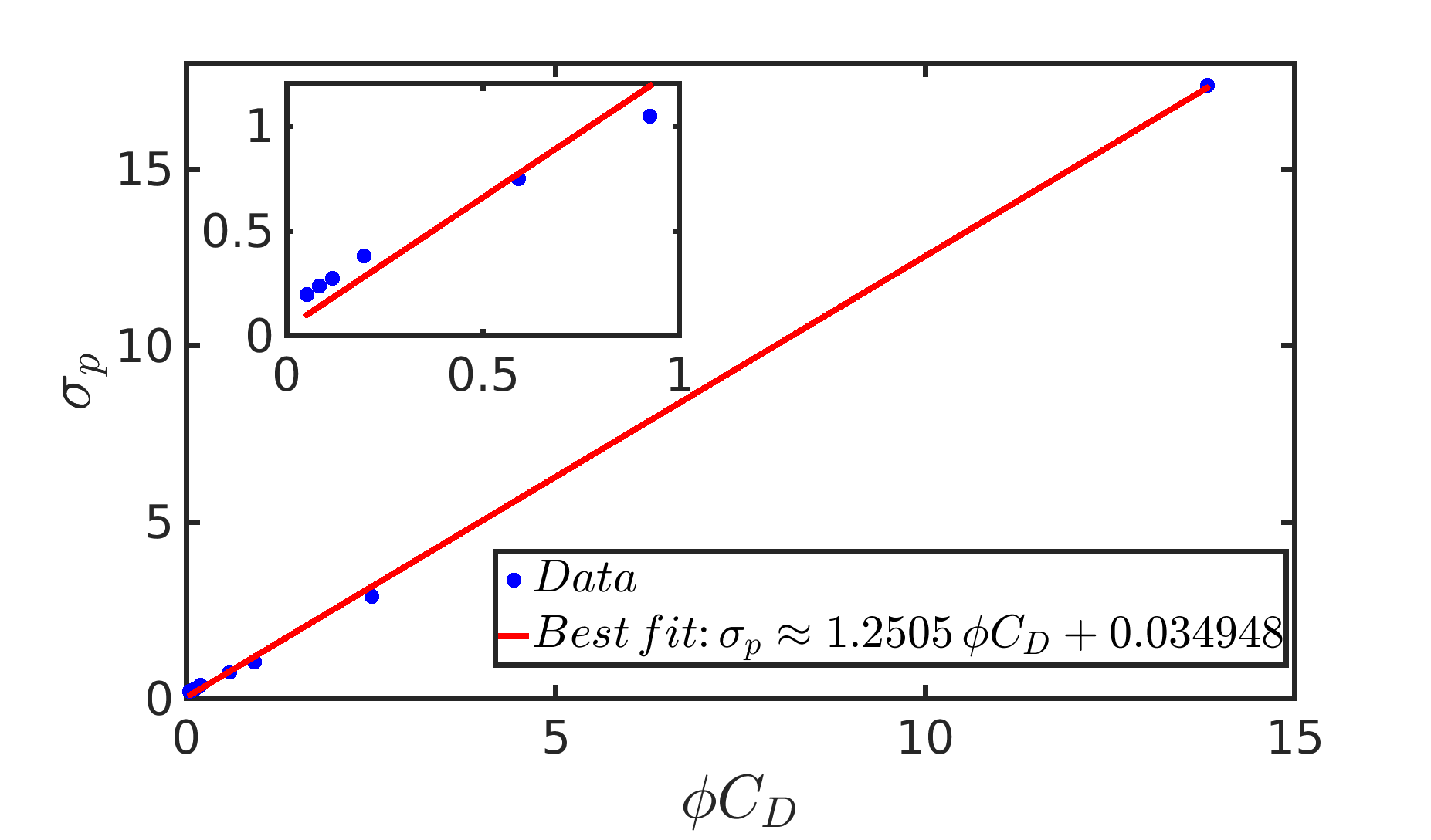}
	\caption{$\sigma_p$ vs $\phi C_{D}$
		for all cases.}
	\label{fig:rmsp}
\end{figure}

For a given particle volume fraction, the rms values of velocities do not show a strong variation to changes in $Re$, however, they tend to increase with volume fraction. 
From Table~\ref{tab:cases}, the rms values of streamwise and transverse velocities can be approximated as follows:
\begin{equation}\label{rms_uv}
\sigma_{u} \approx 0.6444 \, \left[1.4\phi+0.63\right] \, , \quad
\sigma_{v} \approx 0.3075 \, \left[2.6\phi+0.33\right]	 \, .
\end{equation}
It was also observed by \citet{moore2019hybrid} that rms values of pressure for these cases show a very good correlation with $\phi \, C_{D}$, where $C_{D}$ is non-dimensional volume fraction-dependent mean drag on the particles \citep{tenneti2011drag}. This correlation between $p$ and $\phi\,C_{D}$ is given below and depicted in Figure~\ref{fig:rmsp}.
\begin{equation}\label{rms_p_cd}
\sigma_{p} \approx 1.2505 \, \phi C_{D}  + 0.034948 \, ,
\end{equation}
where
\begin{eqnarray}
C_{D} & = & 3\pi\frac{(1-\phi)}{Re_s}\left[\frac{1+0.15Re_{s}^{0.687}}{(1-\phi)^{3}}+\frac{5.81\phi}{(1-\phi)^{3}}+\phi^{3}Re_{s}\left(0.95+\frac{0.61\phi^{3}}{(1-\phi)^{2}}\right)\right]\nonumber\\
Re_{s} & = & \frac{Re}{(1-\phi)} \, .\nonumber
\end{eqnarray}

The mean and rms values presented in Table~\ref{tab:cases} are for the entire inner flow region. We now proceed to calculate the mean and rms for the sub-domains in the following manner: (i) Choose the sub-domain size to be tested; (ii) In the $i^{th}$ DNS realization of a case, define sub-domains centered around each of the $N_i$ particles within the inner particle region. This will result in $N_c = \sum_{i = 1}^{M_c} N_i$ sub-domain information for the training of the ML algorithm, where the sum is over all the $M_c$ realizations of the case; (iii) A grid resolution comparable to that used in the DNS is used to discretize each sub-domain and the flow variables at these grid points are obtained using linear interpolation of the PR-DNS data.   

\begin{figure}
	\begin{center}
		\begin{subfigure}{0.49\textwidth}
			\includegraphics[width=1.\linewidth]{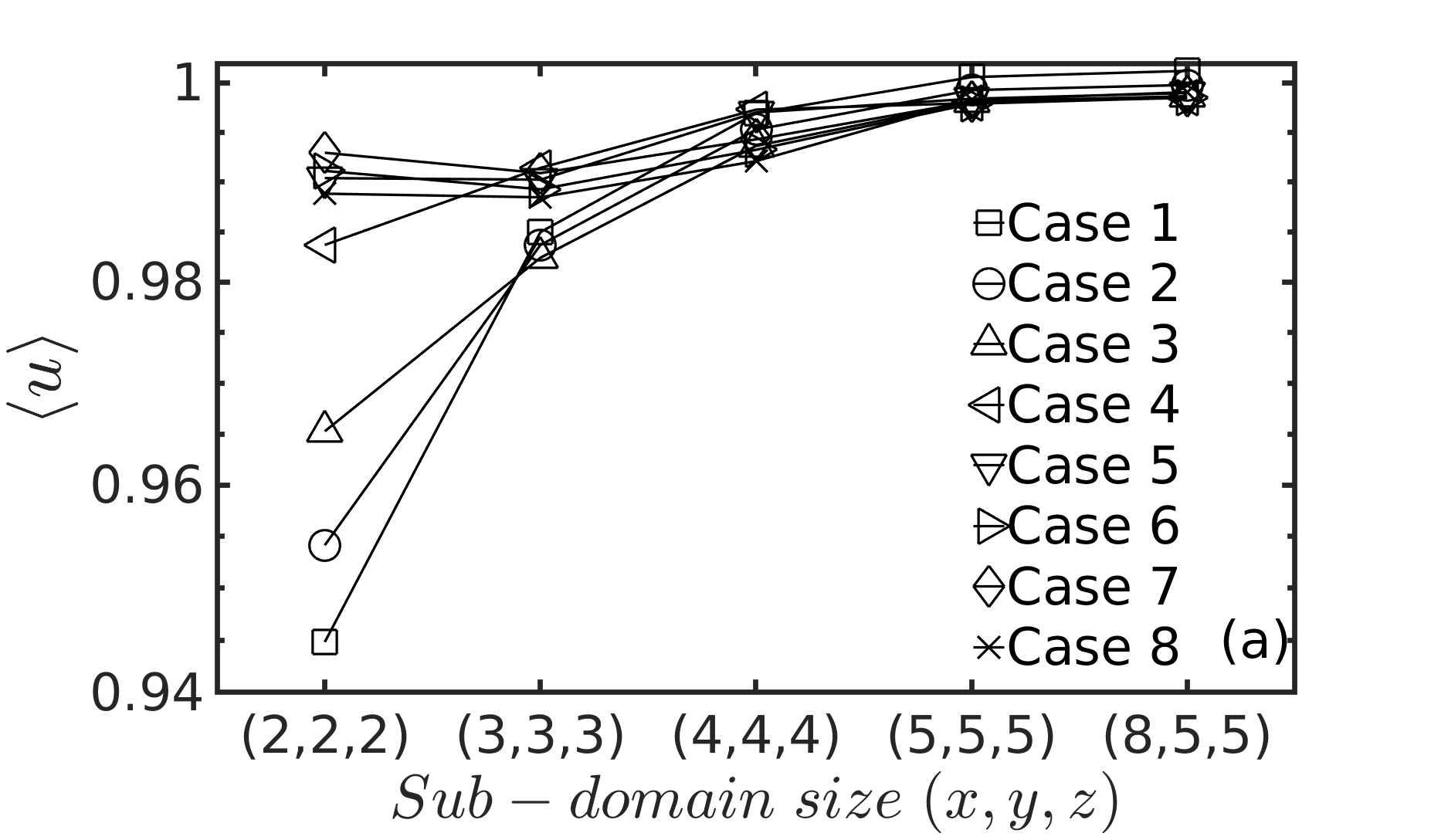}
		\end{subfigure}
		\begin{subfigure}{0.49\textwidth}
			\includegraphics[width=1.\linewidth]{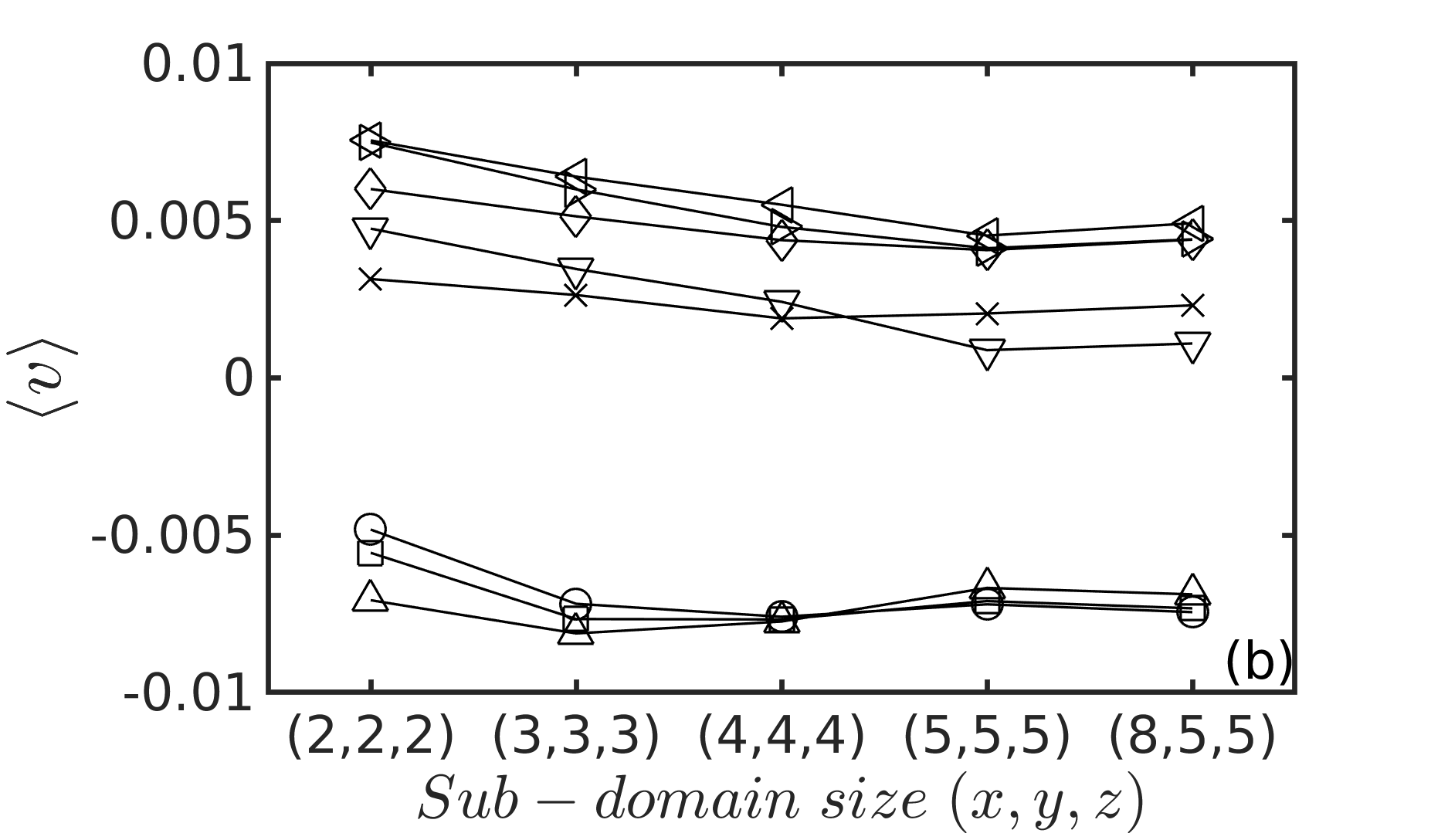}
		\end{subfigure}
	\end{center}
	\caption{Variation of mean (a) streamwise and (b) transverse velocities with sub-domain size}
	\label{fig:u_mn}
\end{figure}

The streamwise and transverse mean velocities ($\langle u \rangle$ \& $\langle v \rangle$, $\langle w \rangle$) computed for the different domain sizes are shown in Figure~\ref{fig:u_mn}. The results of the 8 different cases are shown. It can be seen that for all choices of sub-domain size in every case the transverse mean velocities are less than 1\% of the streamwise mean velocity. Furthermore, domains of size $4 \times 4 \times 4$ and larger are within 1\% of the global average value of unity. Hence, it can be considered that the mean Reynolds number of each sub-domain is primarily determined by the global mean streamwise velocity. A similar conclusion can be drawn for the estimation of mean particle volume fraction within each sub-domain. As a result, in each of the eight cases listed in Table~\ref{tab:cases}, the appropriate Reynolds number and volume fraction of all the  sub-domain data are the same as the macroscale value.    

\begin{figure}
	\begin{center}
		\begin{subfigure}{0.49\textwidth}
			\includegraphics[width=1\linewidth,keepaspectratio=true]{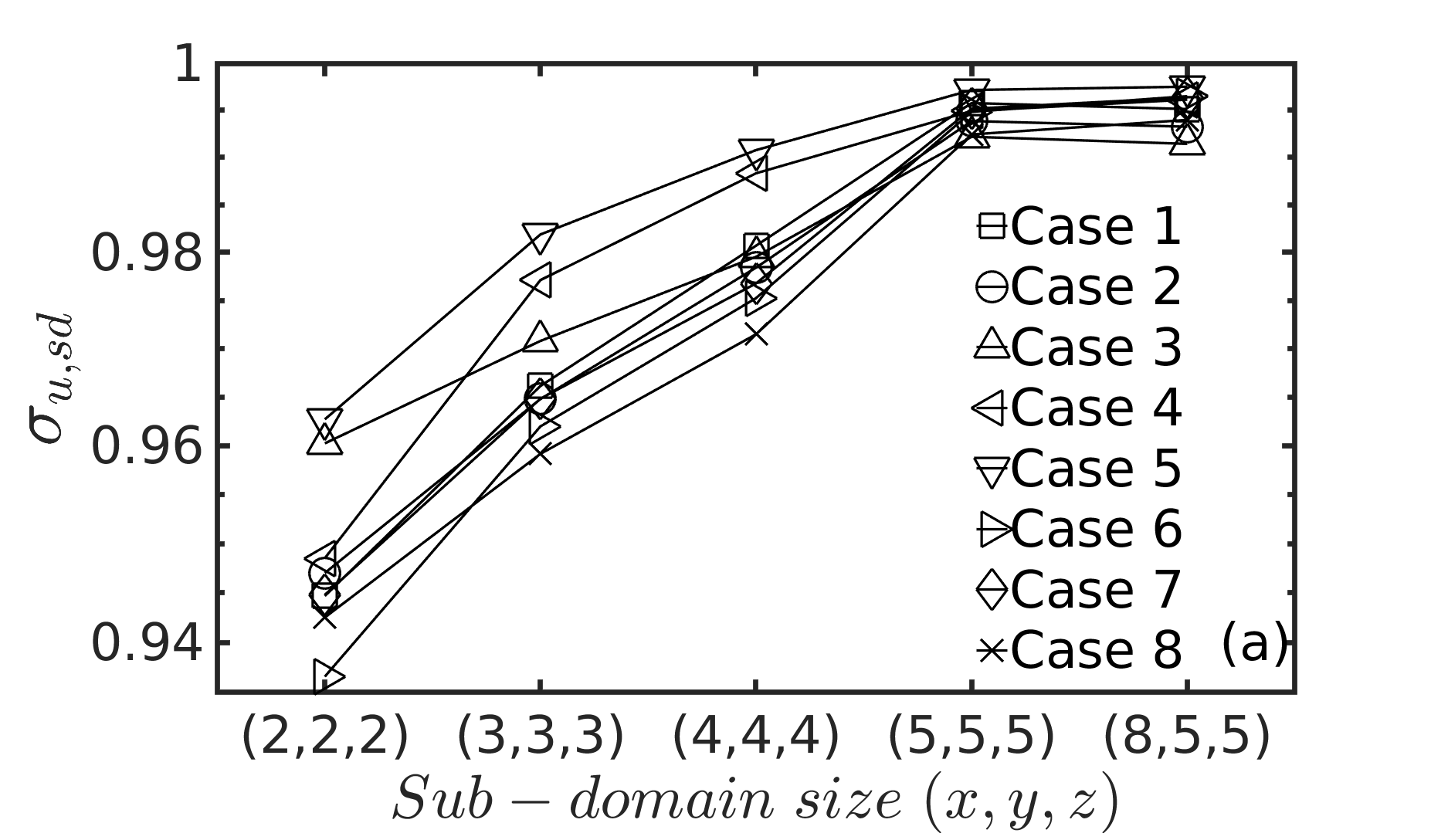}
		\end{subfigure}
		\begin{subfigure}{0.49\textwidth}
			\includegraphics[width=1\linewidth,keepaspectratio=true]{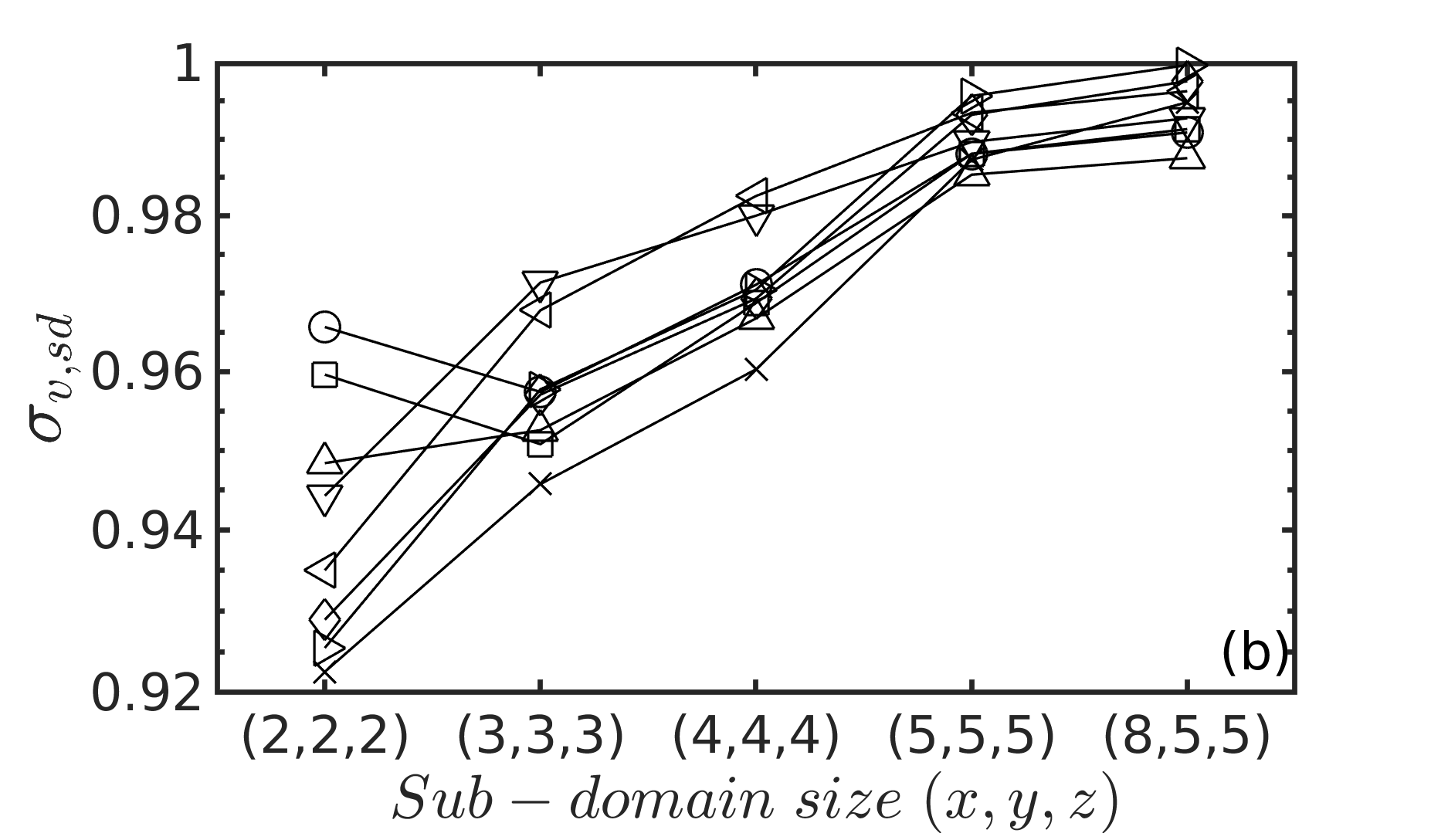}
		\end{subfigure}
		\begin{subfigure}{0.49\textwidth}
			\includegraphics[width=1\linewidth,keepaspectratio=true]{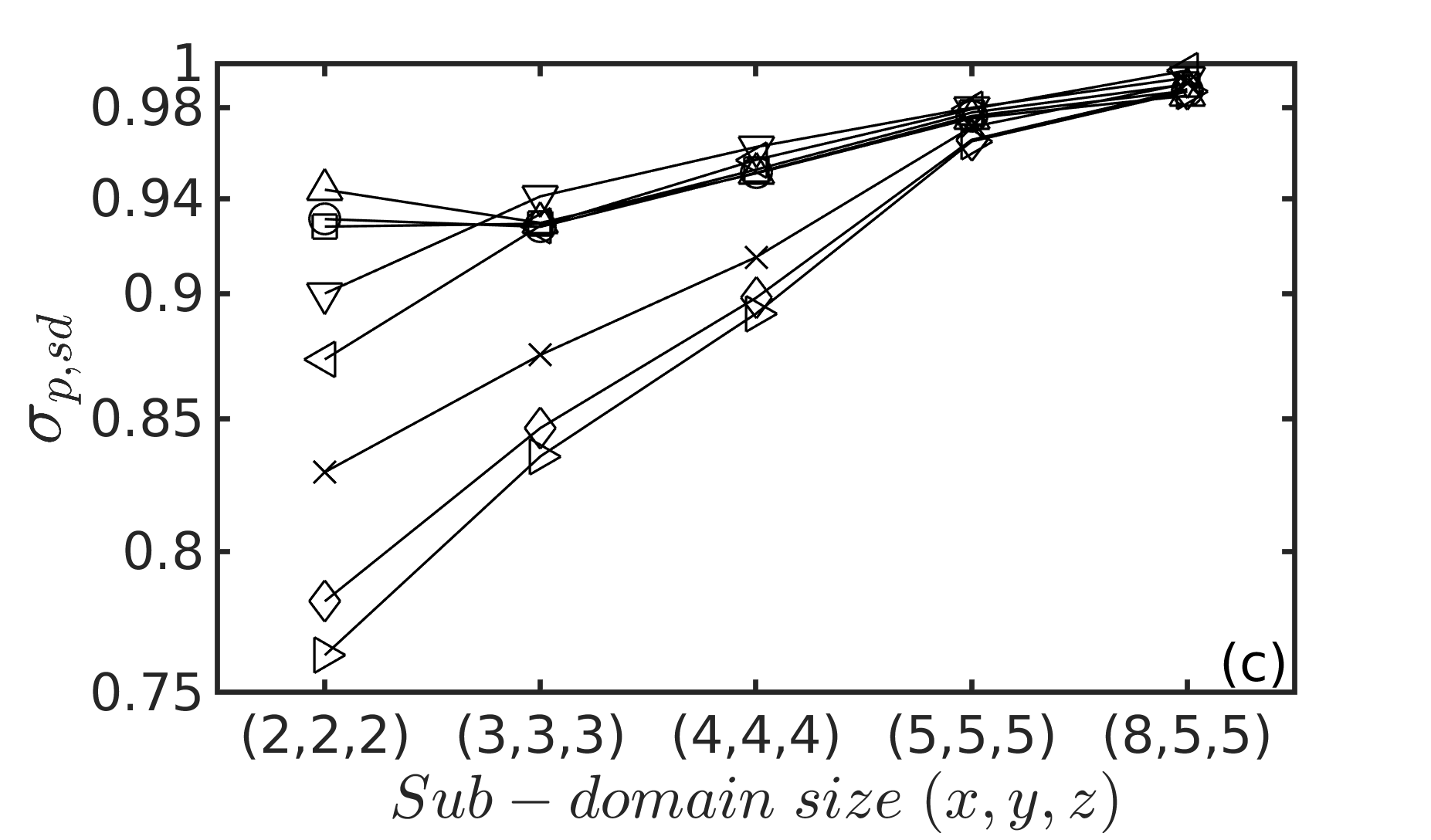}
		\end{subfigure}
		\caption{Effect of sub-domain size on: (a) RMS of \textit{u}, (b) RMS of \textit{v}, and (c) RMS of \textit{p}}
		\label{fig:domain_size}
	\end{center}
\end{figure}

The average value of rms velocity and pressure variation within the sub-domain is defined as follows: 
\begin{equation}\label{a_ds}
\sigma_{u,sd}  =  \frac{\sum\limits_{i=1}^{N_c} \sigma_{u,i} / N_c}{\sigma_{u}}  \quad \mathrm{and} \quad
\sigma_{p,sd}  =  \frac{\sum\limits_{i=1}^{N_c} \sigma_{p,i} / N_c}{\sigma_{p}} \, ,
\end{equation}
where $N_c$ is the total number of sub-domain samples of a case. Thus, $\sigma_{u,sd}$ defines the rms of streamwise velocity computed within each sub-domain, then averaged over all the sub-domains and normalized by the global rms value. Similar definitions apply for $v$ and $w$ velocity components as well. In the above equation $\sigma_{u}$, $\sigma_{v}$, $\sigma_{w}$ and $\sigma_{p}$ are the global rms velocity and pressure fluctuations computed for the entire flow data and are listed in Table~\ref{tab:cases} for all the cases. The effect of sub-domain size on normalized rms of $u$, $v$ and $p$ are shown in Figure~\ref{fig:domain_size}. The general trend is that they increase with increasing sub-domain size. Based on these plots we choose a sub-domain of non-dimensional size $8 \times 5 \times 5$, along the $x$, $y$ and $z$ directions for further investigation, as it satisfies all the requirements discussed above. 

As the final step of data curation we consider normalization of the velocity and pressure fields. Normalization of all inputs/outputs of a neural network to a similar scale is necessary to avoid biased preference to any particular input/output scales by weights in the network. This is achieved by normalizing velocity and pressure perturbations of the $i^{th}$ sample as:
\begin{equation}
u^{\prime} = \frac{u-\langle u \rangle}{3\sigma_{u,i}} \, , \quad
v^{\prime} = \frac{v-\langle v \rangle}{3\sigma_{v,i}} \, , \quad
w^{\prime} = \frac{w-\langle w \rangle}{3\sigma_{v,i}} \, , \quad
p^{\prime} = \frac{p-\langle p \rangle}{3\sigma_{p,i}}
\end{equation}
here, $\sigma_{u,i}, \, \sigma_{v,i}, \, \sigma_{p,i}$ are rms of streamwise and transverse velocities, and pressure respectively of the $i^{th}$ sample obtained using \eqref{rms_uv} \& \eqref{rms_p_cd}. Note that the mean values of streamwise velocity is unity while the mean transverse velocities are zero. By scaling with three sigma, $u^{\prime},\, v^{\prime},\, w^{\prime},\, p^{\prime}$ are essentially normalized to be between $-1$ and $1$.

\section{GAN Methodology}\label{methdlgy}
\begin{figure}
	\centering
	\includegraphics[height=0.36\textheight,keepaspectratio=true]{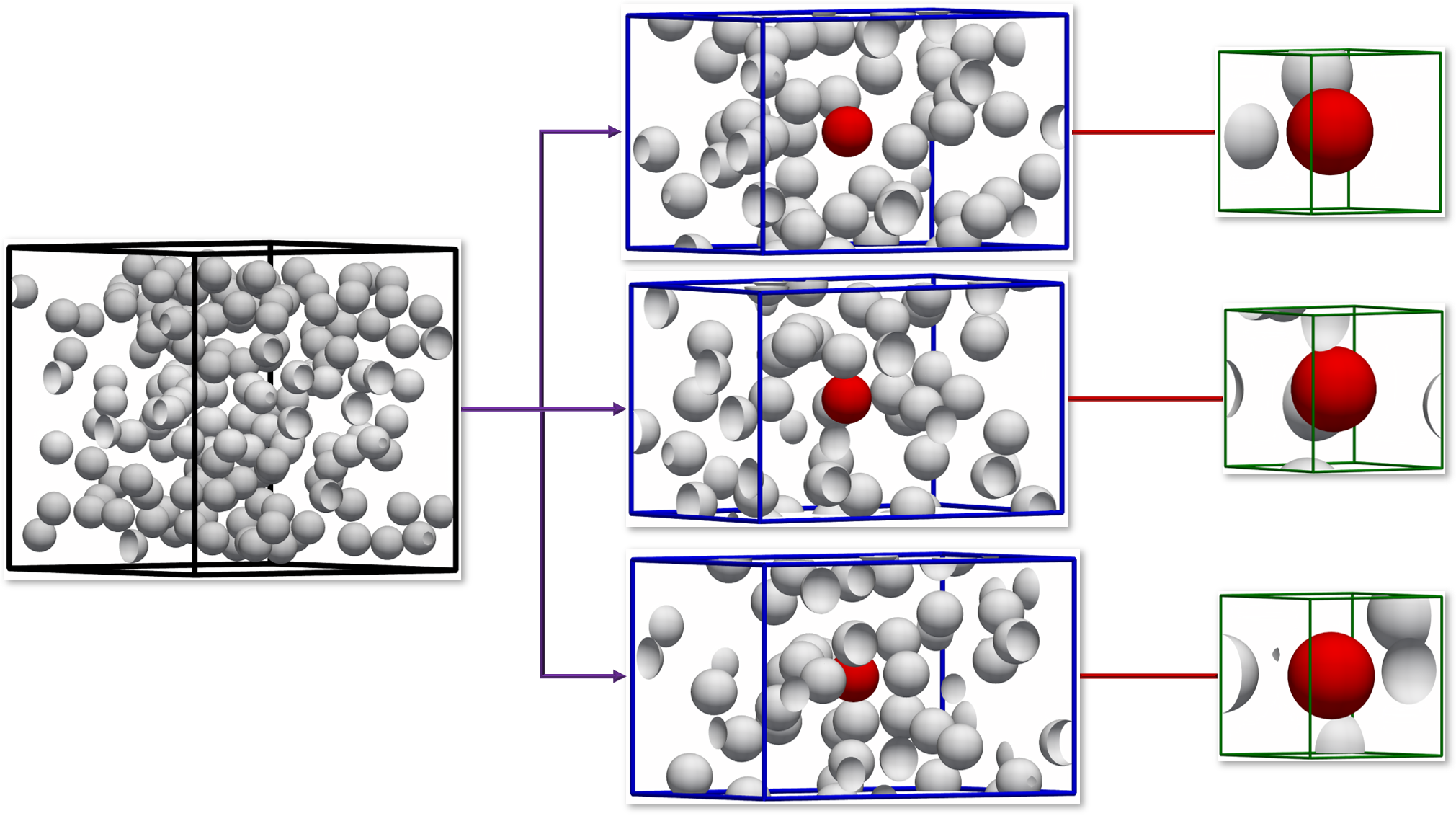}
	\caption{\textit{Model Approach}: The $3\pi$ cubic domain bounded by the black outline on the left column corresponds to a CFD realization of 11\% volume fraction. {\em{sub-domains}} of size $8 \times 5 \times 5$ are defined around every particle within the inner particle region of the cubic domain. Three such sub-domains created around three different particles as the reference particle is shown in the middle column. In each sub-domain (bounded by the blue cuboid) the reference particle is distinctly depicted by the red colored sphere. Coarse-grained mesoscale flow prediction within the sub-domain is obtained using the GAN architecture. This prediction is used to further obtain a highly resolved solution within $2 \times 2 \times 2$ {\em{attention-domain}} centered around the reference particle, which for the three sub-domains are shown on the right column bounded by green outline.}
	\label{fig:data_sampling}
\end{figure}

Our philosophy for the model presented in this paper is as follows. At a high level of spatial resolution, applying the ML algorithm over the entire inner particle region with all the  particles in it as one single data set is not computationally feasible. This motivates the use of sub-domain of size $8 \times 5 \times 5$ which was chosen based on the requirements described in section 2.1. Most importantly, with this choice of sub-domain size, the prediction of the flow around the reference particle will be well informed by all its immediate neighbors. This step also greatly increases the number of data sets to be used in the training and testing processes. Even with the reduced size of the sub-domain, the number of grid points required to obtain resolution similar to that used in PR-DNS is still significantly large. Therefore, the GAN architecture will be used to predict only a coarse-grained mesoscale flow within the sub-domain. The process of improving the accuracy of flow prediction in the immediate neighborhood of the reference particle to the desired spatial resolution is achieved with an attention mechanism. We choose the size of  the innermost domain of high resolution to be of size $2 \times 2 \times 2$ centered around the reference particle and term this the {\em attention-domain} of the particle. This choice is motivated by the fact that such an attention-domain will yield flow information in an immediate region of one diameter around the particle. The above detailed approach of the model has been schematically presented in Figure~\ref{fig:data_sampling}.

Generative Adversarial Networks, abbreviated as GANs, are generative ML models that make use of adversarial learning process to produce artificial data that mimics true data. GANs primarily comprise of two neural networks, namely, \textit{Generator} ($\mathcal{G}$) and \textit{Discriminator} ($\mathcal{D}$). The purpose of a generator is to create synthetic data as close to the true data as possible so that the two cannot be differentiated by the discriminator, while the purpose of a discriminator is to correctly distinguish the true data from the synthetic data created by the generator. From the functioning of these two networks it can be observed that they compete against each other. This adversarial nature with the discriminator helps the generator in learning the target properties of the complex flow to be modeled.

In order to mathematically describe the optimization process of the competition between the generator and the discriminator, here we adopt the following notation \cite{zero_GP}. Let $\bc$ represent a sample input data. In the present problem, the input $\bc$ corresponds to information on the relative position of all the neighbors within the sub-domain centered around the reference particle, along with the mean Reynolds number and volume fraction of the case. Now let $\bxi$ be the true DNS velocity and pressure fields obtained around the particles within the sub-domain. Let $\beta$ be the synthetic velocity and pressure fields generated around the particles within the sub-domain by the generator $\mathcal{G}$. Thus, for the input condition $\bc$, the output of the generator is $\beta$, while the true target output is $\bxi$. We now define the following probability distributions:
\begin{center}
	\begin{tabular}{c@{\hskip 0.2in}l}
		$P_{c}$: & is the distribution of all possible input condition, which in the present\\
		& case corresponds to all possible distribution of neighboring particles.\\
		$P_{\xi}$: & is the distribution of DNS velocity and pressure fields corresponding \\
		& to all possible values of input. \\
		$P_{\eta}$: & is the distribution of generated synthetic velocity and pressure \\
		& fields corresponding to all possible values of input.\\
	\end{tabular}
\end{center}
With this definition, each data set consists of a condition $\bc \in P_c$ and the corresponding flow field $\bxi \in P_\xi$. The generator of the GAN will be trained such that given the condition $\bc$ it should generate the synthetic flow $\beta \in P_\eta$. The goal is then for $\beta$ to as closely mimic $\bxi$ as possible. Note that for a given $\bc \in P_c$, the corresponding true flow $\bxi$ and synthetic flow $\beta$ will also depend on Reynolds number and volume fraction.  

\begin{figure}
	\centering
	\includegraphics[height=0.4\textheight,keepaspectratio=true]{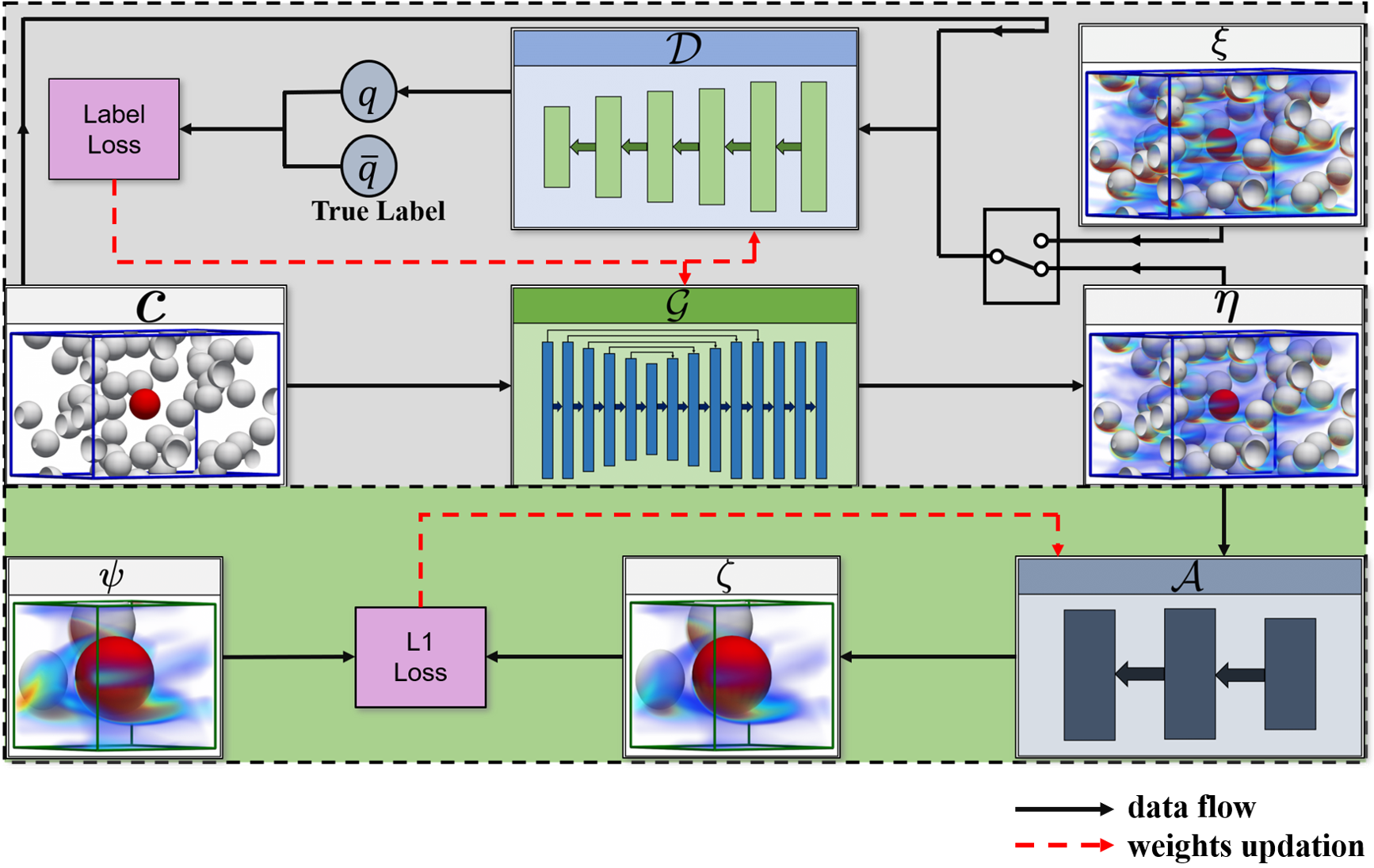}
	\caption{Model Framework.}
	\label{fig:gan_frame}
\end{figure}

The working of GAN algorithm has been pictorially represented in Figure~\ref{fig:gan_frame} \cite{goodfellow2016nips,Mehta_2019}. The output of the discriminator ($\mathcal{D}$) is a scalar value, $q$, between 0 and 1, with 0 being the expected output, $\bar{q}$, if the discriminator identifies its input to be synthetic and 1 being the expected output if it identifies the input as real. Thus, as shown in the figure, $\mathcal{D}$ tries to optimize its weights to produce a value close to one if its input is $\bxi \in P_\xi$ and a value near zero for $\beta \in P_\eta$ as input. On the contrary, $\mathcal{G}$ optimizes its coefficients in such a way that its output $\beta$ will fool the discriminator to yield $\mathcal{D}(\beta)$ closer to one. This entire process can be represented using objective function ($V(\mathcal{D},
\mathcal{G})$) given by
\begin{equation}\label{org_OF}
\min_{\mathcal{G}} \max_{\mathcal{D}} V(\mathcal{D},
\mathcal{G}) = \mathbb{E}_{\scriptsize{\bxi} \in P_{\xi}} \left[\log \mathcal{D}(\bxi)\right]+ \mathbb{E}_{\scriptsize{\bc} \in P_{c}}\left[\log (1-\mathcal{D}(\mathcal{G}(\bc)))\right] \, ,
\end{equation}
where $\mathbb{E}$ stands for expectation. The first term on the right is the contribution to the objective function for a true field $\bxi \in P_\xi$ as input, which the discriminator tries to maximize. The second term on the right is the contribution to the objective function for a synthetic field $\beta \in P_\eta$ as input, which the discriminator tries to maximize, while the generator tries to minimize. The adversarial networks $\mathcal{D}$ and $\mathcal{G}$ are pitted against each other and upon iteration, they approach what is known as Nash equilibrium in game theory. The training process thus theoretically achieves convergence when discriminator cannot differentiate a true sample from a generator generated synthetic sample, i.e., $\mathcal{D}(\bxi) = \mathcal{D}(\beta) = 0.5$. This also implies that the generator has efficiently learnt the characteristics of the true flow given the condition $\bc$.

However, \citet{mescheder2018training} show that a GAN training with objective function (\ref{org_OF}) does not always converge for data whose distribution is not absolutely continuous. They also show that instance noise and zero-centered gradient penalty methods stabilize GAN training even for discontinuous data distributions. \citet{zero_GP} proposed a zero-centered gradient penalty method that along with local convergence improves generalization and prevents gradient explosion in the discriminator. Gradient explosion in a discriminator leads to mode collapse in a generator. Mode collapse can be described as a process where the generator is not sufficiently sensitive to its input information \citep{che2016mode, NIPS2017_6923, salimans2016improved}. Therefore, the GAN methodology presented in this paper was trained using the following objective function \citep{zero_GP}:
\begin{equation}\label{D_OF}
\min_{\mathcal{G}} \max_{\mathcal{D}} V(\mathcal{D},\mathcal{G}) = \mathbb{E}_{\scriptsize{\bxi} \in P_{\xi}} \left[\log \mathcal{D}({\bxi})\right]+\mathbb{E}_{\scriptsize{\bc} \in P_{c}}\left[\log (1-\mathcal{D}(\mathcal{G}(\bc)))\right]-\lambda \mathbb{E}_{\scriptsize{\widetilde{\bxi}}}\left[ {\|(\nabla \mathcal{D})_{\scriptsize{\widetilde{\bxi}}}\|}_{2}^{2} \right] \, ,
\end{equation}
where $\widetilde{\bxi} = \alpha\,\bxi+(1-\alpha)\beta$ and $\alpha \sim \mathcal{U}(0,1)$, which represents that $\alpha$ is a random number from a uniform distribution on the interval $[0,1)$. In equation \eqref{D_OF} ${\|\,\cdot\,\|}_{2}$ stands for $L2$-norm, $(\nabla \mathcal{D})_{\scriptsize{\widetilde{\bxi}}\def\bbeta {\pmb {$\beta$}}}$ is the gradient of $\mathcal{D}(\widetilde{\bxi})$ with respect to $\widetilde{\bxi}$ and parameter $\lambda$ is a measure of discriminative power of $\mathcal{D}$. A $\lambda$ value of 0 makes the $\mathcal{D}$ only focus on maximizing its discriminative power. $\mathcal{D}$ has maximum generalization capability and no discriminative power if $\lambda$ approaches infinity. It can be observed that the original GAN objective function (\ref{org_OF}) can be obtained for $\lambda = 0$. Details of the loss functions for generator and discriminator, which are based on \eqref{D_OF}, are elaborated in \ref{appA}.  

\subsection{GAN Architecture}
The generator and the discriminator used in the current work are built using 3D convolutional layers, whose schematic is shown in Figure~\ref{fig:gan_frame}. As shown in the figure, the generator comprises of 14 convolutional layers and the discriminator contains 5 convolutional layers which are followed by a single linear layer. The architectural details of these neural networks have been detailed in Tables \ref{tab:g_arch} \& \ref{tab:d_arch} respectively in \ref{appA}.

The input to the generator is the 3D \textit{Indicator Function} ($I_{f}$)
\begin{equation} \label{eq3.3}
I_f(x,y,z) = \left\{
\begin{array}{ll}
0, & if\;(x,y,z)\;is\;inside\; a\;particle.\\[2pt]
1, & if\;(x,y,z)\;is\;outside\; a\;particle.
\end{array} \right.
\end{equation}  
which clearly demarcates the regions of the sub-domain occupied by the reference particle (which is located at the center of the sub-domain) and its neighbors. Note that with the above definition of indicator function, portions of neighboring particles that happen to lie within the sub-domain are taken into account, even when that neighbor's center happen to fall outside the sub-domain. In addition, the input information includes the Reynolds number of the flow. The local volume fraction information has already been provided in terms of the indicator function. The indicator function and the Reynolds number information can be combined together by redefining the indicator function to equal the Reynolds number at every point within the fluid region. The redefined indicator function is then the condition $\bc$ that serves as input to the generator. The output of the generator are the velocity and pressure fields within the sub-domain in regions occupied by the fluid. 

The input to the discriminator is both the redefined indicator function $\bc$ and also the three components of velocity and pressure within the sub-domain in the region outside the particles. As shown in Figure~\ref{fig:gan_frame}, this 3D velocity and pressure inputs to $\mathcal{D}$ can be either in the form of true data $\bxi$ or synthetic data $\beta$ outputted by the generator.

The 3D ML inputs and output fields described above have to be discretized and prepared in terms of 3D voxels. The computer memory and computational time required for training and testing of the GAN scale as the number of voxels. A finely discretized input and output fields may be desirable for accurate representation of the flow, but available memory on a GPU, where the GAN is implemented, places important restriction on the number of voxels. In this work, we limit the number of equi-spaced grid points along each direction to be 64, and thus, $(64)^3$ voxels are used to represent the indicator function and the velocity and pressure fields within the sub-domain. Since the sub-domain is of size $8 \times 5 \times 5$, the resolution is slightly lower along the streamwise direction than along the transverse direction. Clearly the resolution of the sub-domain with $(64)^3$ voxles is somewhat lower than the resolution of the DNS. However, the purpose of predicting the flow within the sub-domain is to only obtain a coarse-grained solution around the reference particle taking into account a somewhat larger neighborhood of particles. As we will see below, once the coarse-grained solution is obtained within the sub-domain, a much finer solution will be obtained in the immediate neighborhood of the reference particle through attention mechanism.

\subsection{Attention Mechanism CNN}
The attention-domain of a sample corresponds to the innermost $16 \times 26 \times 26$ voxels of the generator's output. Increasing the resolution to $(64)^3$ within the attention-domain will provide high enough resolution that is comparable to DNS. This can be achieved by a simple interpolation from the coarse grid to a finer grid. However, such an interpolation will not improve upon the accuracy of the GAN solution that was obtained on the coarse grid.  

Here a more refined solution in the attention-domain is obtained using a simple three layered convolutional neural network $\mathcal{A}$ (schematically shown in figure~\ref{fig:gan_frame} and architectural details are given by \ref{tab:cnn_arch}), whose goal is to improve the accuracy of the solution within the attention-domain. As shown in figure~\ref{fig:gan_frame}, the input to $\mathcal{A}$ is the generator's output $\beta$ for an input condition $\bc$. The objective of $\mathcal{A}$ is to make its output $\bzeta$ as close as possible to $\bpsi$, which is the corresponding actual DNS solution on $(64)^3$ grid points in that attention-domain. The synthetic flow prediction $\beta$ of GAN is first interpolated onto a $(16)^3$ voxel spanning an inner cube of side 2 centered around the reference particle. This information is passed as input to $\mathcal{A}$ and the expected output is $u^{\prime},\, v^{\prime},\, w^{\prime},\, p^{\prime}$ fields in the $2 \times 2 \times 2$ attention-domain around the reference particle discretized by $(64)^3$ uniformly distributed voxels. The functioning of $\mathcal{A}$ can be perceived as super-resolution in the inner $2 \times 2 \times 2$ attention-domain. 

Neural network training approach adopted for this model is that all the three networks, namely Generator, Discriminator and Attention Mechanism CNN, are trained together. This mode of training indicates that $\beta$ which is passed as input to $\mathcal{A}$ is varying throughout the entire training process. 
It has to be mentioned that this is not the only approach that can be used in implementing the attention mechanism. For example, training of  $\mathcal{A}$ can be started after the completion of GAN training. This will ensure that $\mathcal{A}$ is always trained using the optimum version of $\beta$. But this may not guarantee optimal value of $\bzeta$, which is the final desired output from the attention mechanism. Here we train all three CNNs together in order to obtain an optimal $\bzeta$.  Similar to the existence of different training approaches, there also exists a choice in network architecture for the attention mechanism. A simple three layer CNN was used in this work, however, a separate GAN architecture can also be implemented for this purpose \cite{tempoGAN,subramaniam2020turbulence}.           

It should be pointed out that the above described two step process of first predicting the coarse grained flow in the sub-domain using a GAN followed by the second step of improving the accuracy with a fine-grained solution in the attention-domain with a CNN is essential. It is not possible to directly go to the fine-grained solution in the attention-domain, since the smaller domain will not contain information on many of the neighbors of the reference particle. As a summary, the input and output of the three networks $\mathcal{G}$, $\mathcal{D}$ and $\mathcal{A}$ have been concisely presented in Table~\ref{tab:tab2}.
\begin{table}
	\begin{center}
		\def~{\hphantom{0}}
		\begin{tabular}{|c@{\hskip 0.5in}c@{\hskip 0.5in}c|}
			\hline
			Network & Input & Output\\
			\hline
			$\mathcal{G}$ & $\bc$ & $\beta = $ {sub-domain} $(u^{\prime},v^{\prime},w^{\prime},p^{\prime})$\\[3mm]
			$\mathcal{D}$ & $\bc \cup \{ \bxi$  or $\beta \}$ & $q = $ Prediction value ($\in \left[0,1\right]$) \\[3mm]
			$\mathcal{A}$ & $\beta$ & {$\bzeta =$ attention-domain} ($u^{\prime},v^{\prime},w^{\prime},p^{\prime}$)\\
			\hline 
		\end{tabular}
		\caption{Inputs and outputs of neural networks used in this work} \label{tab:tab2}
	\end{center}
\end{table}

\subsection{Data Augmentation With Symmetry}
Since the mean flow of each sub-domain is along the $x$-direction, this is the only preferred direction of the problem. We desire the GAN networks $\mathcal{G}$, $\mathcal{D}$, and $\mathcal{A}$ to satisfy these symmetries. In other words, given a condition $\bc$ as input if the generator $\mathcal{G}$ outputs the synthetic field $\beta$, then if a rotated (or reflected) condition $\bc'$ were to be provided then the generator $\mathcal{G}$ must output a synthetic field $\beta'$ that is a correspondingly rotated (or reflected) copy of $\beta$. An unbounded or a cylindrical domain presents axisymmetry and allows continuous rotation about the $x$-axis. However, the present cuboid domain allows only discrete rotations of $90^{\circ}$, $180^{\circ}$, $270^{\circ}$ and reflections about the $y$ and $z$ axis. 

\begin{center}
	\begin{figure}
		\includegraphics[width=1.\textwidth,keepaspectratio=true]{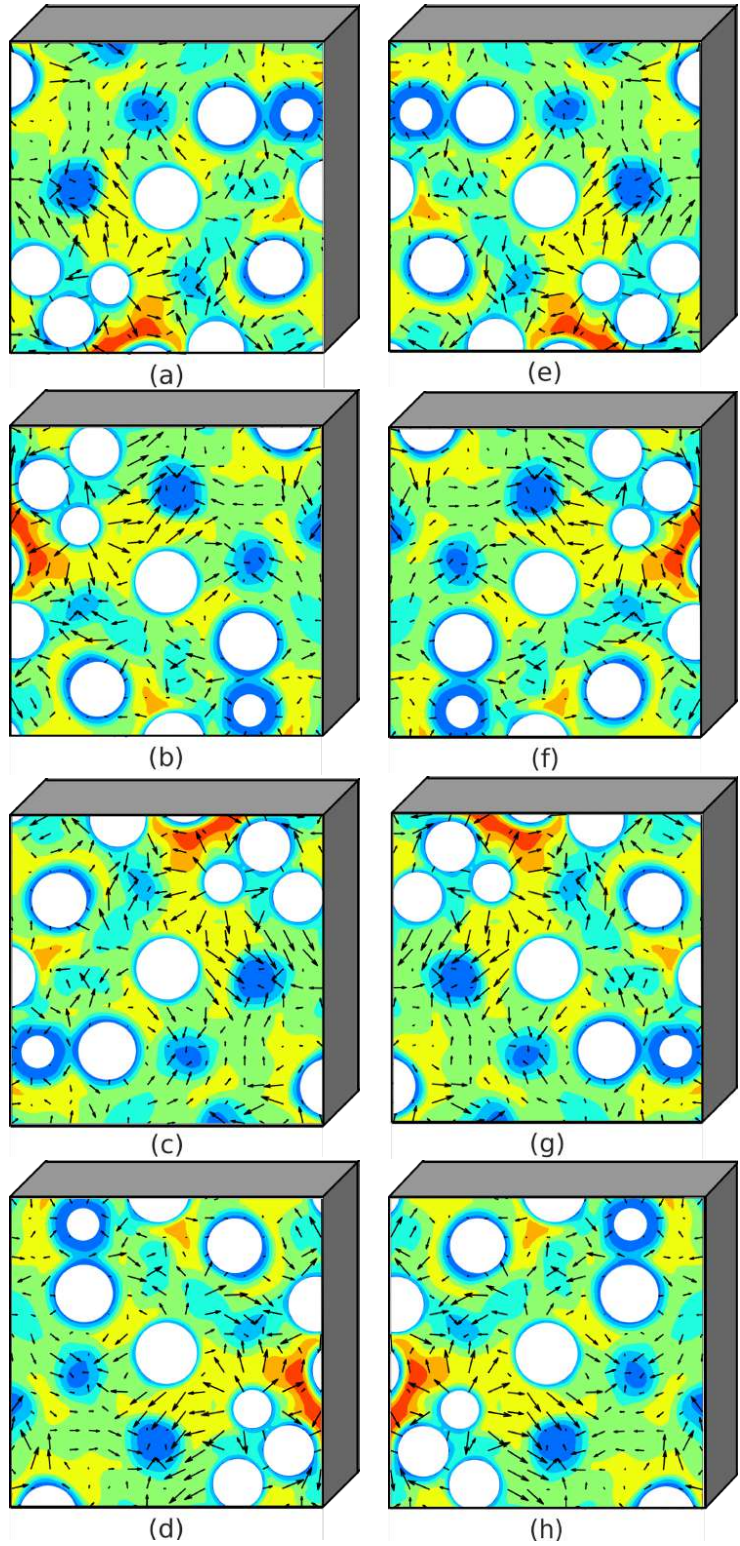}
		\caption{Data Augmentation using rotations and corresponding reflections shown for a \textit{y-z} plane of a sample from case  $Re = 172.96, \, \phi=0.11$.}
		\label{fig:data_aug}
	\end{figure}
\end{center}

The symmetry requirement of the GAN networks $\mathcal{G}$, $\mathcal{D}$, and $\mathcal{A}$ is weakly enforced by data augmentation. For every training data represented by the redefined indicator function $\bc$ and the associated velocity/pressure fields which form the corresponding true data $\bxi$, eight equivalent conditions and associated flow fields can be generated by applying the discrete rotations and reflections. A schematic of this data augmentation is shown in figure \ref{fig:data_aug}. Only one $y-z$ plane of the sub-domain is shown where the intersection of the plane with the particles appears as circles of varying radius, the color contours show local value of streamwise velocity, and the arrows in the plane correspond to the in-plane velocity vector at that location. In the figure, frame (a) corresponds to the original data as obtained from PR-DNS and frames (b)-(d) are discrete rotations with an increment of $90^{\circ}$ about $x$-axis. Frames (e)-(h) are reflections of (a)-(d) about the $z$ coordinate. As expected a rotation/reflection leads to a corresponding rotation/reflection of independent scalar quantities like locations of particles, streamwise  velocity, and pressure. On the other hand, rotation/reflection causes not only a change in the location of in-plane velocity vector but also a change in its direction. This change in quantities due to rotation/reflection can be observed in figure~\ref{fig:data_aug}. Thus, the imposition of symmetry results in eight fold increase in the training and testing data.
\section{Results and Discussion}
The neural networks were trained separately for every case mentioned in Table \ref{tab:cases}. In each case, segregation of the entire dataset into {\em training data} and {\em testing data} was achieved by leaving one realization as testing data and all other realizations are used as training data. This type of data segregation ensures that there is no overlap between training and testing data, thereby, leading to a genuine network performance evaluation on test dataset. However, it is not clear a-priori which realization should serve as the testing data with others serving as training data. 

It should be also mentioned that in a given case (specified by Reynolds number and particle volume fraction) there does not exist a small and practical set of parameters that can be used to uniquely classify the arrangement of neighboring particles around the reference particle. This suggests that the level of similarity between testing and training datasets cannot be determined. Hence, to properly establish the uncertainty in the performance of the GAN networks we must quantify error by selecting every single realization as \textit{testing realization} one at a time. I.e., in the example of case 1, where there are 10 realizations (see Table~\ref{tab:cases}), training, testing and error analysis will be repeated 10 times, and in each repetition a different realization will be the testing data, while all others form the training data. The overall error to be reported is an average of these. In each of the Reynolds number, volume fraction cases considered, the number of training data $N_{tr}$ and the number of testing data $N_{te}$ are listed in Table~\ref{tab:cases}.

Furthermore, it is very well known that generalization capabilities of a neural network increase with the amount of new training data given to it. However, the number of training samples required to achieve desired performance on test data is problem-specific. To evaluate the rate of increase of test performance with increase in training samples the networks are trained on different-sized subsets of the training dataset. This will help in getting an estimate of the required number of training samples to achieve a desired level of test performance. Therefore, the number of training data used for training the GAN and attention-CNN was varied from $N_{sub,tr} = N_{tr}/6$ to $N_{tr}$. These subsets are chosen randomly from the total training dataset. Entire test dataset is used in evaluating test performance of these networks that are trained on different sized subsets to ensure a valid and consistent measure of increase in generalization with increase in training data. 

\subsection{Sample Display}
\begin{figure}
	\begin{center}
		\begin{subfigure}{1\textwidth}
			\includegraphics[width=1\linewidth,keepaspectratio=true]{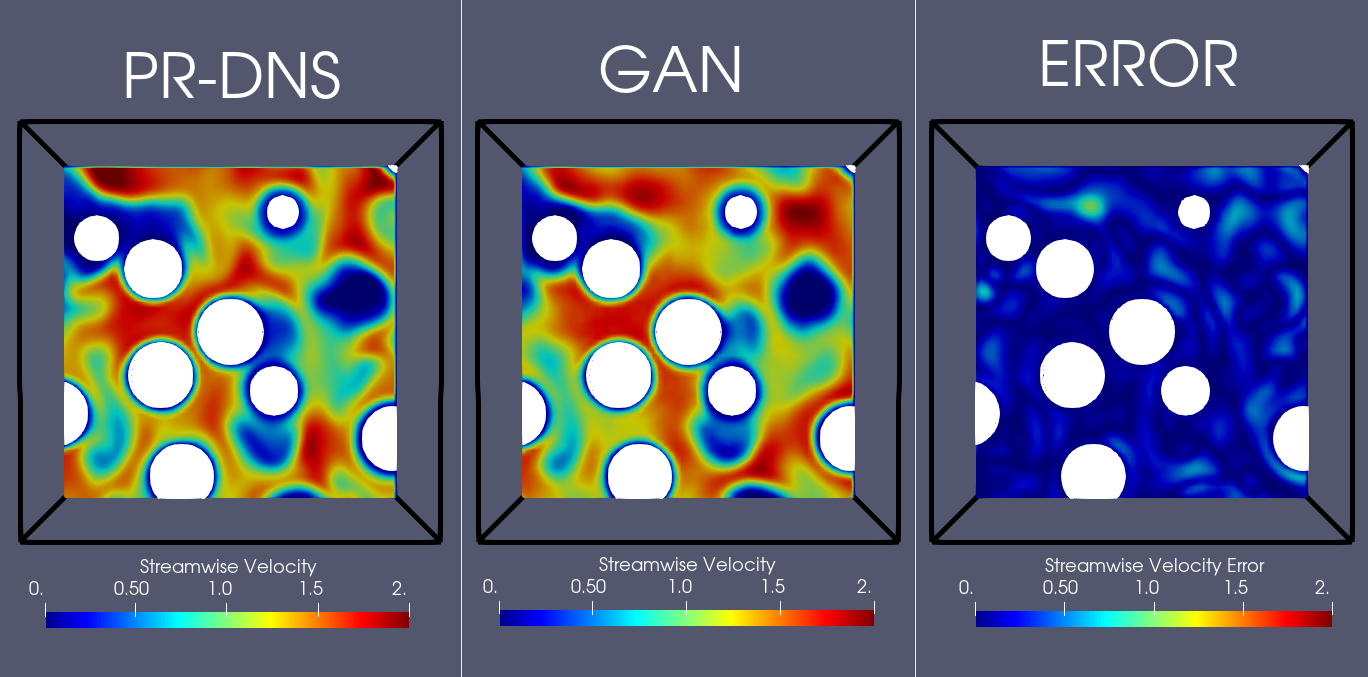}
			\caption{}
		\end{subfigure}
		\begin{subfigure}{1\textwidth}
			\includegraphics[width=1\linewidth,keepaspectratio=true]{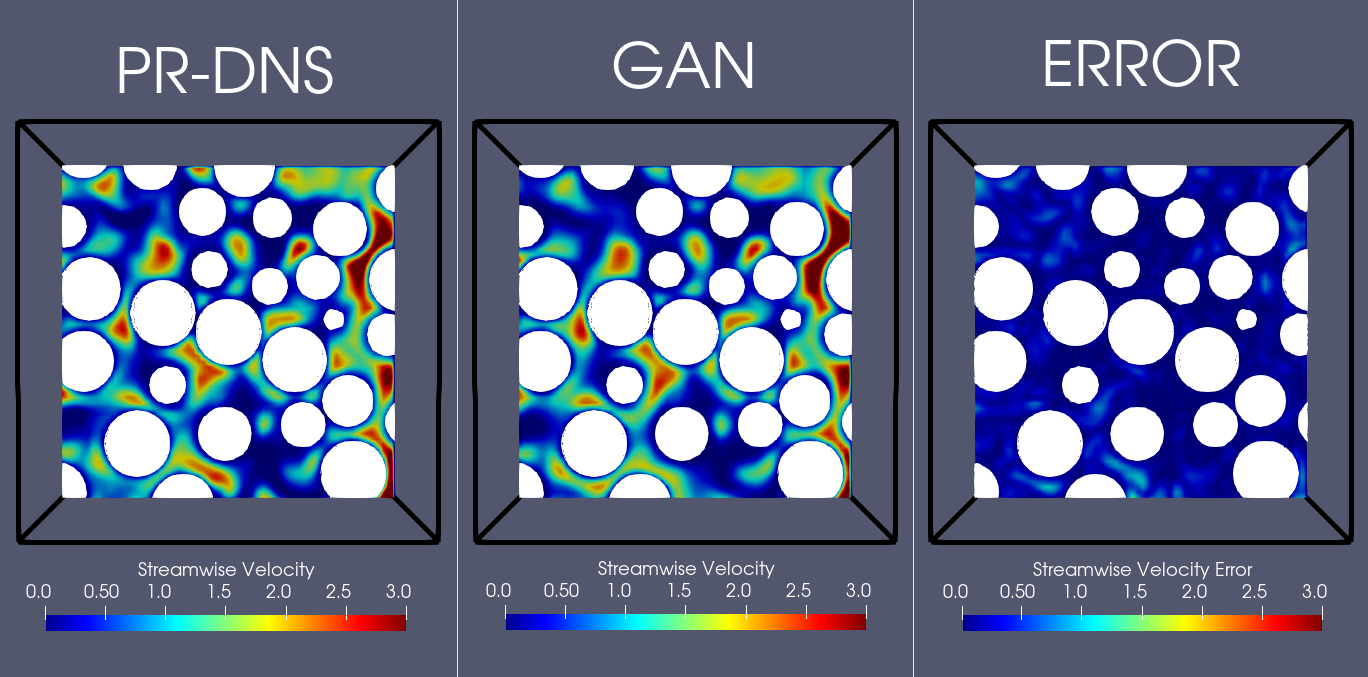}
			\caption{}
		\end{subfigure}
		\caption{Comparison of GAN model output with PR-DNS using streamwise velocity for a sample from (a) Case 3 and (b) Case 8.}
		\label{fig:smpl_disp1}
	\end{center}
\end{figure}

\begin{figure}
	\begin{center}
		\begin{subfigure}{1\textwidth}
			\includegraphics[width=1\linewidth,keepaspectratio=true]{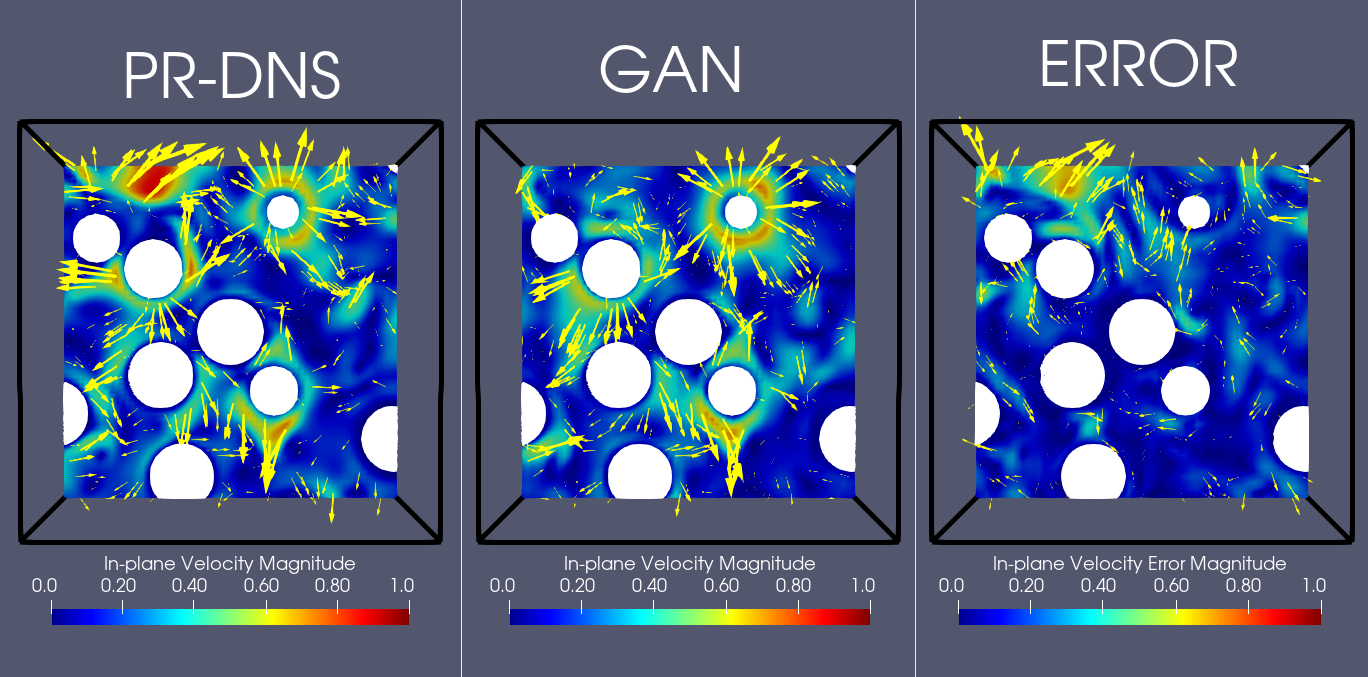}
			\caption{}
		\end{subfigure}
		\begin{subfigure}{1\textwidth}
			\includegraphics[width=1\linewidth,keepaspectratio=true]{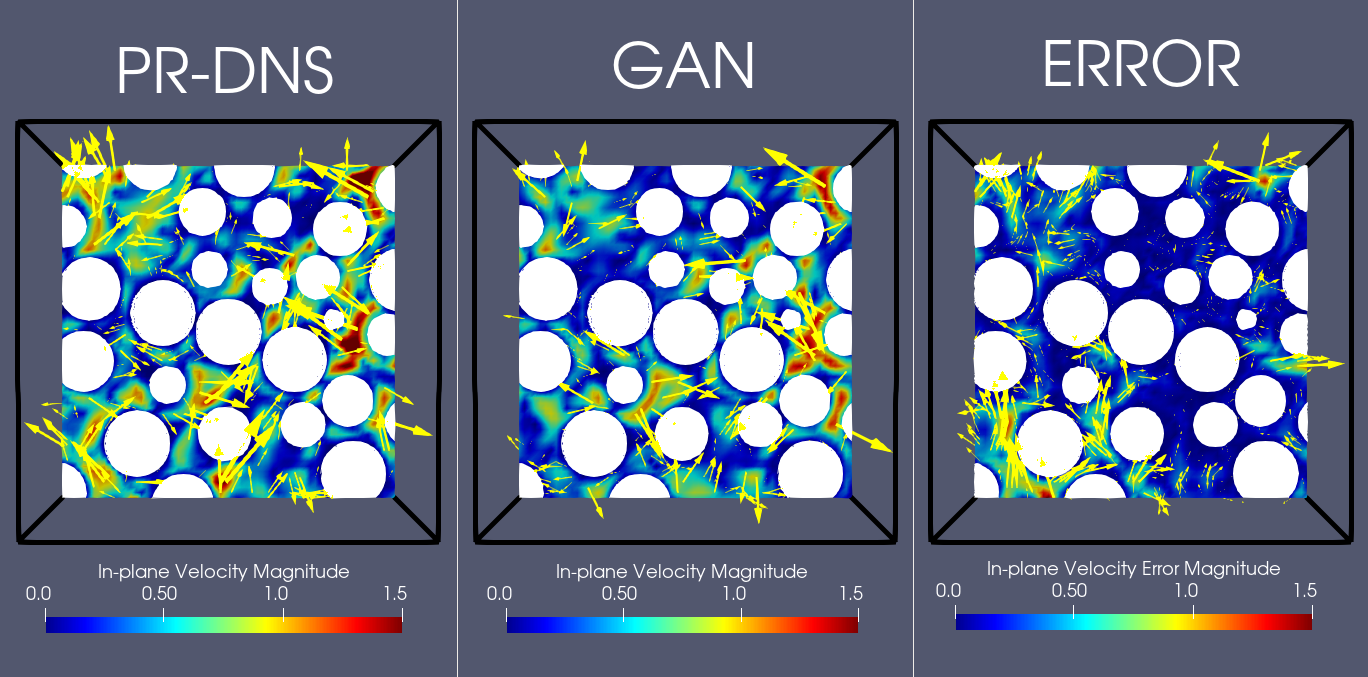}
			\caption{}
		\end{subfigure}
		\caption{Comparison of GAN model output with PR-DNS using in-plane velocity for a sample from (a) Case 3 and (b) Case 8.}
		\label{fig:smpl_disp2}
	\end{center}
\end{figure}

\begin{figure}
	\begin{center}
		\begin{subfigure}{1\textwidth}
			\includegraphics[width=1\linewidth,keepaspectratio=true]{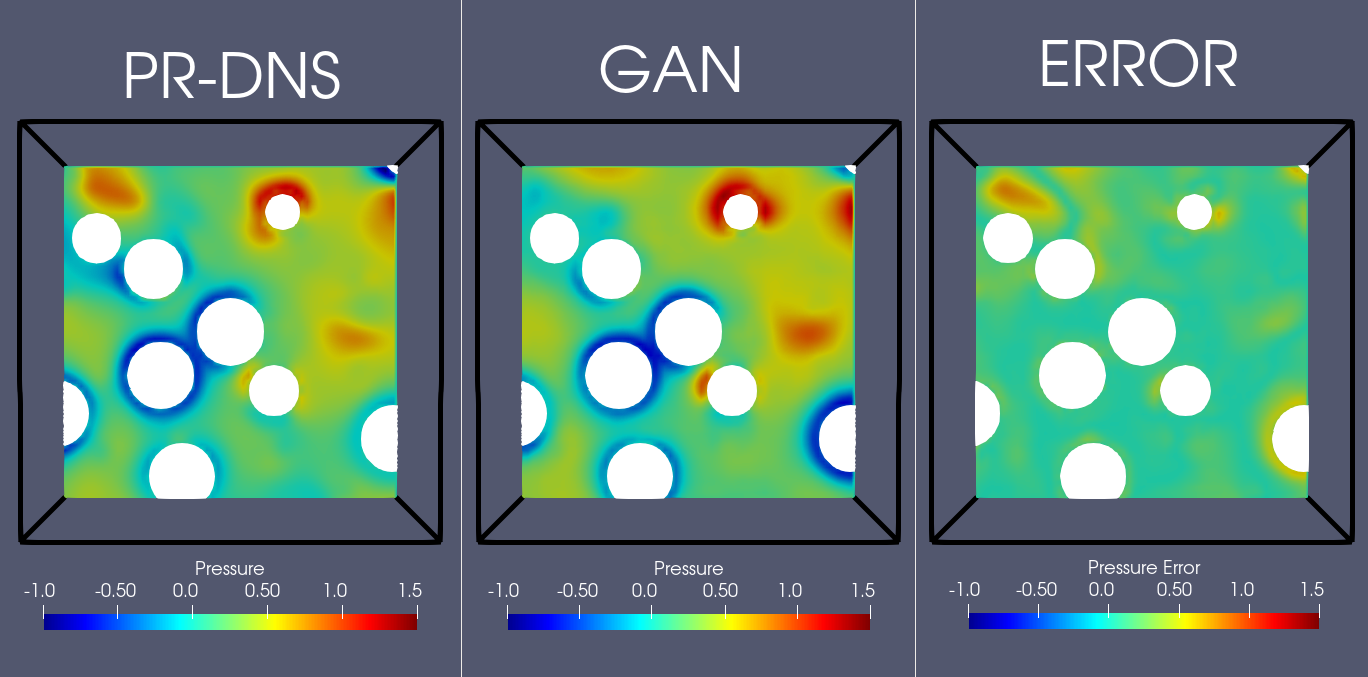}
			\caption{}
		\end{subfigure}
		\begin{subfigure}{1\textwidth}
			\includegraphics[width=1\linewidth,keepaspectratio=true]{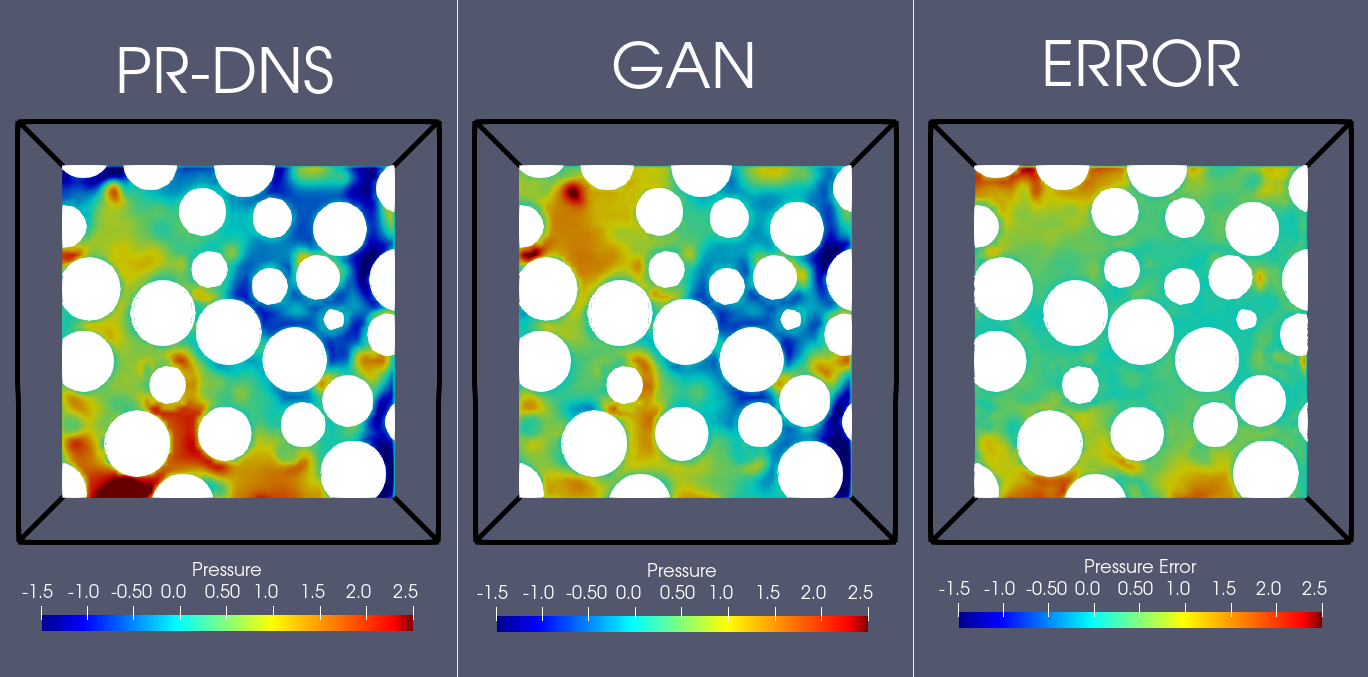}
			\caption{}
		\end{subfigure}
		\caption{Comparison of GAN model output with PR-DNS using pressure for a sample from (a) Case 3 and (b) Case 8.}
		\label{fig:smpl_disp3}
	\end{center}
\end{figure}

To provide a visual image of flow field generated by the GAN model, sample results from case 3 and case 8 are presented here in Figures ~\ref{fig:smpl_disp1}, \ref{fig:smpl_disp2} and \ref{fig:smpl_disp3}. These samples have been randomly chosen from the model's testing dataset when the model is trained on the entire training dataset, i.e., $N_{sub,tr} = N_{tr}$. GAN architecture output $\beta$ is of particular interest as it produces flow field information in the sub-domain by taking location of particles and volume-averaged Reynolds number as input. For appropriate visualization and compact presentation of the output, velocity and pressure have been transformed back to their original scales, i.e., $u^{\prime},\, v^{\prime},\, w^{\prime}$, and $p^\prime$ are transformed to $u,\,v,\,w$ and $p$. Central \textit{y-z} plane of the sub-domain for both samples have been shown in the figures. In Figure~\ref{fig:smpl_disp1} contours of true PR-DNS streamwise velocity and synthetic streamwise velocity from GAN are shown (streamwise velocity is the normal velocity on this plane) along with contours of the difference as a measure of error. Figure~\ref{fig:smpl_disp2} shows plots of in-plane velocity vector from PR-DNS and GAN along with the error and Figure~\ref{fig:smpl_disp3}  shows the corresponding pressure contours. Videos of several \textit{y-z} planes spanning over the streamwise length of sub-domain for these two samples can be seen in supplementary data. It can be seen from the figures and the corresponding videos that the GAN is able to capture the velocity and pressure fields reasonably accurately. In particular, it can be noted that the errors are quite low in the region immediately surrounding the central reference particle. The higher errors mostly occur closer to the boundaries of the sub-domain. This is to be expected, since for the central reference particle, its entire neighborhood is contained within the sub-domain so that the local flow is well predicted by GAN. In contrast, as we approach the boundaries of the sub-domain, the arrangement of particles that are outside the sub-domain are unknown as they are not contained in the condition $\bc$. So as a result GAN prediction goes down in accuracy as we move away from the reference particle towards the sub-domain boundaries. Fortunately, in the first place, the idea of sub-domain was to get accurate flow field around the reference particle, which is being achieved to a very good extent. This will be quantified below in succeeding results.

Flow information ($u,\, v, \, w,\, p$) for all grid points that fall inside of particles have been zeroed out in every PR-DNS sample. 
The generator which is trained using the loss function given by (\ref{g_loss}) takes only points outside of particles into consideration through $L1$-norm. Thus, flow variable data generated by $\mathcal{G}$ for grid points inside of particles is not optimized through $L1$-norm but may be learned from adversarial learning against $\mathcal{D}$. From Figures~\ref{fig:smpl_disp1}, \ref{fig:smpl_disp2}, \ref{fig:smpl_disp3} and corresponding videos, it can be seen that the model fairly predicts velocities and pressure outside of particles for both cases.  
      
\subsection{Performance Evaluation Metric}
Coefficient of determination, $R^2$, is used to evaluate test performance of neural networks $\mathcal{G}$ and $\mathcal{A}$. $R^2$ is defined as
\begin{equation} \label{eq8}
R^{2}_{u^{\prime},\,box} = 1-\dfrac{\sum\limits_{i=1}^{N_{te}}\sum\limits_{box}(u_{DNS}^{\prime}-u_{network}^{\prime})^{2}I_{f}}{\sum\limits_{i=1}^{N_{te}}\sum\limits_{box}(u_{DNS}^{\prime})^{2}I_{f}} \, ,
\end{equation}
where, $N_{te}$ is the number of test samples. Here the pair ($box,\, network$) corresponds to either ($sub-domain, \, \mathcal{G}$) or ($attention-domain,\,\mathcal{A}$). Therefore, if the box corresponds to sub-domain, then $u^{\prime}_{DNS}$ represents normalized perturbation streamwise velocity ($u^{\prime}$) obtained from PR-DNS within the $8 \times 5 \times 5$ sub-domain discretized on $(64)^3$ voxels. $u^{\prime}_{network}$ then corresponds to the synthetic solution of the network $\mathcal{G}$. Similarly, if box is attention-domain, then  $u^{\prime}_{DNS}$ is DNS solution within the $2 \times 2 \times 2$ region discretized on $(64)^{3}$ voxels and $u^{\prime}_{network}$ is the predicted solution of the attention network $\mathcal{A}$. An $R^2_{u^{\prime},\,box}$ value of one is equivalent to a perfect model that exactly predicts the DNS solution. Hence, the performance of GAN and the Attention-CNN can be measured by how close their $R^2_{u^{\prime}, box}$ value is to unity. Similar performance metrics are defined for network's transverse velocities and pressure components, which are given by $R^2_{v^{\prime},\,box}$ and $R^2_{p^{\prime},\,box}$ respectively.

\subsection{Generator Performance}
\begin{figure}
	\centering
	\begin{subfigure}{0.49\textwidth}
		\includegraphics[width=1\linewidth,keepaspectratio=true]{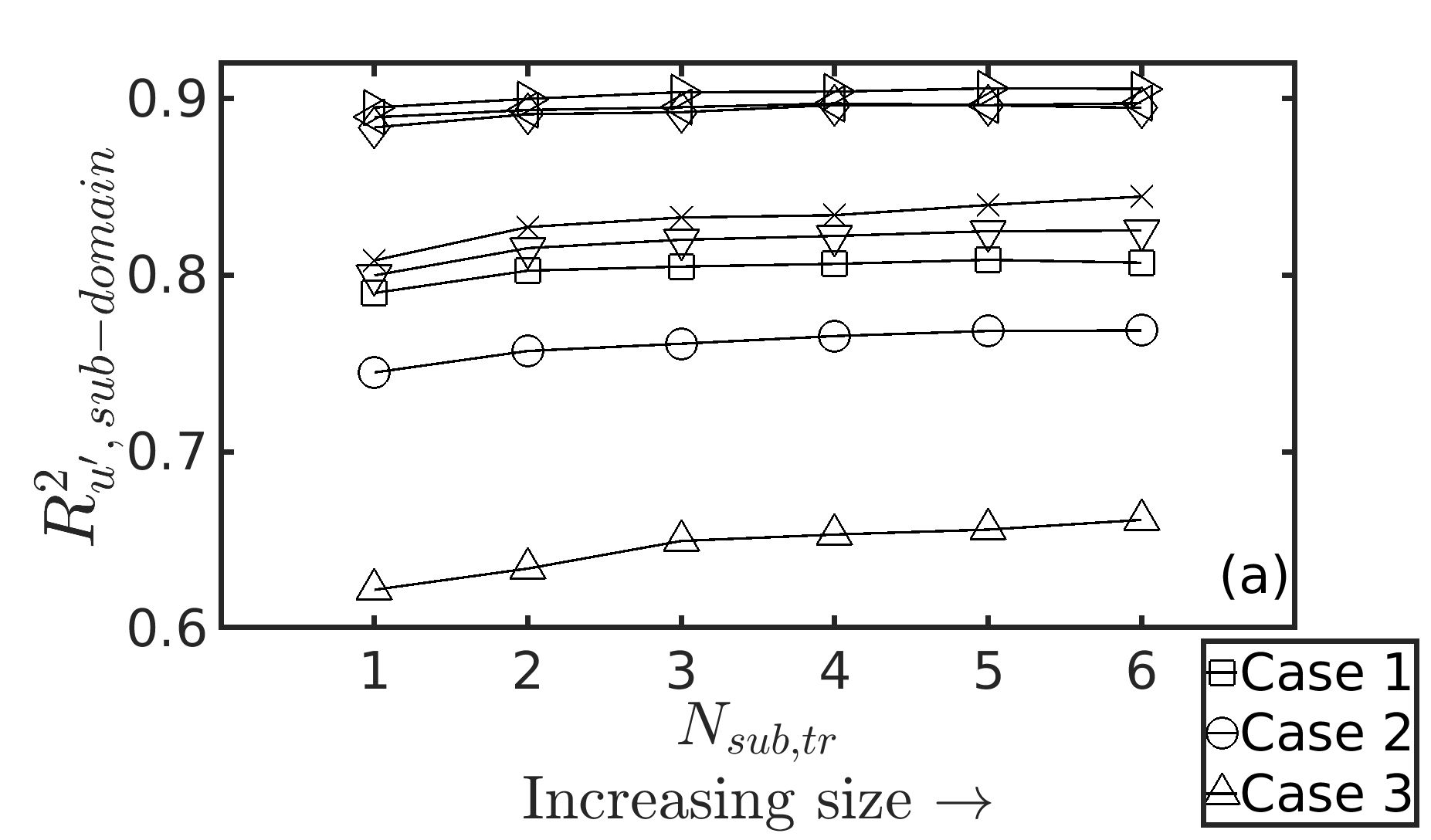}
	\end{subfigure}
	\begin{subfigure}{0.49\textwidth}
		\includegraphics[width=1\linewidth,keepaspectratio=true]{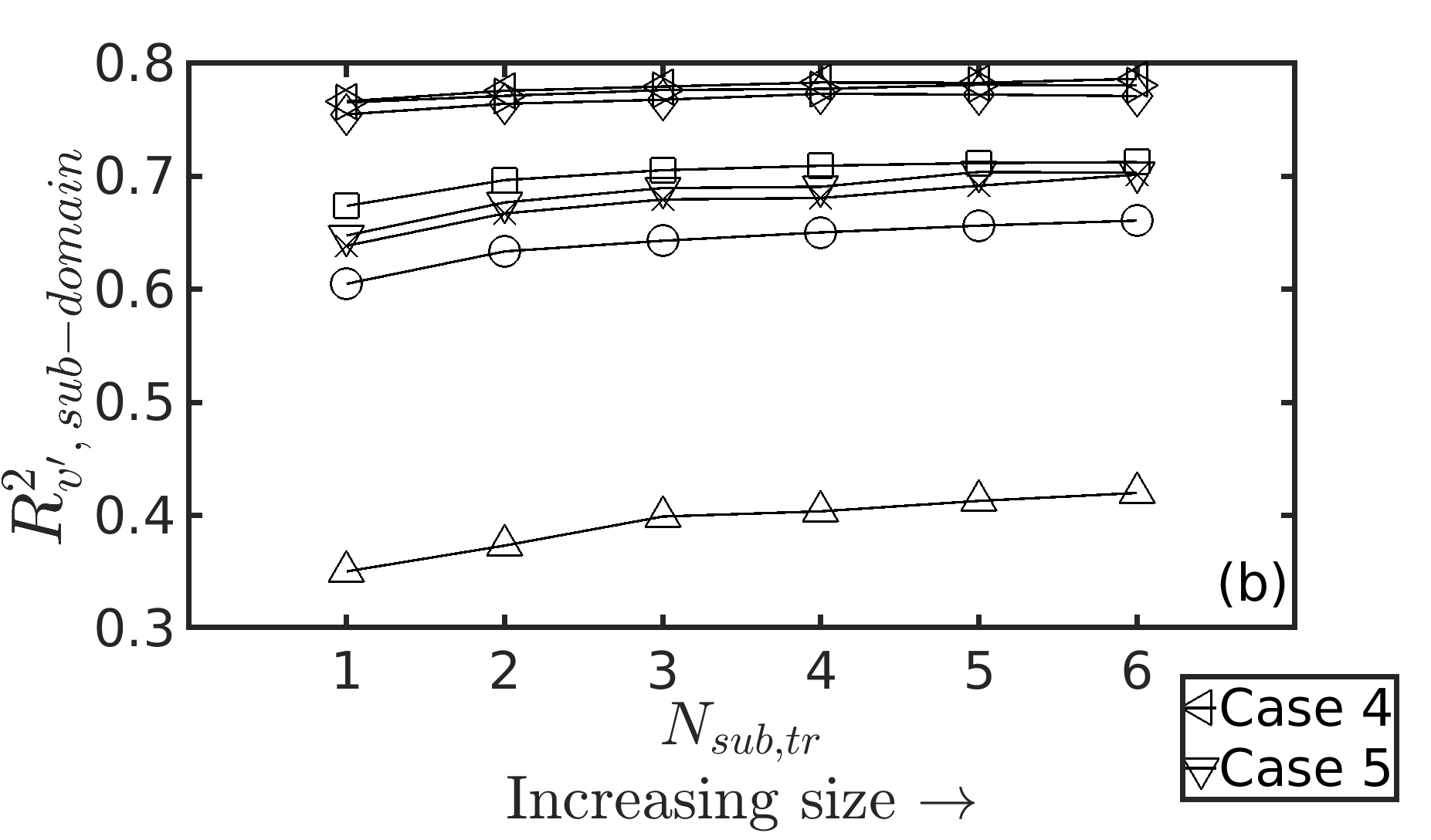}
	\end{subfigure}
	\begin{subfigure}{0.49\textwidth}
		\includegraphics[width=1\linewidth,keepaspectratio=true]{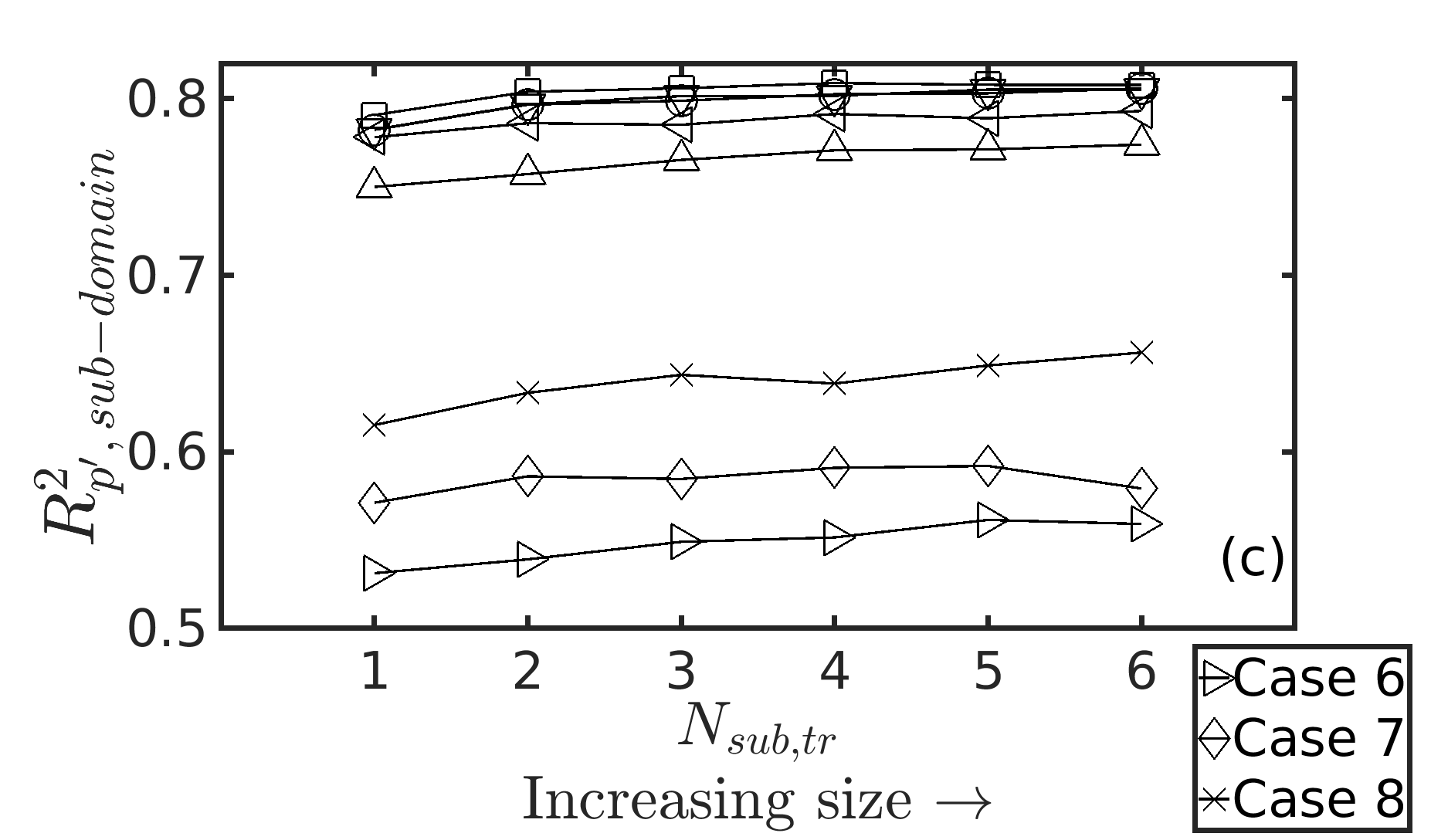}
	\end{subfigure}
	\caption{Performance evaluation of Generator ($\mathcal{G}$) for (a) Streamwise Velocity, (b) Transverse Velocities, and (c) Pressure}
	\label{fig:gen_perf}
\end{figure}

As explained earlier in this section, performance within a sub-domain is essentially evaluation of generator's capability to predict normalized perturbations inside the sub-domain. Figure~\ref{fig:gen_perf} depicts the average performance of generator when trained on different subset sizes. For a given particle volume fraction the performance of velocity prediction with the GAN model decreases with increasing Reynolds number. The performance of transverse velocity components is lower than corresponding streamwise component in all of the cases. This can be explained using the reason that axisymmetric nature of the problem about streamwise direction has only been weakly built into the model through inclusion of discrete rotations and reflections. As previously discussed in Figure~\ref{fig:data_aug}, rotation (or reflection) leads to a scalar rotation (or reflection) for streamwise velocity, however, it leads to a vector rotation (or reflection) for transverse velocity suggesting that the latter task is more involved compared to the former. Furthermore, the highest and least performances of streamwise component are associated with the smallest and largest Reynolds number cases respectively. There is a larger difference in the performance of velocity components and pressure for $\phi = 0.45$ cases compared to lower volume fraction cases.

It can also be observed from Figure~\ref{fig:gen_perf} that the training set is sufficient in all cases to obtain reasonably converged GAN model. The rate of increase in performance with increase in $N_{sub,tr}$ is different for every case. A noticeable increase in a model's performance with increase of training dataset size occurs only in cases that have low performance value when trained on smaller sized datasets. It can also be seen that the performance of a model when $N_{sub,tr} = N_{tr}/6$ in some cases is higher than that when a model is trained with $N_{sub,tr} = N_{tr}$ in other cases. This indicates that the generalization capabilities of the presented architecture vary with the considered case. Hence, it requires minor optimization of the GAN model in terms of architecture, training dataset size, loss terms and training approach in order to achieve a desired level of test performance for a particular case. 

\subsection{Attention-CNN Performance}
\begin{figure}
	\centering
	\begin{subfigure}{0.49\textwidth}
		\includegraphics[width=1\linewidth,keepaspectratio=true]{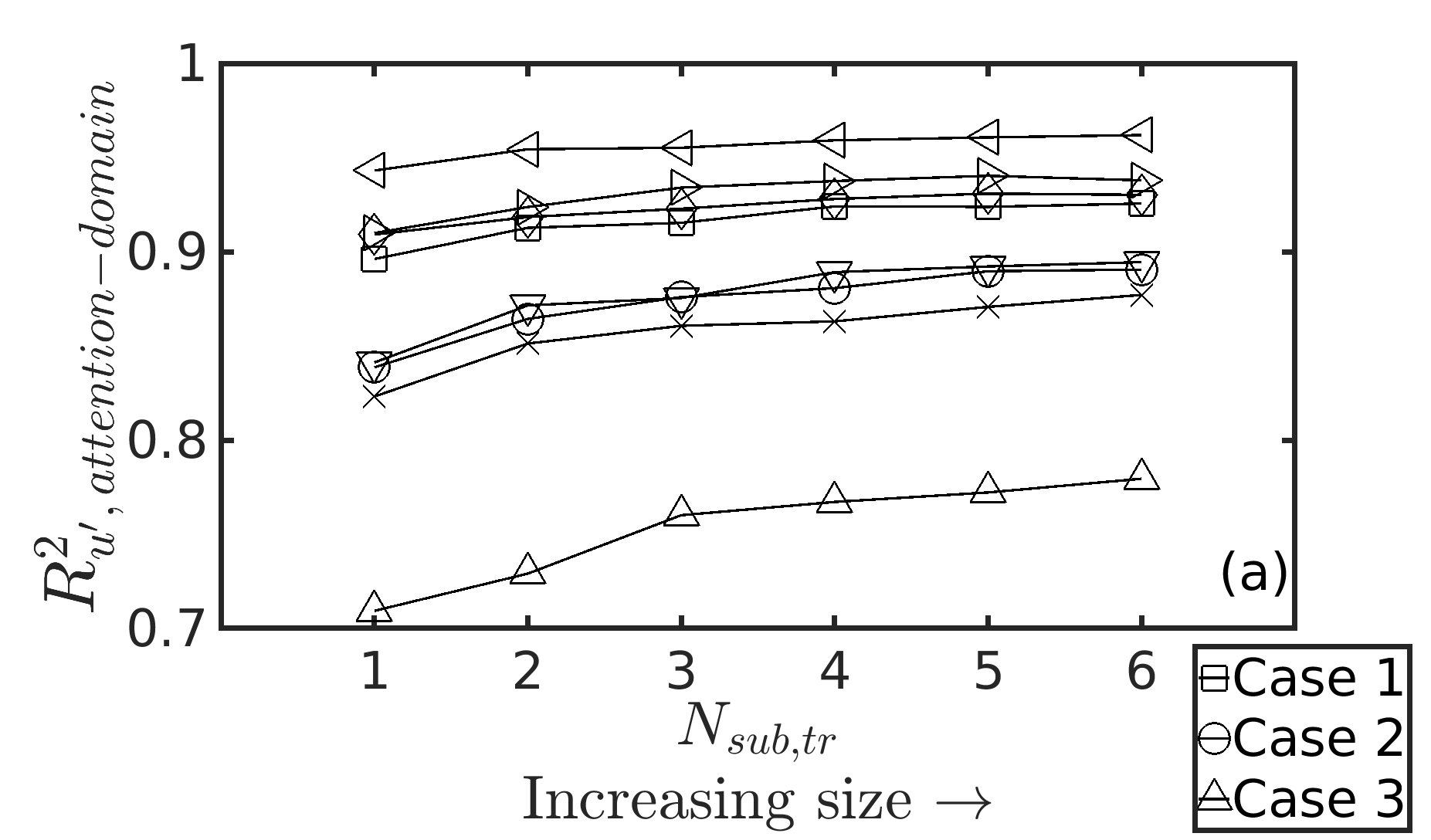}
	\end{subfigure}
	\begin{subfigure}{0.49\textwidth}
		\includegraphics[width=1\linewidth,keepaspectratio=true]{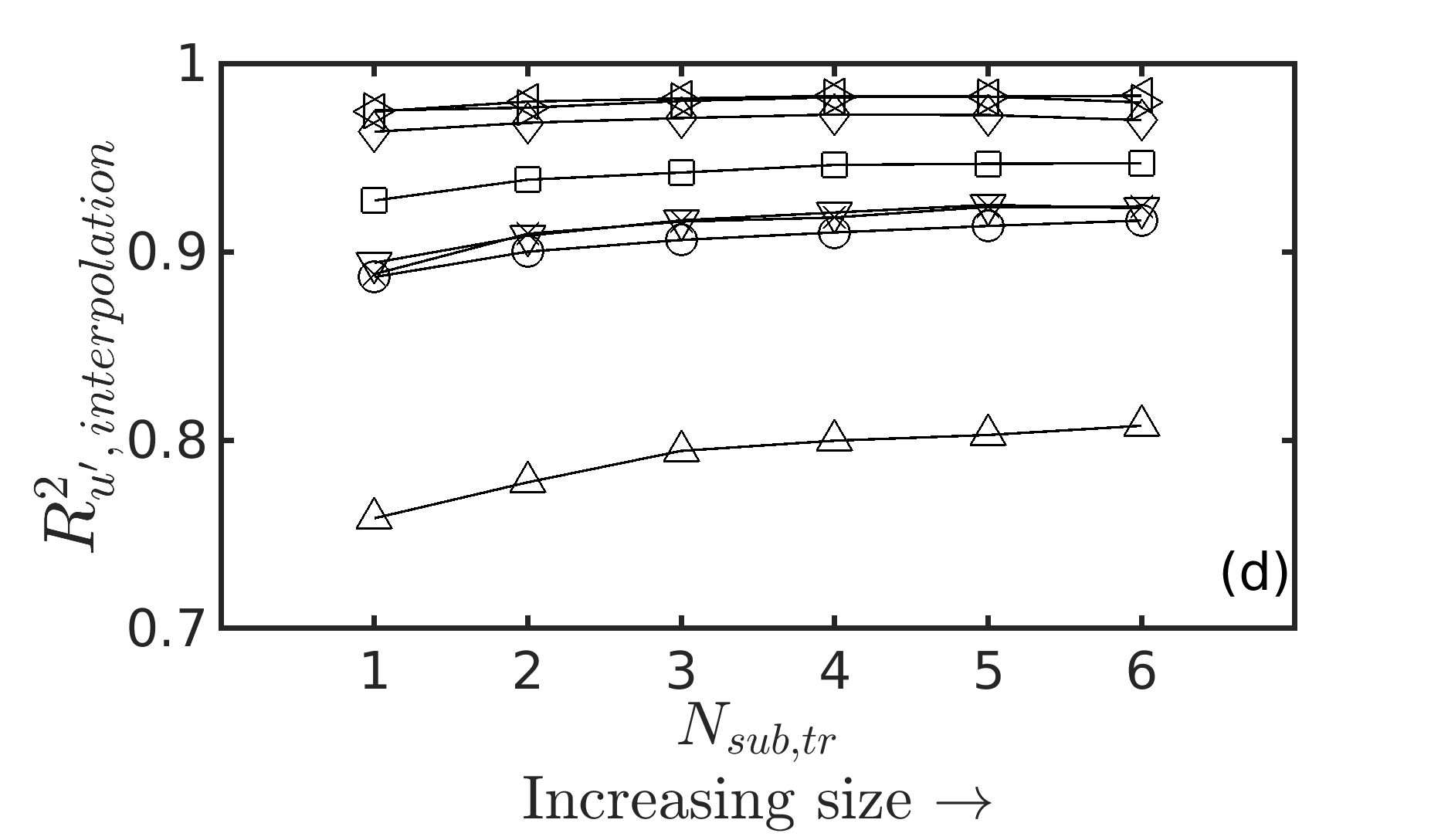}
	\end{subfigure}
	\begin{subfigure}{0.49\textwidth}
		\includegraphics[width=1\linewidth,keepaspectratio=true]{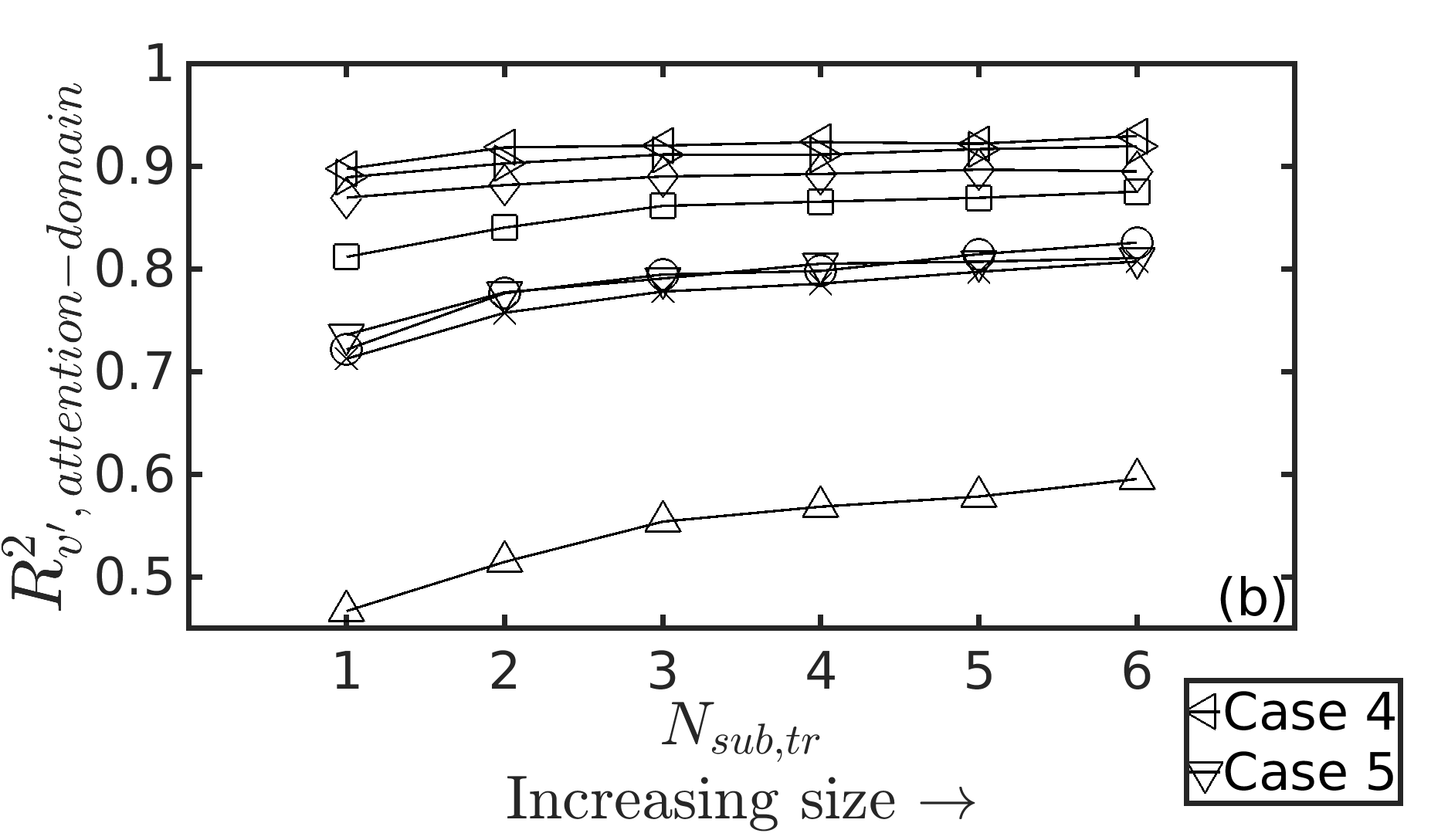}
	\end{subfigure}
	\begin{subfigure}{0.49\textwidth}
		\includegraphics[width=1\linewidth,keepaspectratio=true]{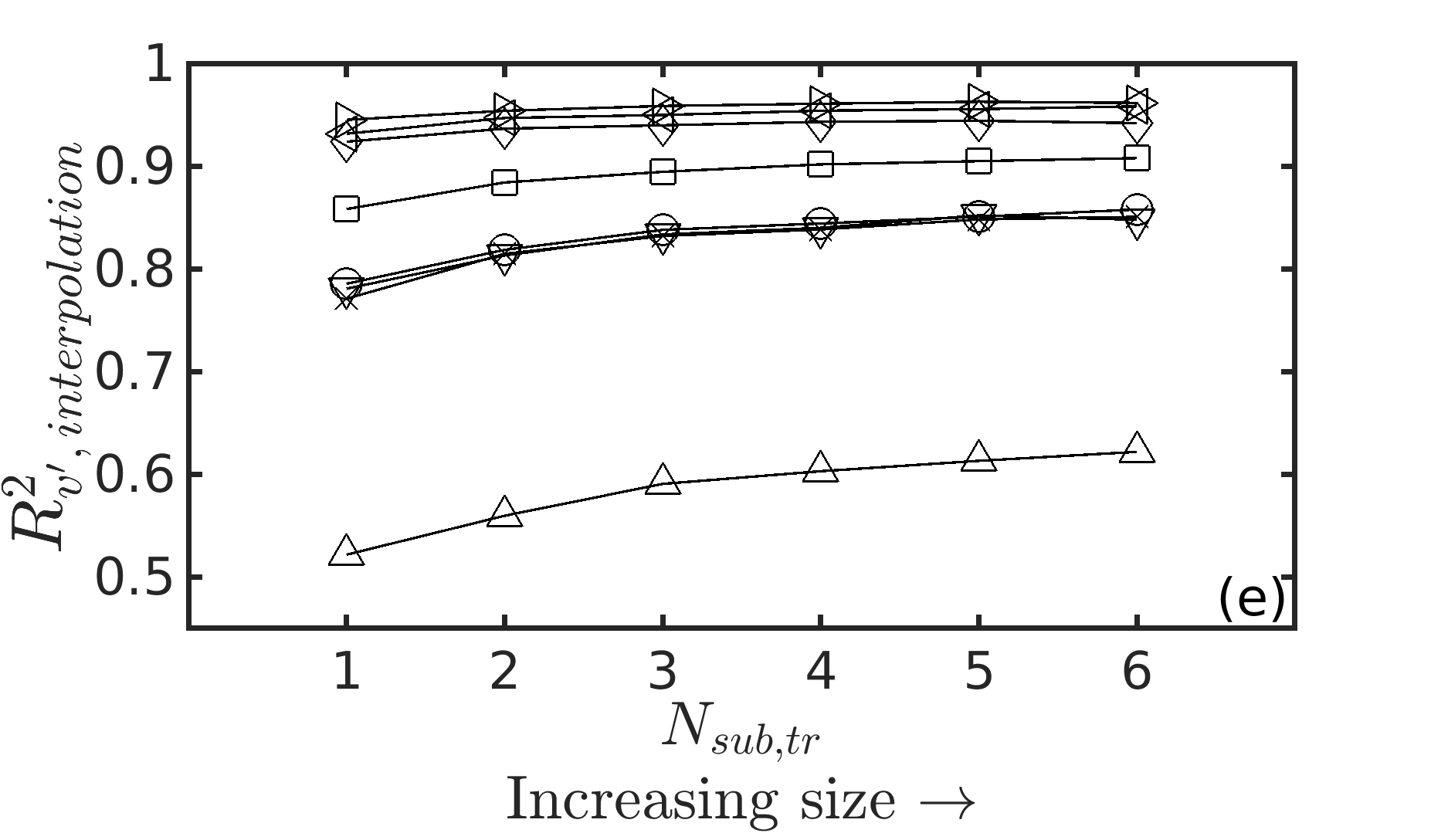}
	\end{subfigure}
	\begin{subfigure}{0.49\textwidth}
		\includegraphics[width=1\linewidth,keepaspectratio=true]{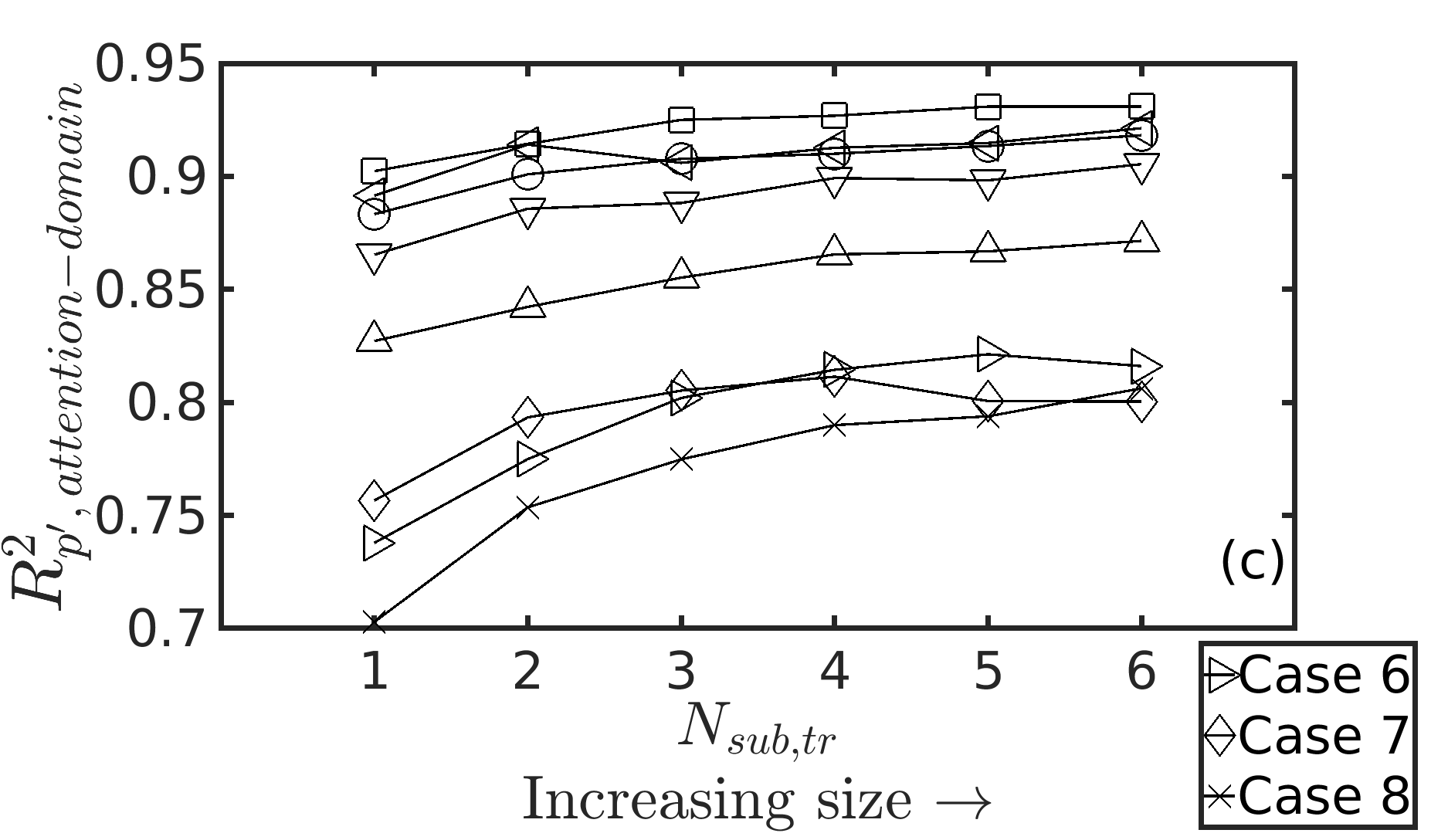}
	\end{subfigure}
	\begin{subfigure}{0.49\textwidth}
		\includegraphics[width=1\linewidth,keepaspectratio=true]{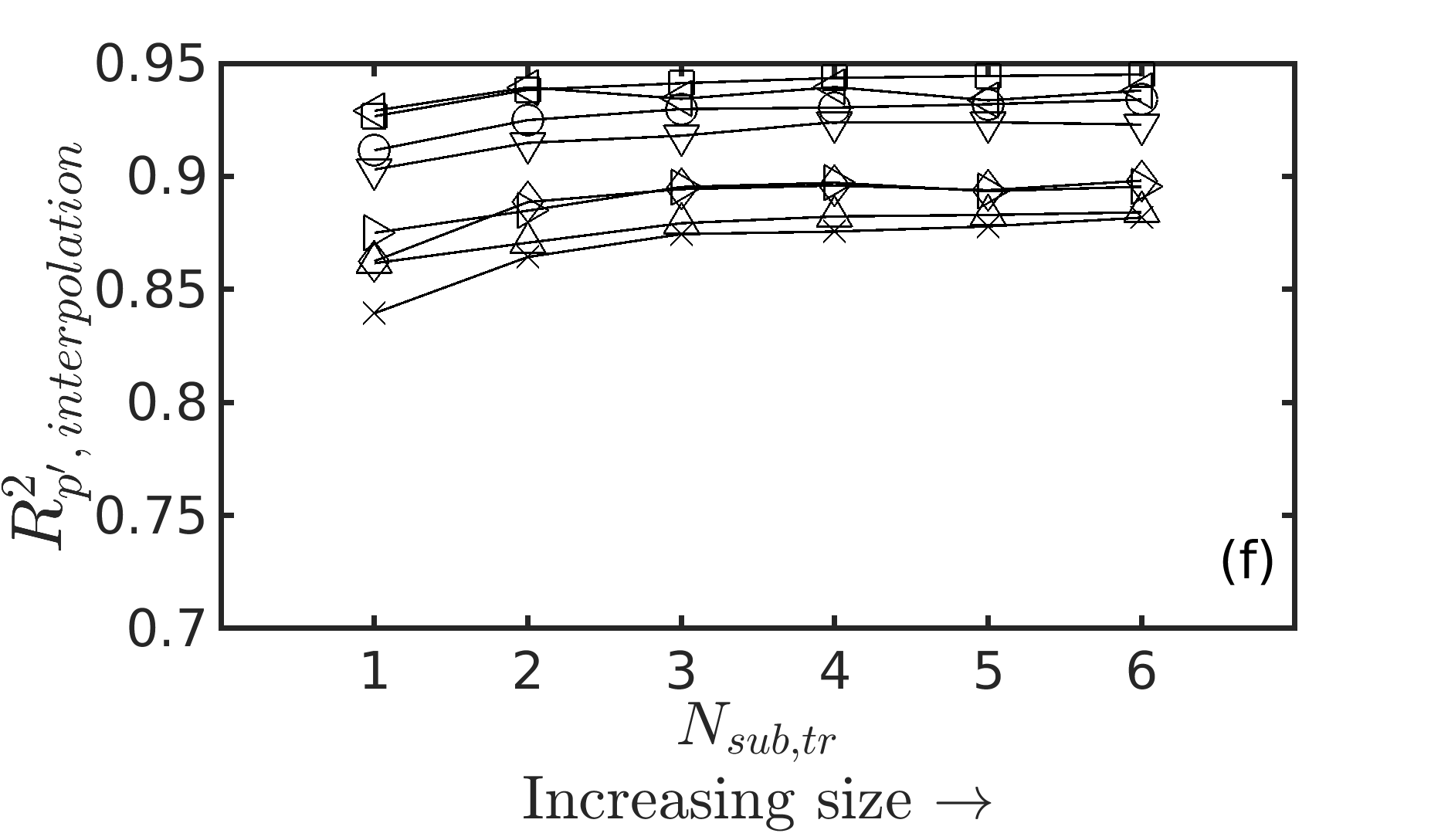}
	\end{subfigure}
	\caption{Performance of $\mathcal{A}$ for (a) Streamwise Velocity, (b) Transverse Velocities, (c) Pressure, and Linear interpolation for (d) Streamwise Velocity, (e) Transverse Velocities, (f) Pressure}
	\label{fig:cnn_perf}
\end{figure}

Unlike the task of a generator which has to produce flow data from a condition, $\mathcal{A}$ has to perform super-resolution of flow data variables. This task can be also perceived as performing an abstract  non-linear interpolation from the coarse-grain output of $\mathcal{G}$ to the finely-resolved spatial points. Thus, the performance of $\mathcal{A}$ is compared with that of a linear interpolation performed on generator's output. The performance of interpolation is equivalent to the generator's performance in the attention-domain. Average performance of $\mathcal{A}$ and its comparison with linear interpolation have been shown in figure~\ref{fig:cnn_perf}. First and foremost, the performance of attention-cnn in all cases is higher compared to their respective generator's performance, with the improvement being significant in the higher volume fraction cases and modest in the other cases.

It can be seen from the figure that performance of a simple three layered $\mathcal{A}$ for flow fields is similar or slightly lower to that of linear interpolation. This suggests that $\mathcal{A}$ is not necessarily improving upon the generator's output in the attention-domain. However, these results do not rule out the option of using an attention mechanism but only indicate that this architecture and methodology needs improvements like passing in an indicator function also as an input along with coarse-grain output. Furthermore, it should also be taken into consideration that performing linear interpolation using only the fluid grid points is not computationally straightforward.
 
\subsection{Networks Performance Uncertainty}
\begin{figure}
	\centering
	\begin{subfigure}{0.49\textwidth}
		\includegraphics[width=1\linewidth,keepaspectratio=true]{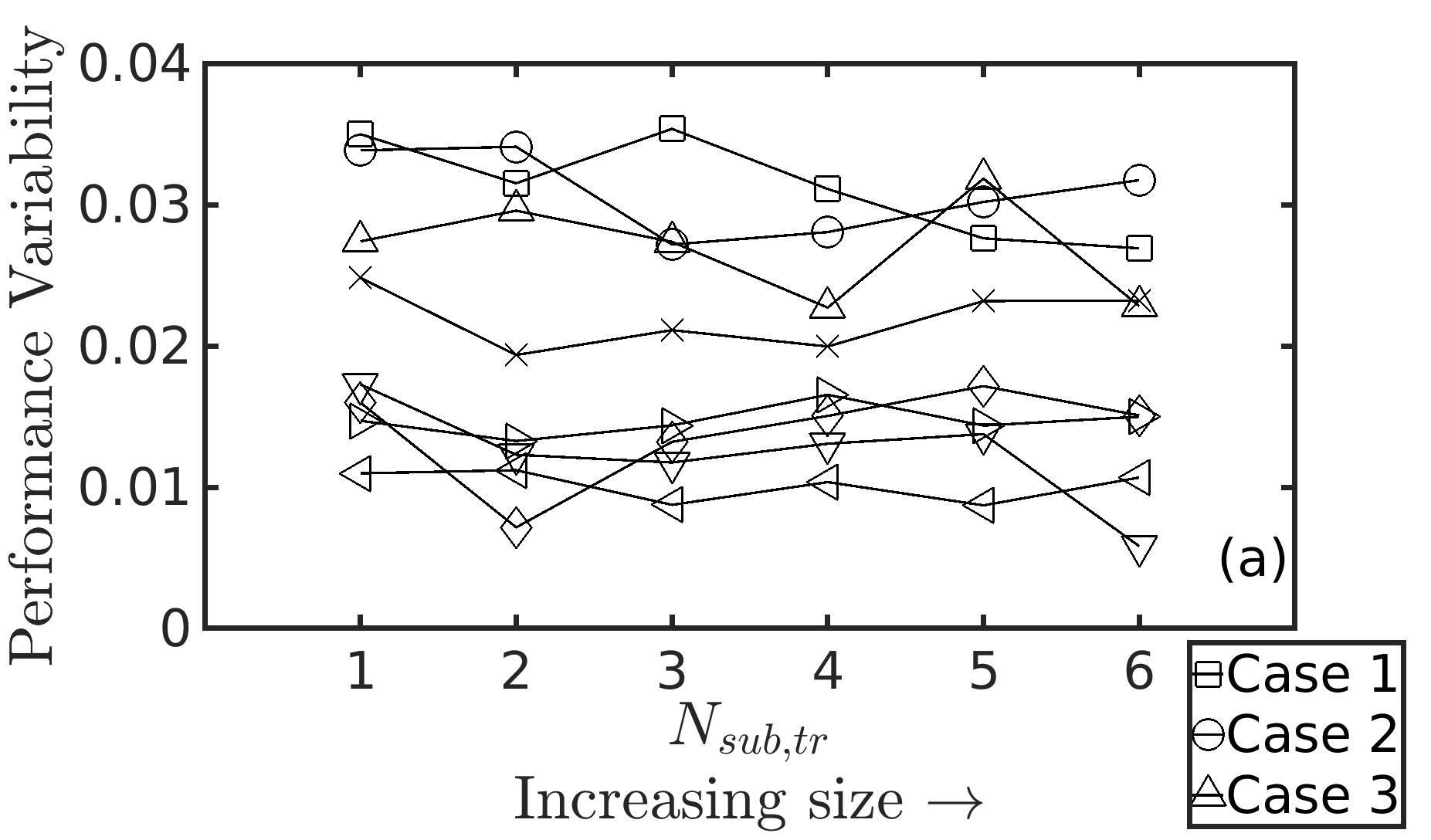}
	\end{subfigure}
	\begin{subfigure}{0.49\textwidth}
		\includegraphics[width=1\linewidth,keepaspectratio=true]{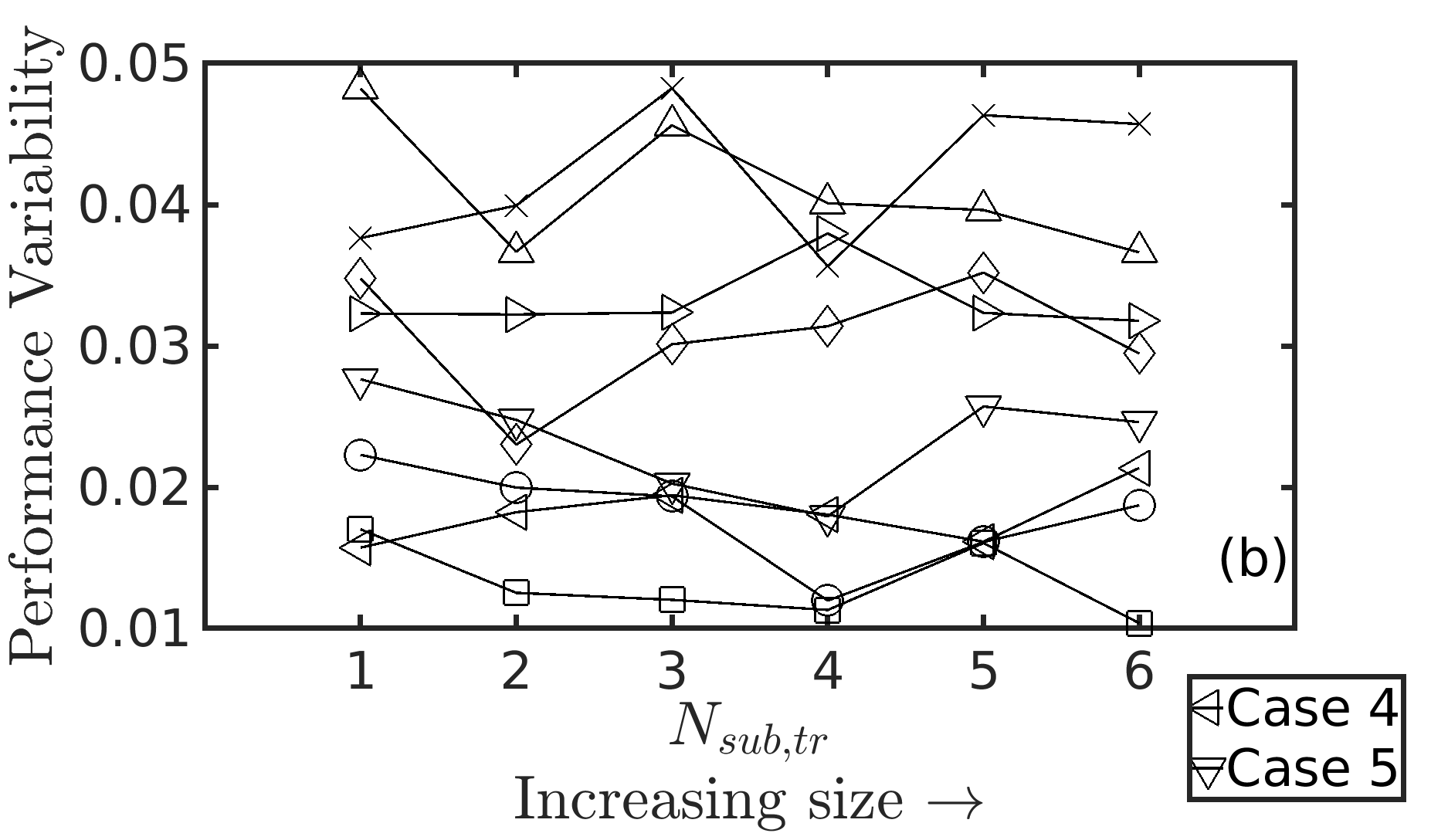}
	\end{subfigure}
	\begin{subfigure}{0.49\textwidth}
		\includegraphics[width=1\linewidth,keepaspectratio=true]{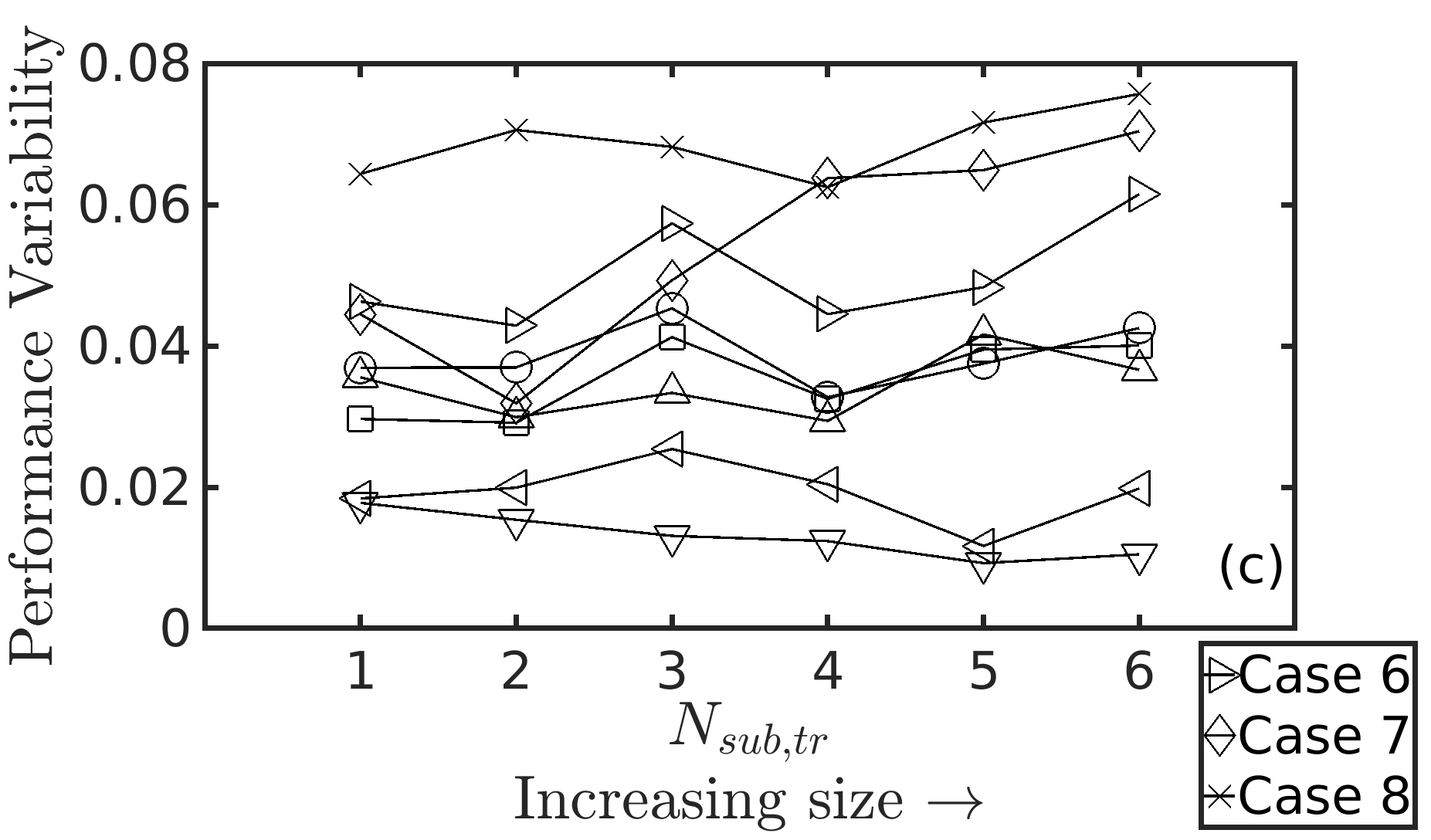}
	\end{subfigure}
	\caption{Performance variability of Generator ($\mathcal{G}$) for (a) Streamwise Velocity, (b) Transverse Velocities, and (c) Pressure}
	\label{fig:gen_var}
\end{figure}

\begin{figure}
	\centering
	\begin{subfigure}{0.49\textwidth}
		\includegraphics[width=1\linewidth,keepaspectratio=true]{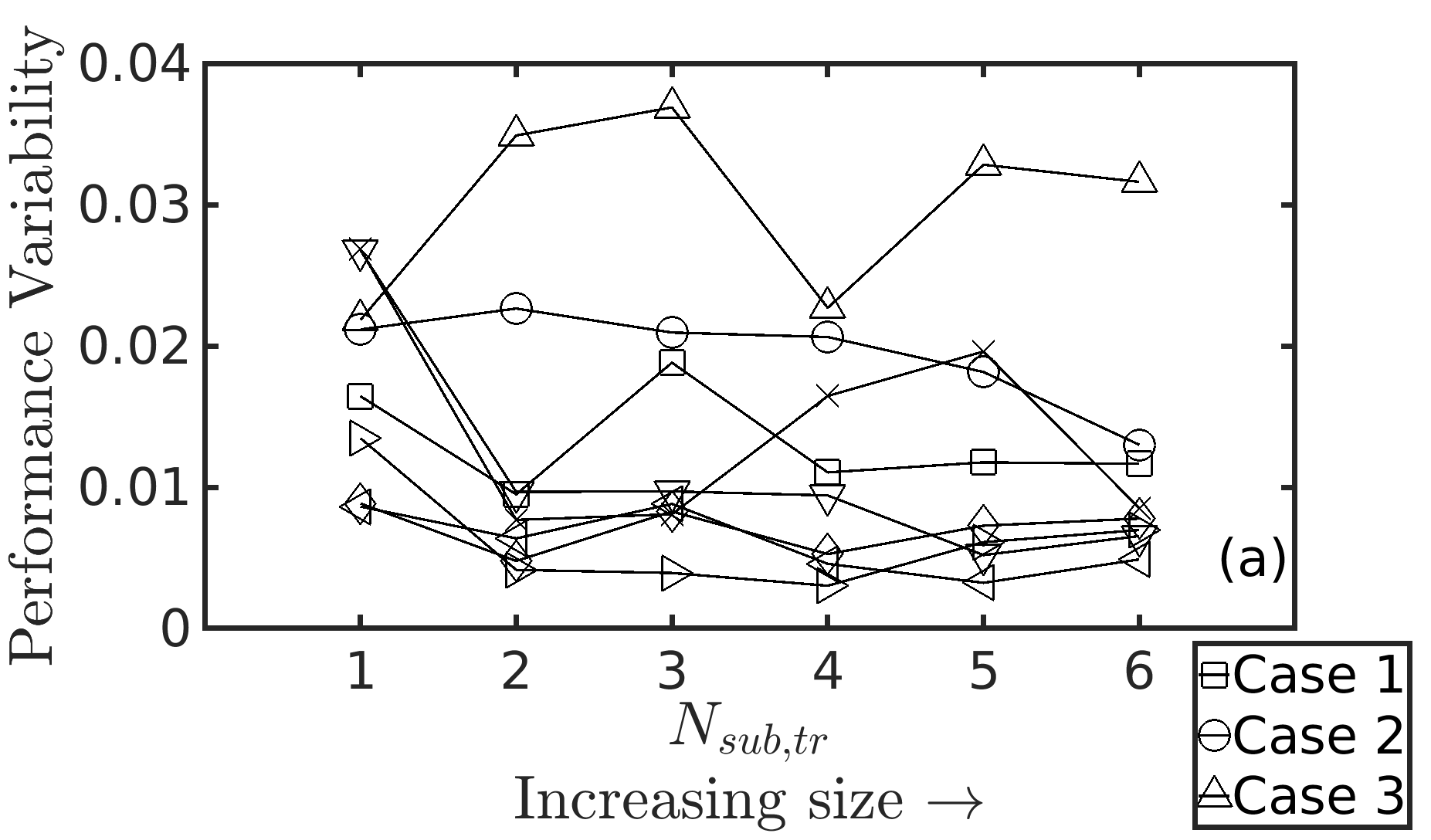}
	\end{subfigure}
	\begin{subfigure}{0.49\textwidth}
		\includegraphics[width=1\linewidth,keepaspectratio=true]{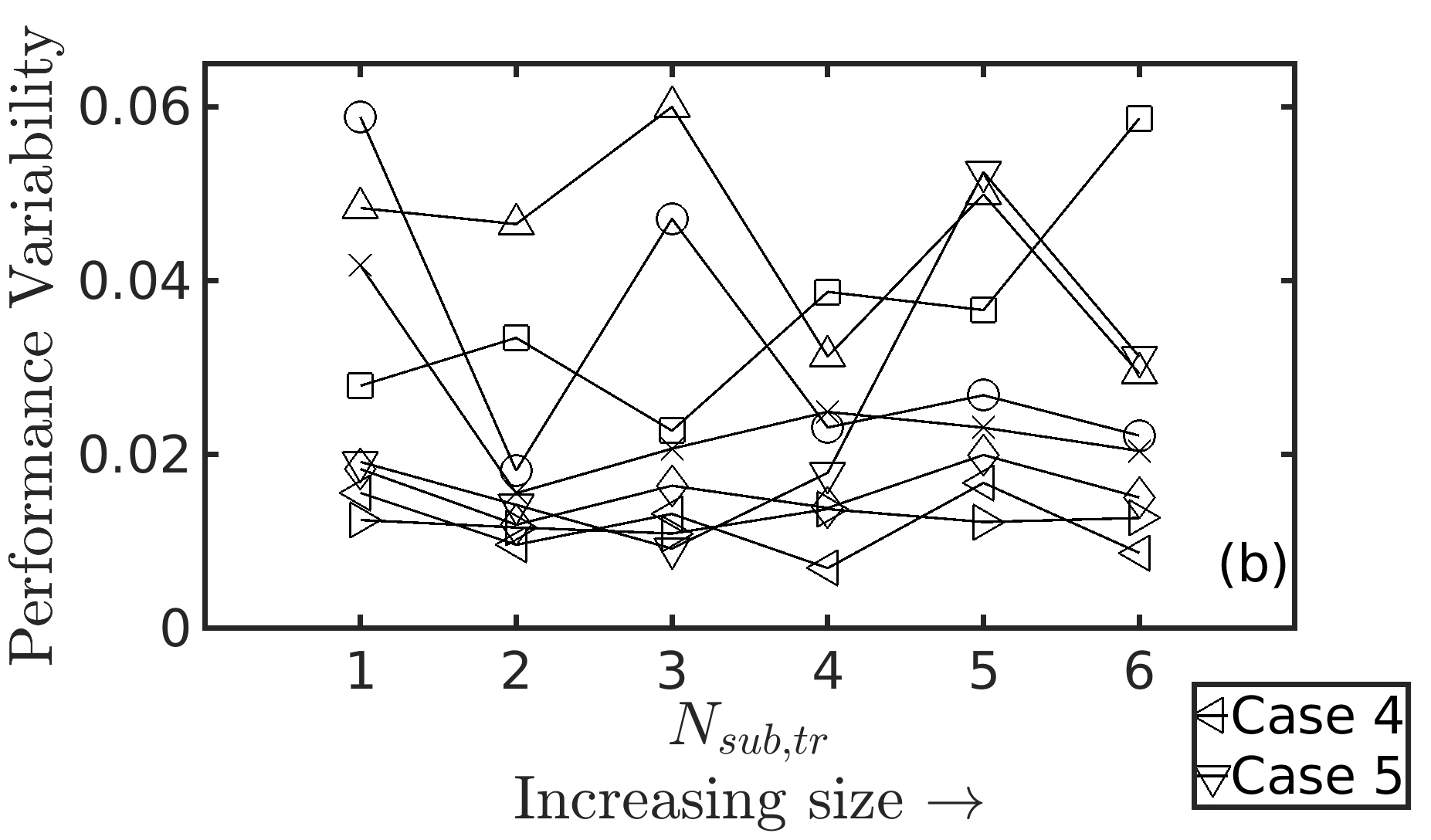}
	\end{subfigure}
	\begin{subfigure}{0.49\textwidth}
		\includegraphics[width=1\linewidth,keepaspectratio=true]{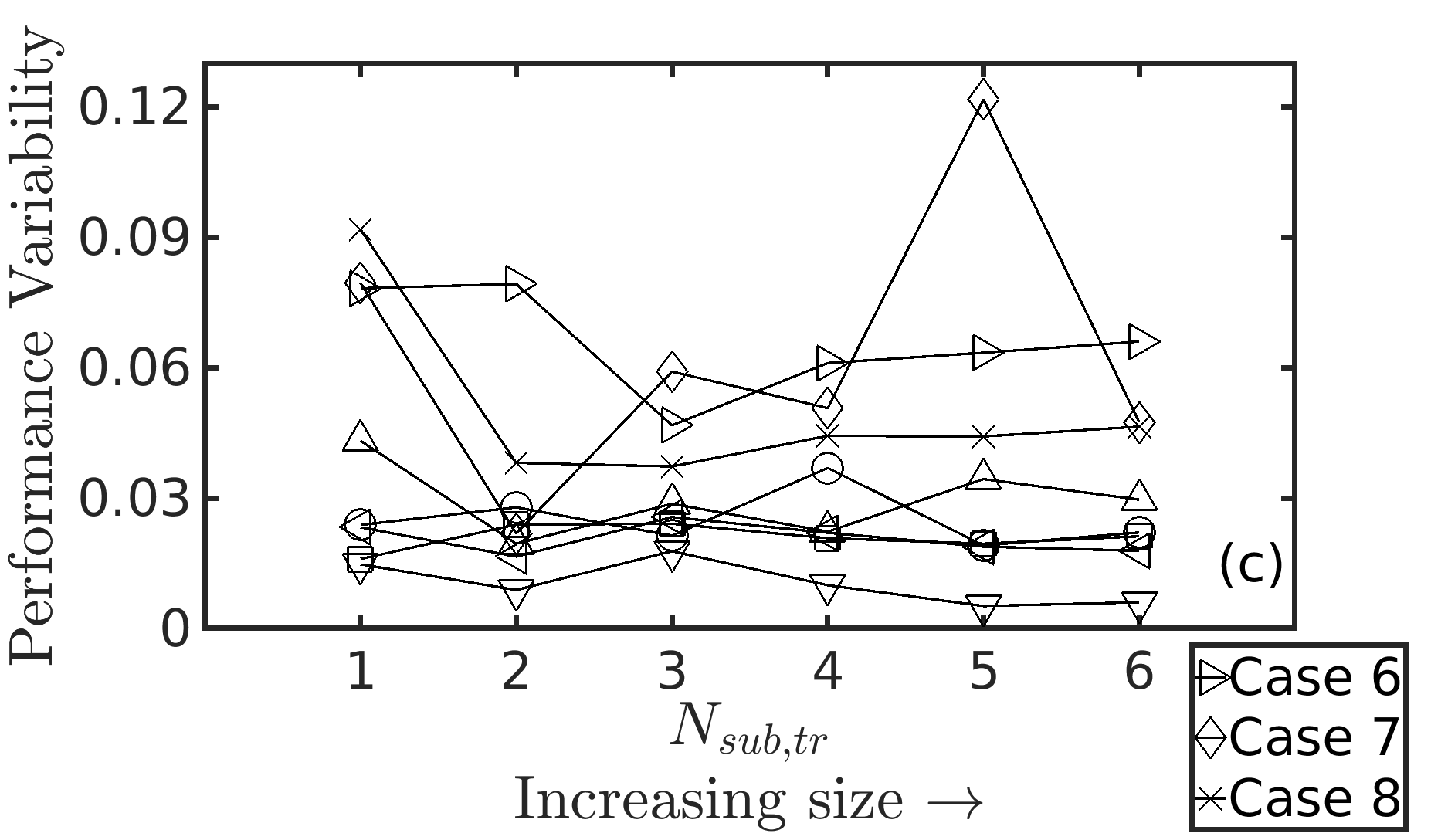}
	\end{subfigure}
	\caption{Performance variability of Attention-CNN ($\mathcal{A}$) for (a) Streamwise Velocity, (b) Transverse Velocities, and (c) Pressure}
	\label{fig:att_var}
\end{figure}

Here we discuss uncertainty in the performance of $\mathcal{G}$ and $\mathcal{A}$. We evaluate uncertainty in terms of how much the performance varies when each realization of a case is considered as the test realization. For example, in case 1 there are ten realizations and $R^2$ value can be computed as given in \eqref{eq8} using only the $i^{th}$ realization as the test realization and all other nine realizations to be used in the training process. As the test realization $i$ is varied from 1 to $M_c = 10$, the computed $R^2$ value will vary and this variation is denoted as performance variability (or uncertainty) and is defined as
\begin{equation}
 \| R^2_i - R^2 \|_{\infty} \, ,
\end{equation}
where $R^2_i$ is the performance evaluated using only the $i^{th}$ realization as the test realization, while $R^2$ is the corresponding average performance, averaged over $i=1, \cdots M_c$ and $\| \, \cdot \,\|_{\infty}$ is $L$-infinity-norm. Since $R^2$ is already normalized further normalization of its variability is not necessary. A low value of performance variability would indicate that the GAN model is likely to yield close to average performance for any new random condition $\bc$. The performance variability of $\mathcal{G}$ and $\mathcal{A}$ for all cases are shown in figures \ref{fig:gen_var} \& \ref{fig:att_var} respectively. Performance variability does not seem to depend strongly on the size of the training dataset. The largest pressure variability in $\mathcal{G}$ and $\mathcal{A}$ among all of the different test runs for the presented cases is approximately twice the largest variability for velocity components.  
 
In general, low variation in the performance of $\mathcal{G}$ and $\mathcal{A}$, especially for velocity components, is certainly an indication that the realizations of a case are not completely unique. This low variability of the networks is an optimistic sign for their performance on unknown datasets. 

\subsection{Flow Field Reconstruction}
The GAN methodology has been designed to predict the flow field in a sub-domain around the reference particle. Here we present a simple approach that uses the sub-domain results to predict the flow field over the entire computational domain. The entire domain around the particles is divided into voronoi cells \cite{voro++} to achieve this particular task. Flow field at all the grid points that are inside a particle's voronoi volume is taken to be the GAN output of the sub-domain with that particle as the reference at the center. The entire flow field is created by assembling these individual solutions within all the voronoi volumes together. It was observed in earlier sections that the generator's predictions are more accurate closer to the reference particle. Hence, it is expected that the performance of the reconstructed flow field in the entire domain is even better than $\mathcal{G}$'s performance (see Figure~\ref{fig:gen_perf}) and closer to the performance of linear interpolation shown in Figure~\ref{fig:cnn_perf}.

\begin{figure}
	\centering
	\begin{subfigure}{\textwidth}
		\includegraphics[width=0.328\textwidth,keepaspectratio=true]{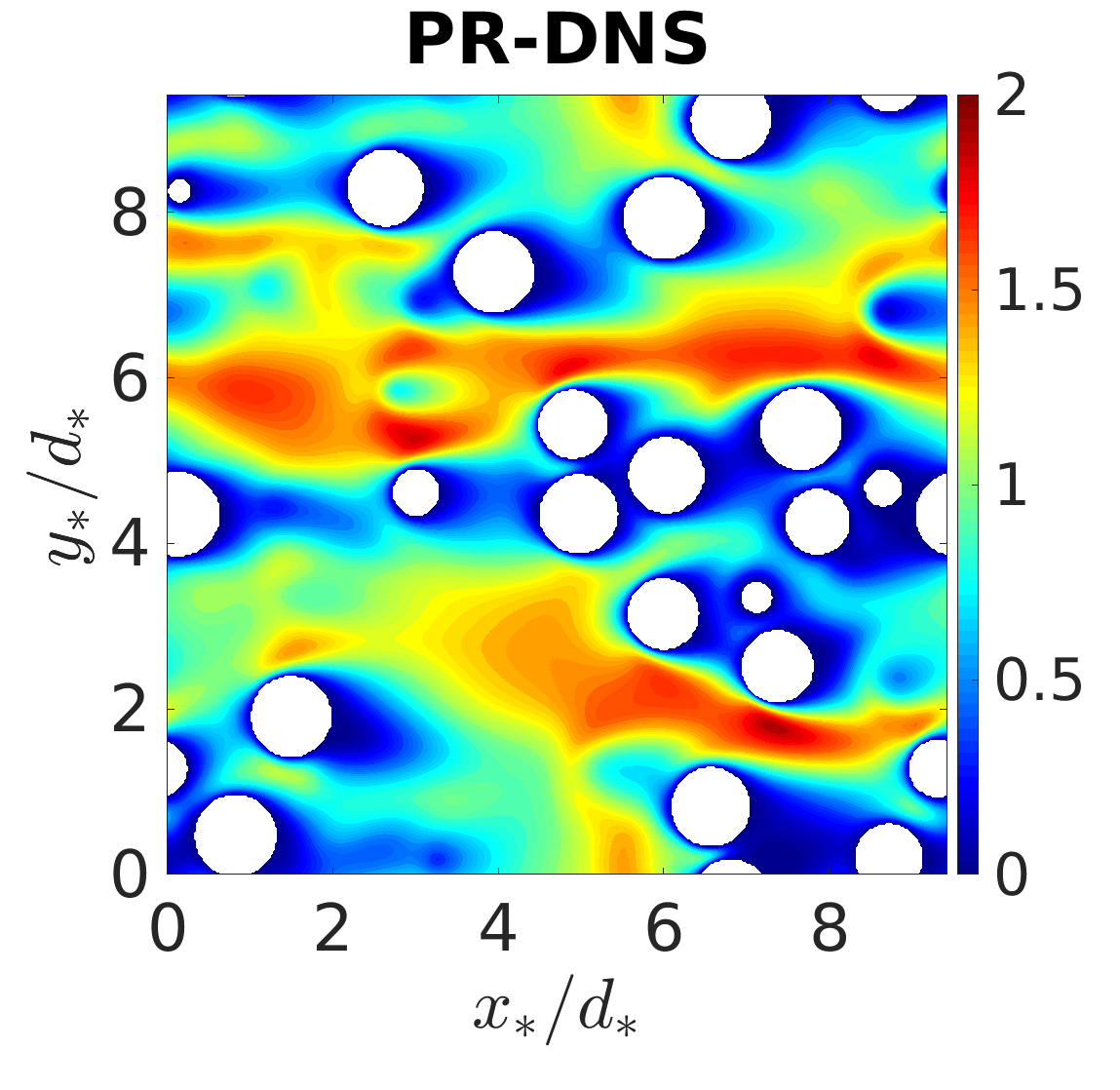}
		\includegraphics[width=0.328\textwidth,keepaspectratio=true]{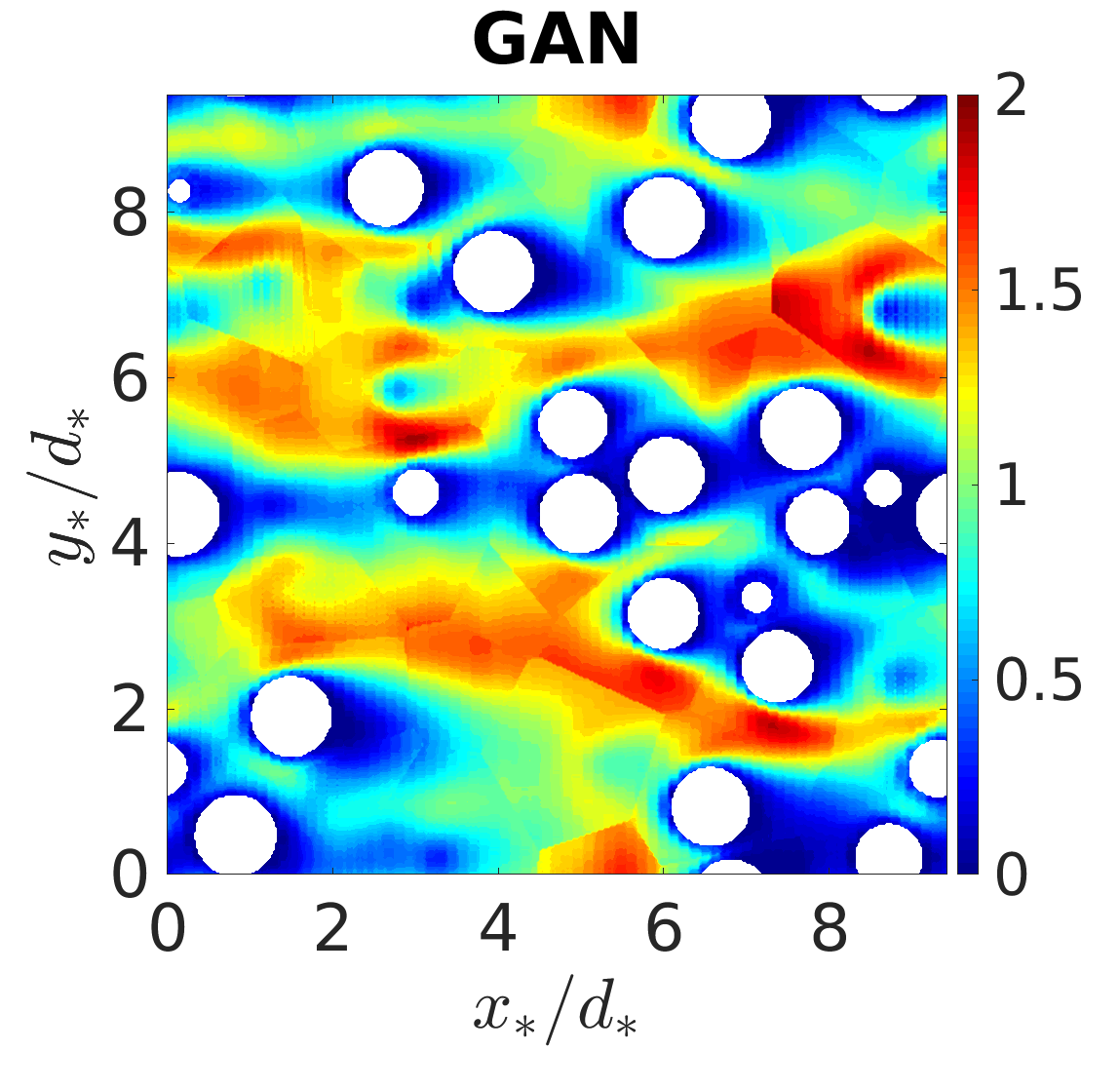}
		\includegraphics[width=0.328\textwidth,keepaspectratio=true]{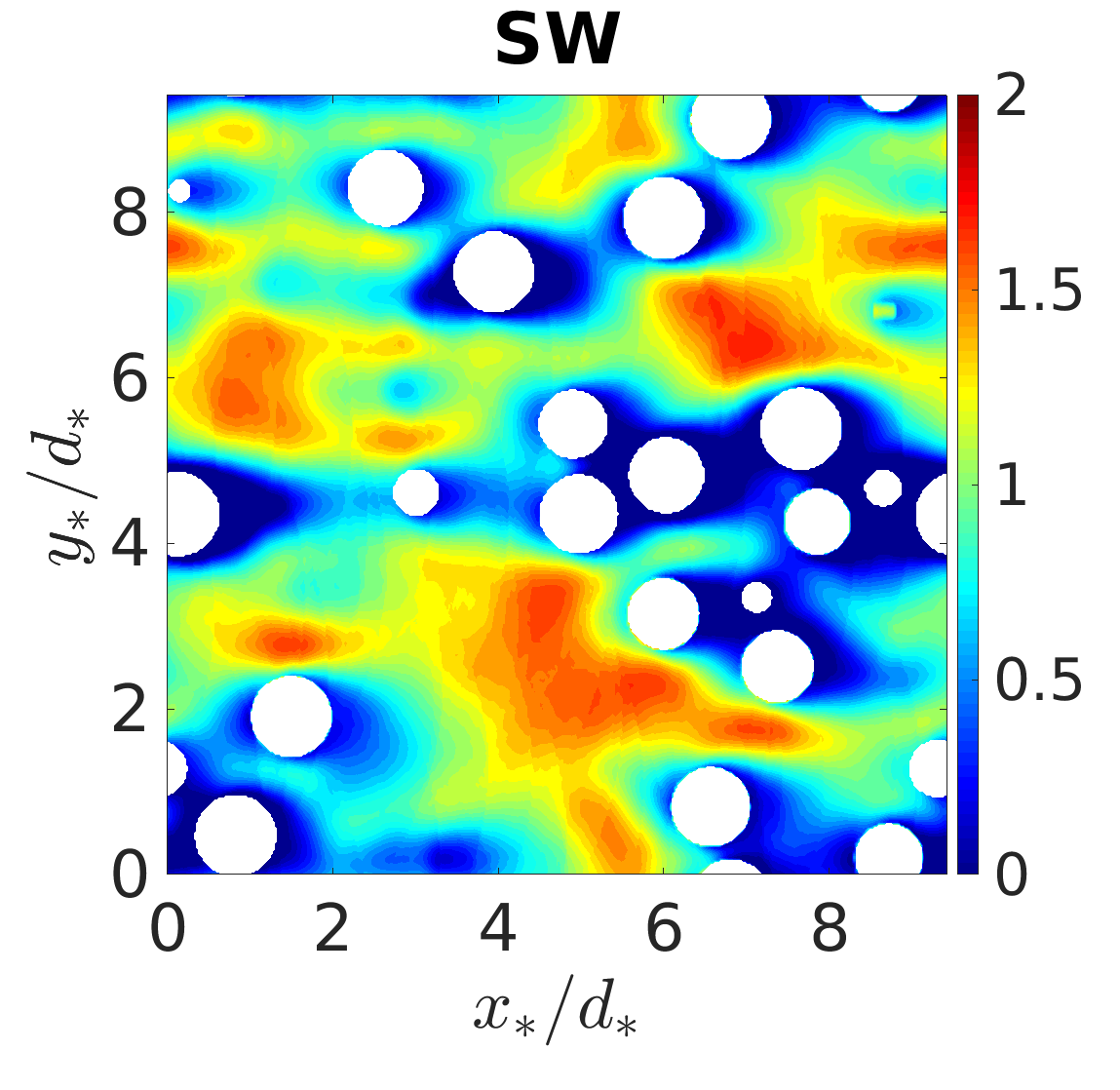}
		\caption{}
	\end{subfigure}
    \begin{subfigure}{\textwidth}
    	\includegraphics[width=0.328\textwidth,keepaspectratio=true]{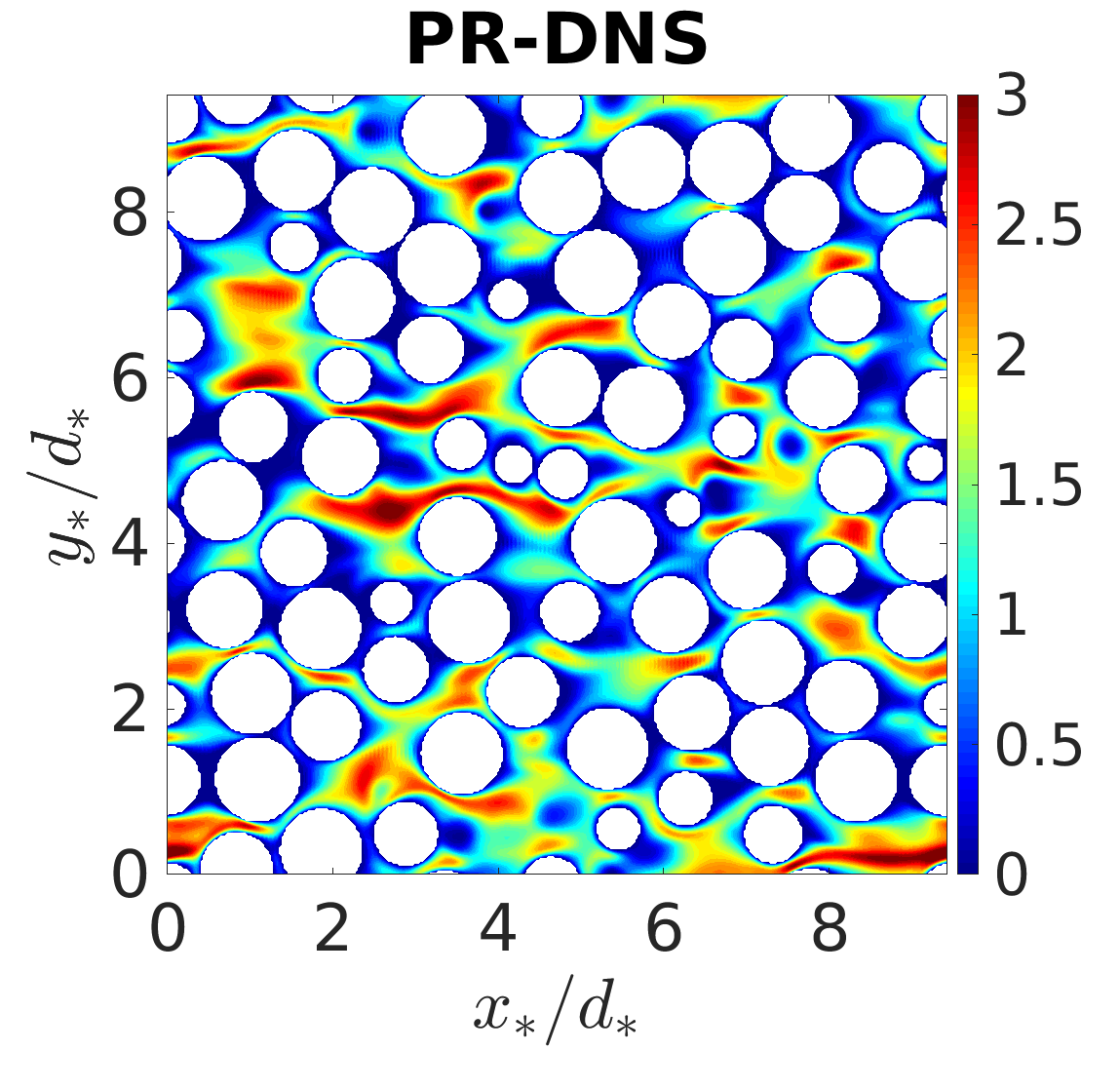}
    	\includegraphics[width=0.328\textwidth,keepaspectratio=true]{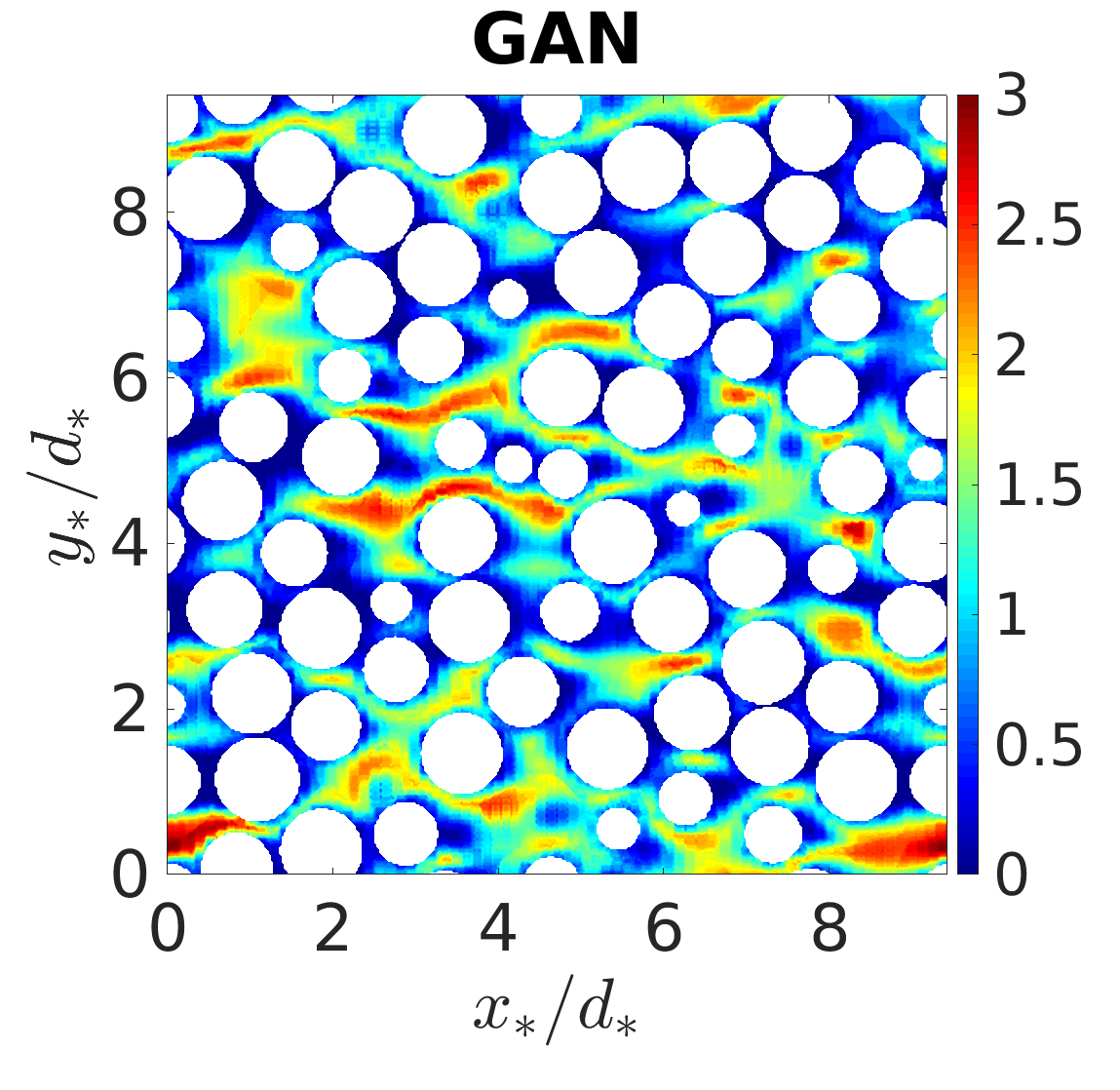}
    	\includegraphics[width=0.328\textwidth,keepaspectratio=true]{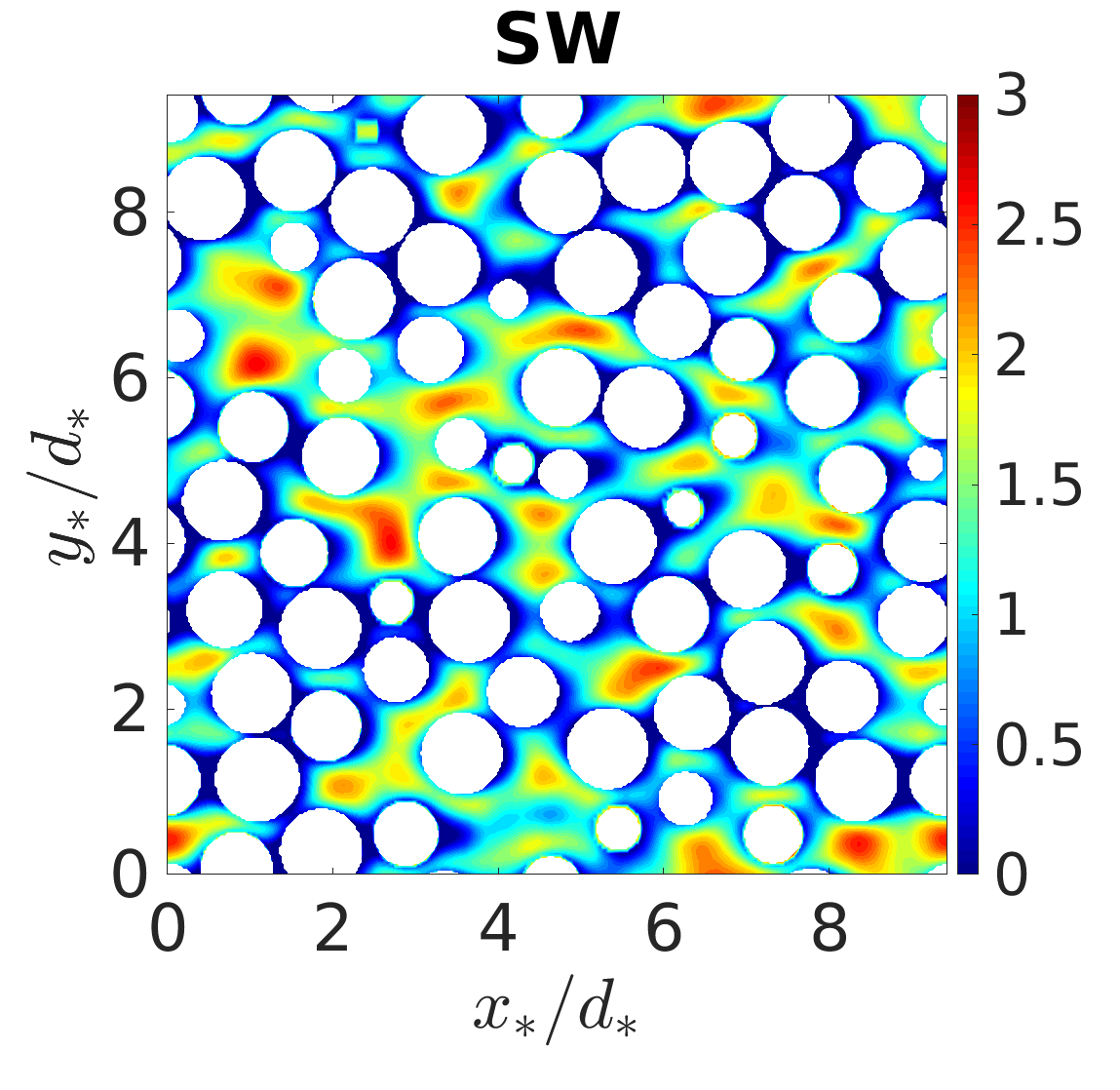}
    	\caption{}
    \end{subfigure}
	\caption{Streamwise velocity for central $x-y$ plane of a test realization from  (a) Case 1 and (b) Case 8}
	\label{fig:recon_u}
\end{figure}

\begin{figure}
	\centering
	\begin{subfigure}{\textwidth}
		\includegraphics[width=0.328\textwidth,keepaspectratio=true]{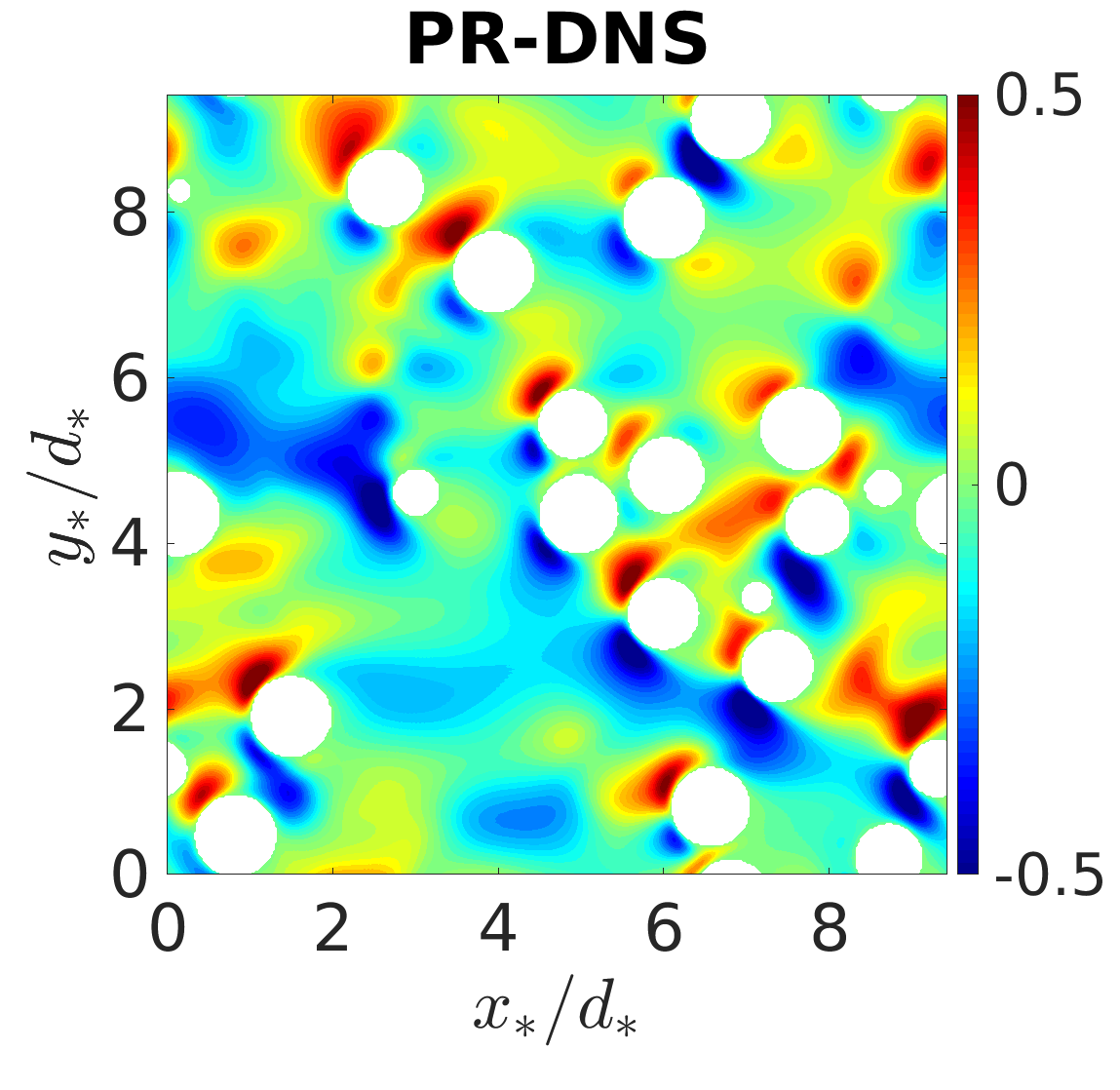}
		\includegraphics[width=0.328\textwidth,keepaspectratio=true]{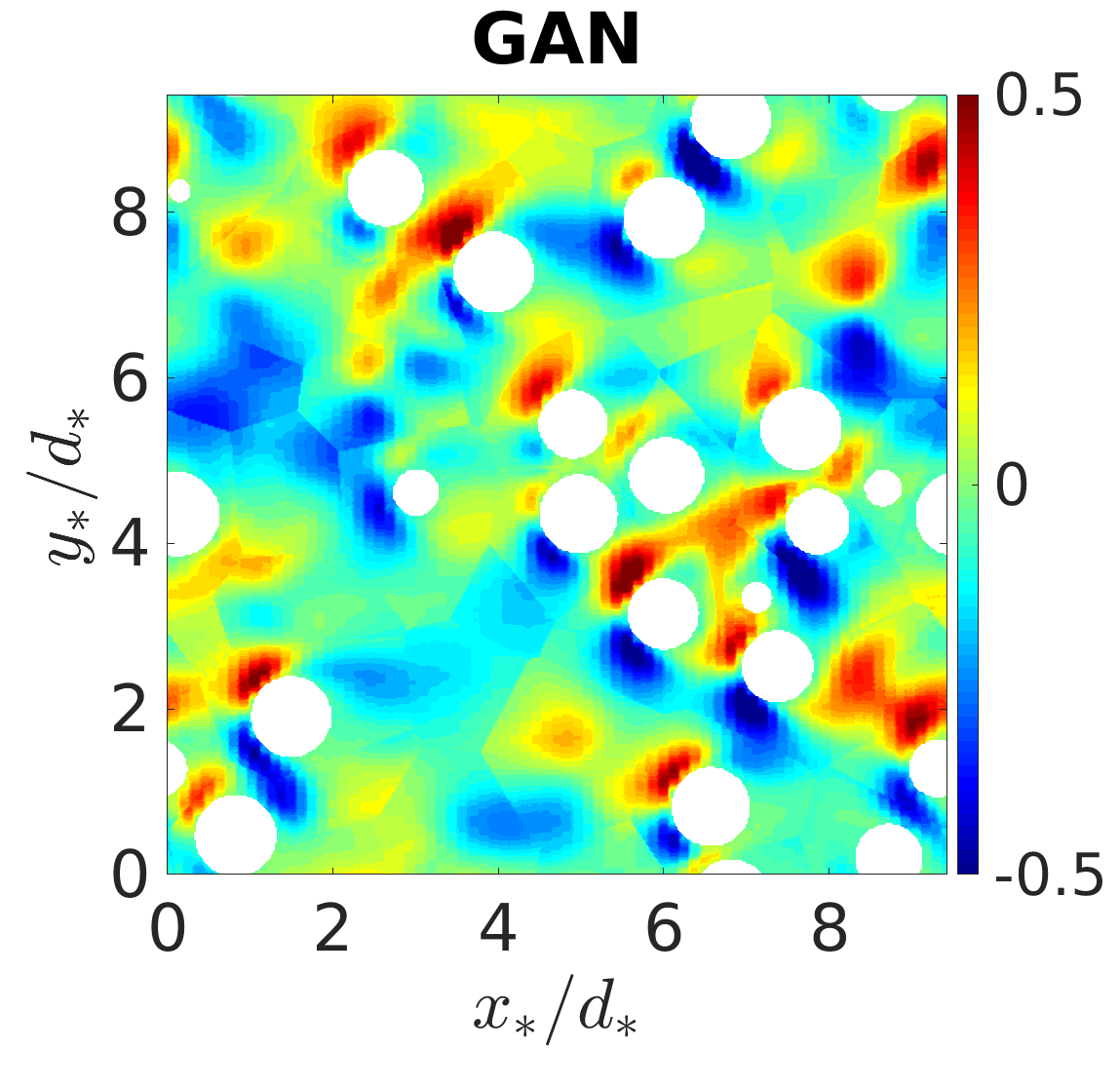}
		\includegraphics[width=0.328\textwidth,keepaspectratio=true]{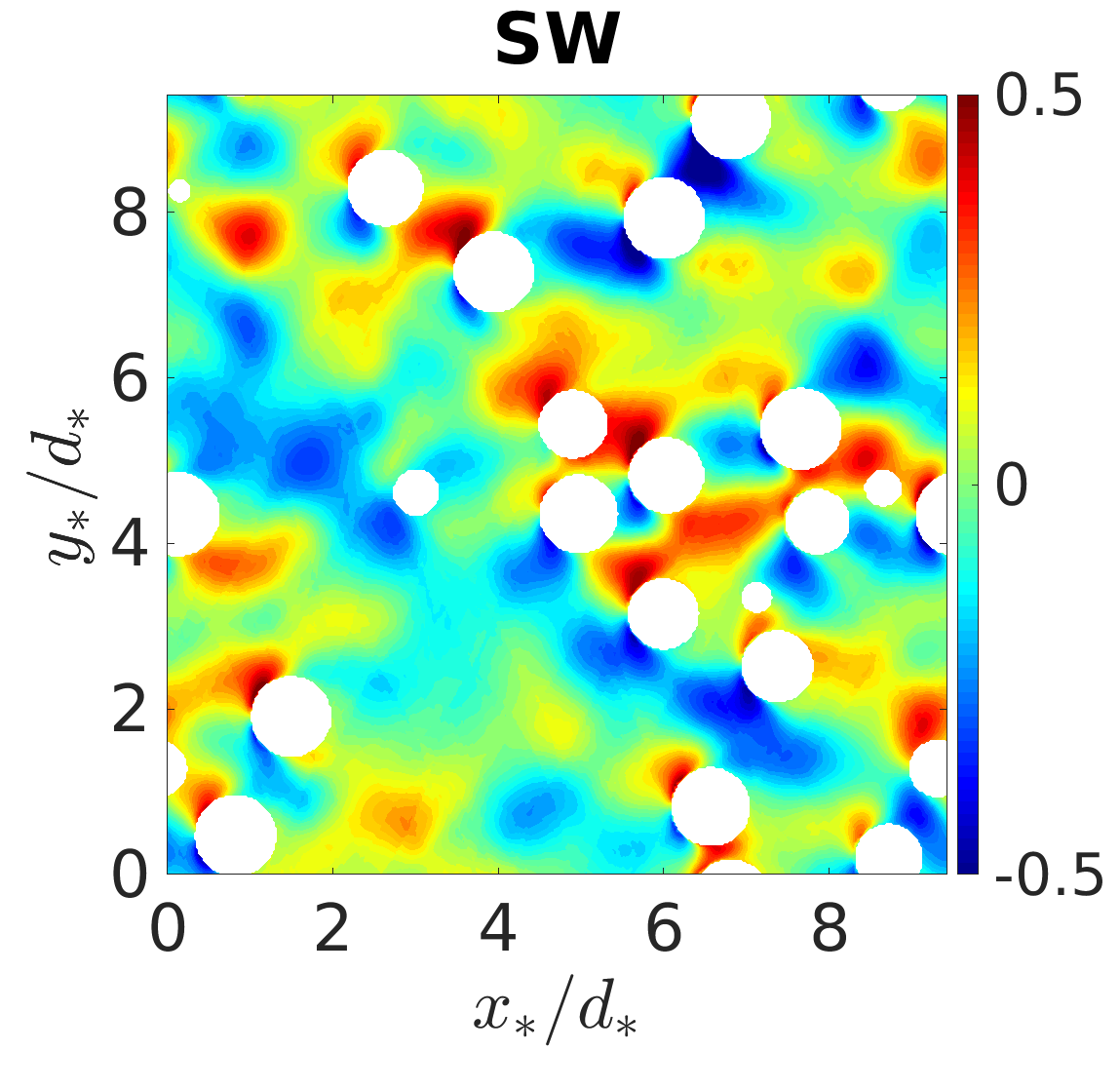}
		\caption{}
	\end{subfigure}
	\begin{subfigure}{\textwidth}
		\includegraphics[width=0.328\textwidth,keepaspectratio=true]{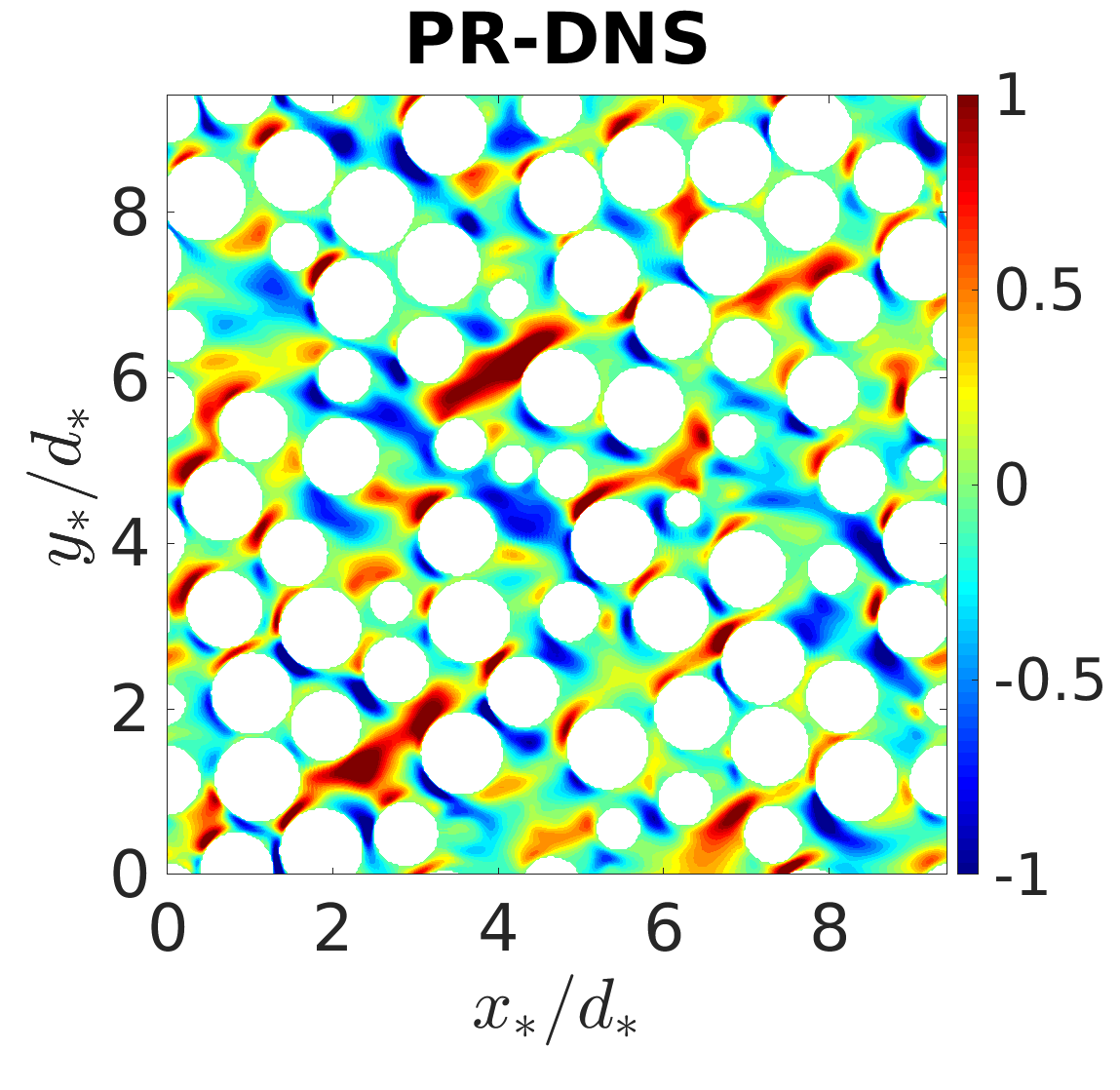}
		\includegraphics[width=0.328\textwidth,keepaspectratio=true]{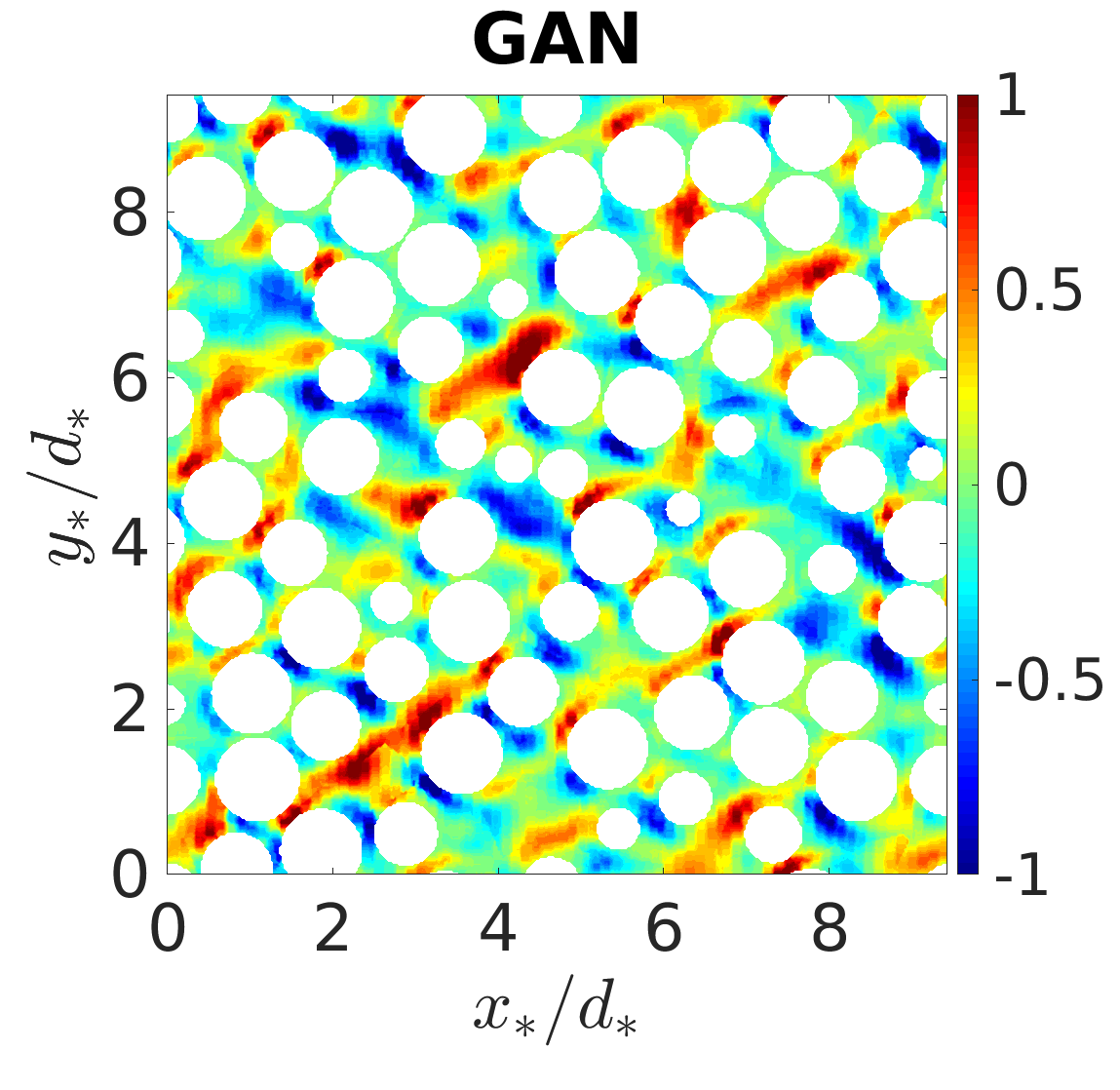}
		\includegraphics[width=0.328\textwidth,keepaspectratio=true]{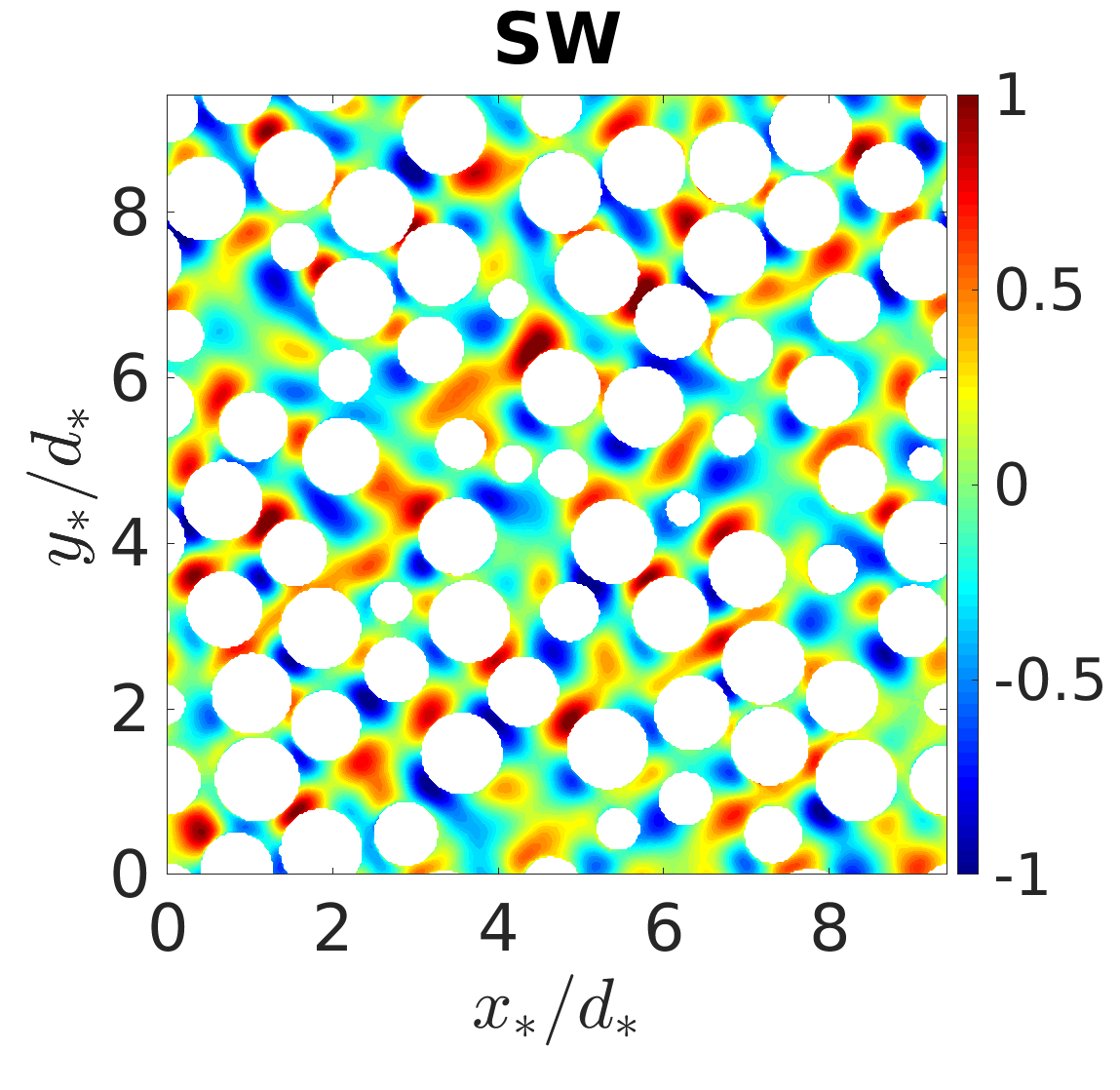}
		\caption{}
	\end{subfigure}
	\caption{In-plane transverse velocity component ($v$) for central $x-y$ plane of a test realization from  (a) Case 1 and (b) Case 8}
	\label{fig:recon_v}
\end{figure}

\begin{figure}
	\centering
	\begin{subfigure}{\textwidth}
		\includegraphics[width=0.328\textwidth,keepaspectratio=true]{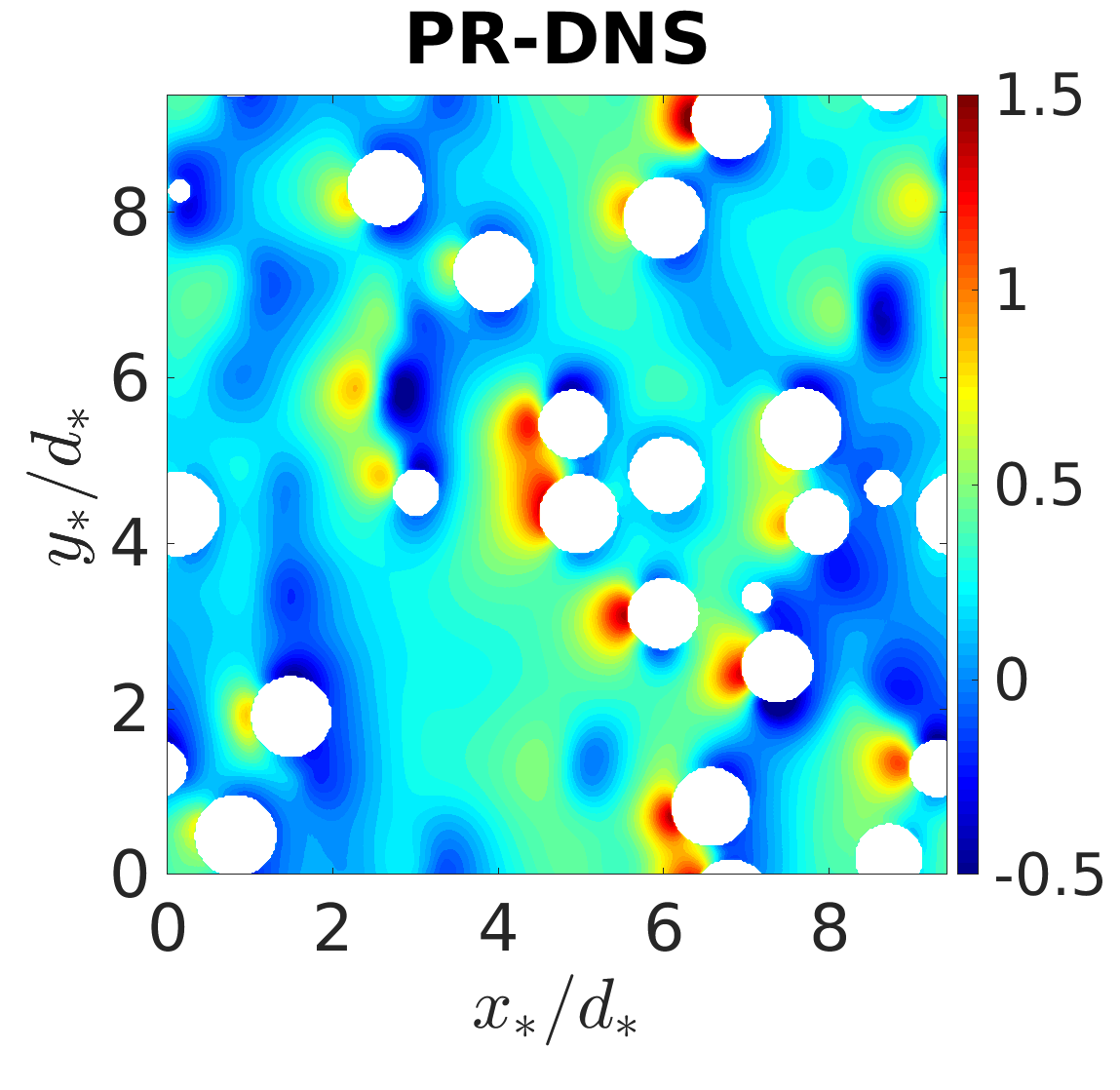}
		\includegraphics[width=0.328\textwidth,keepaspectratio=true]{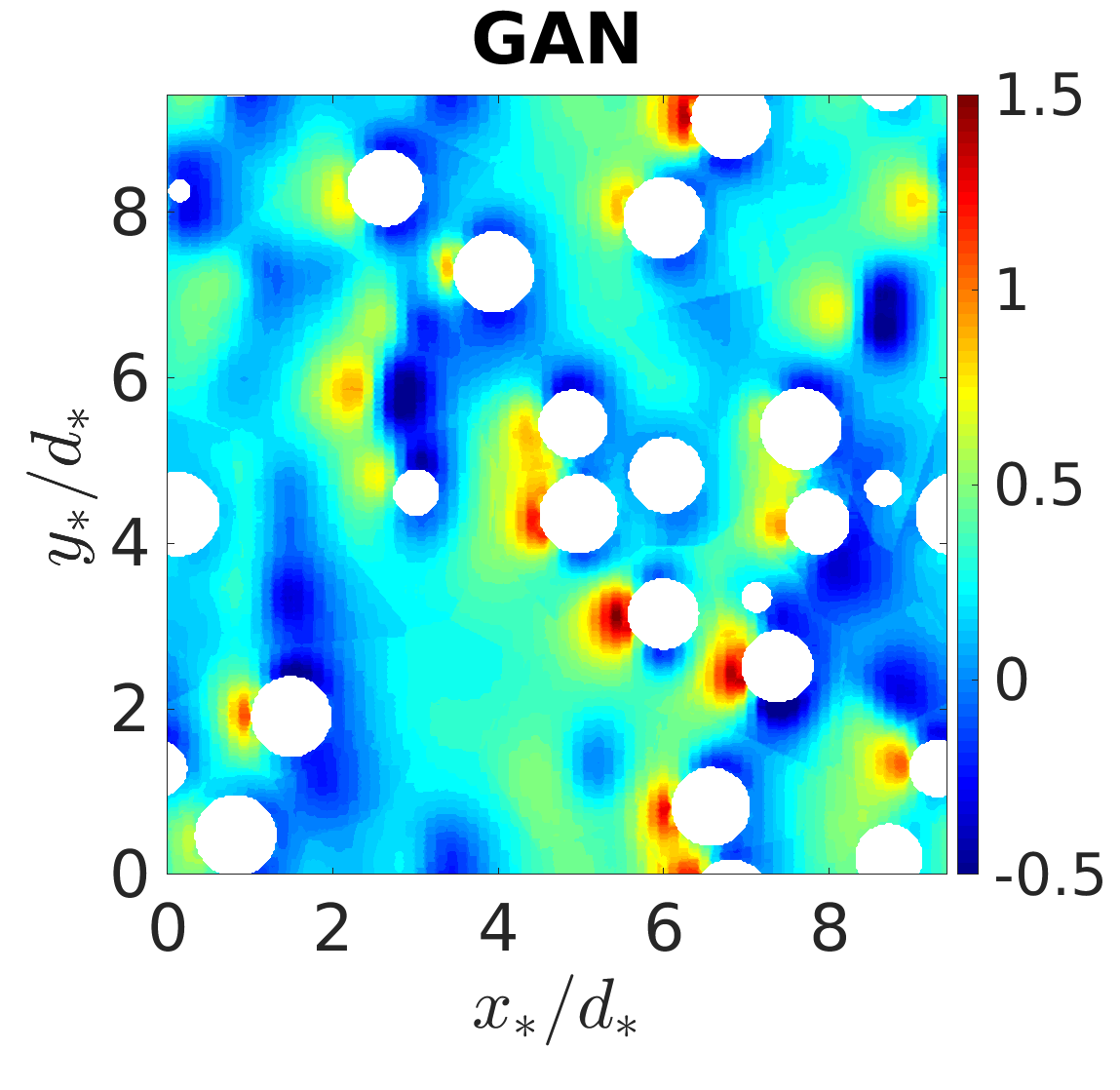}
		\includegraphics[width=0.328\textwidth,keepaspectratio=true]{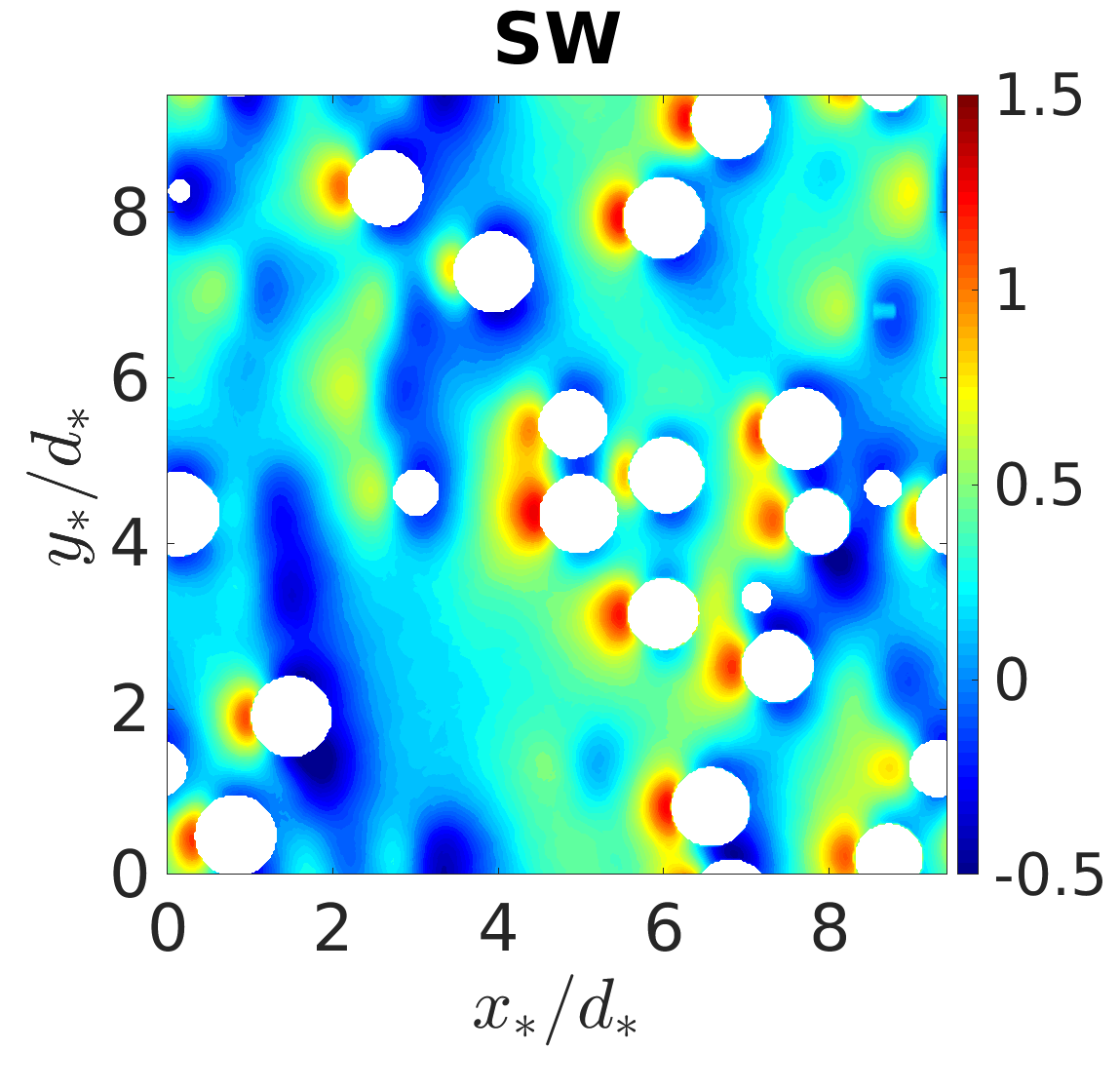}
		\caption{}
	\end{subfigure}
	\begin{subfigure}{\textwidth}
		\includegraphics[width=0.328\textwidth,keepaspectratio=true]{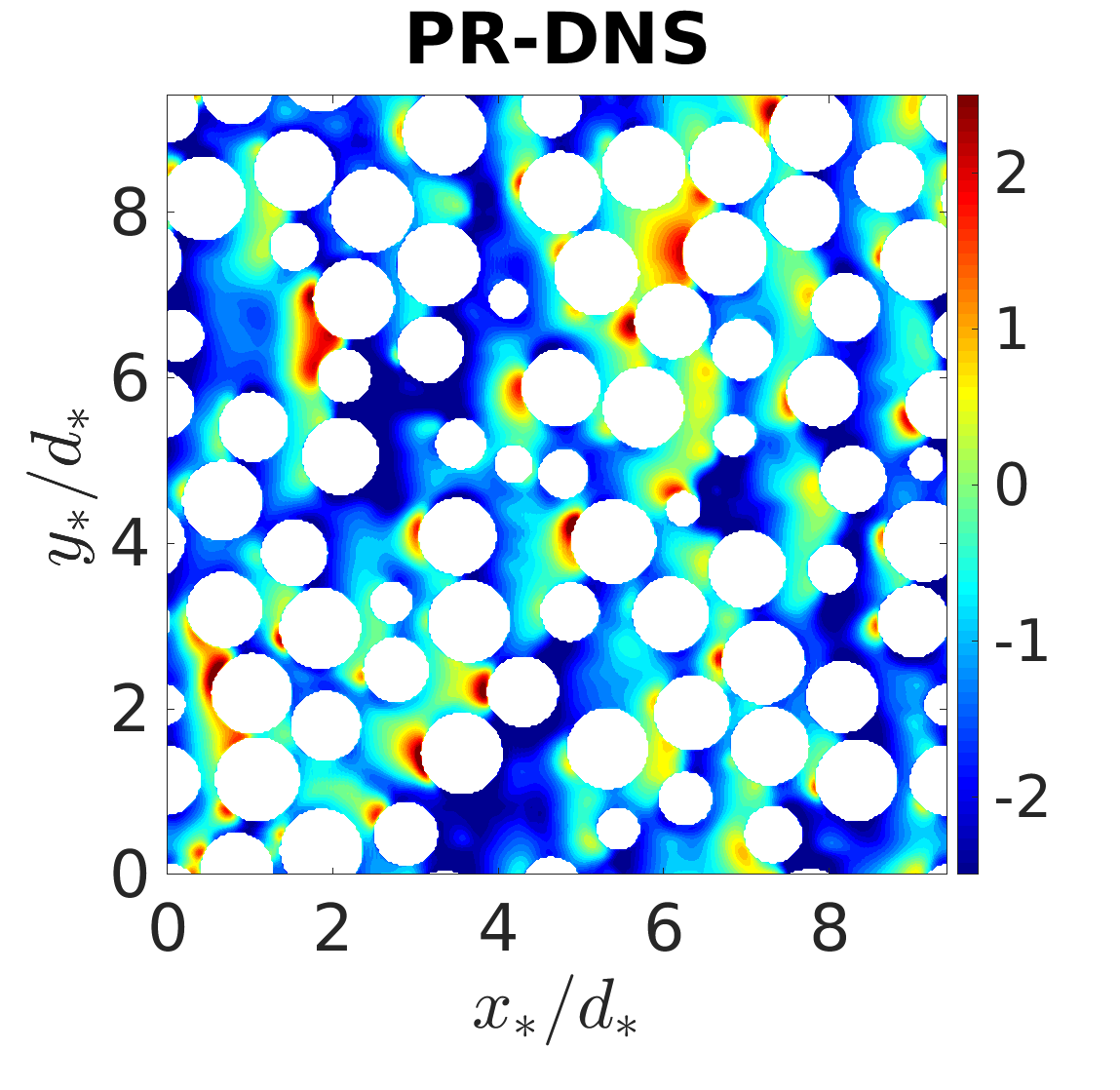}
		\includegraphics[width=0.328\textwidth,keepaspectratio=true]{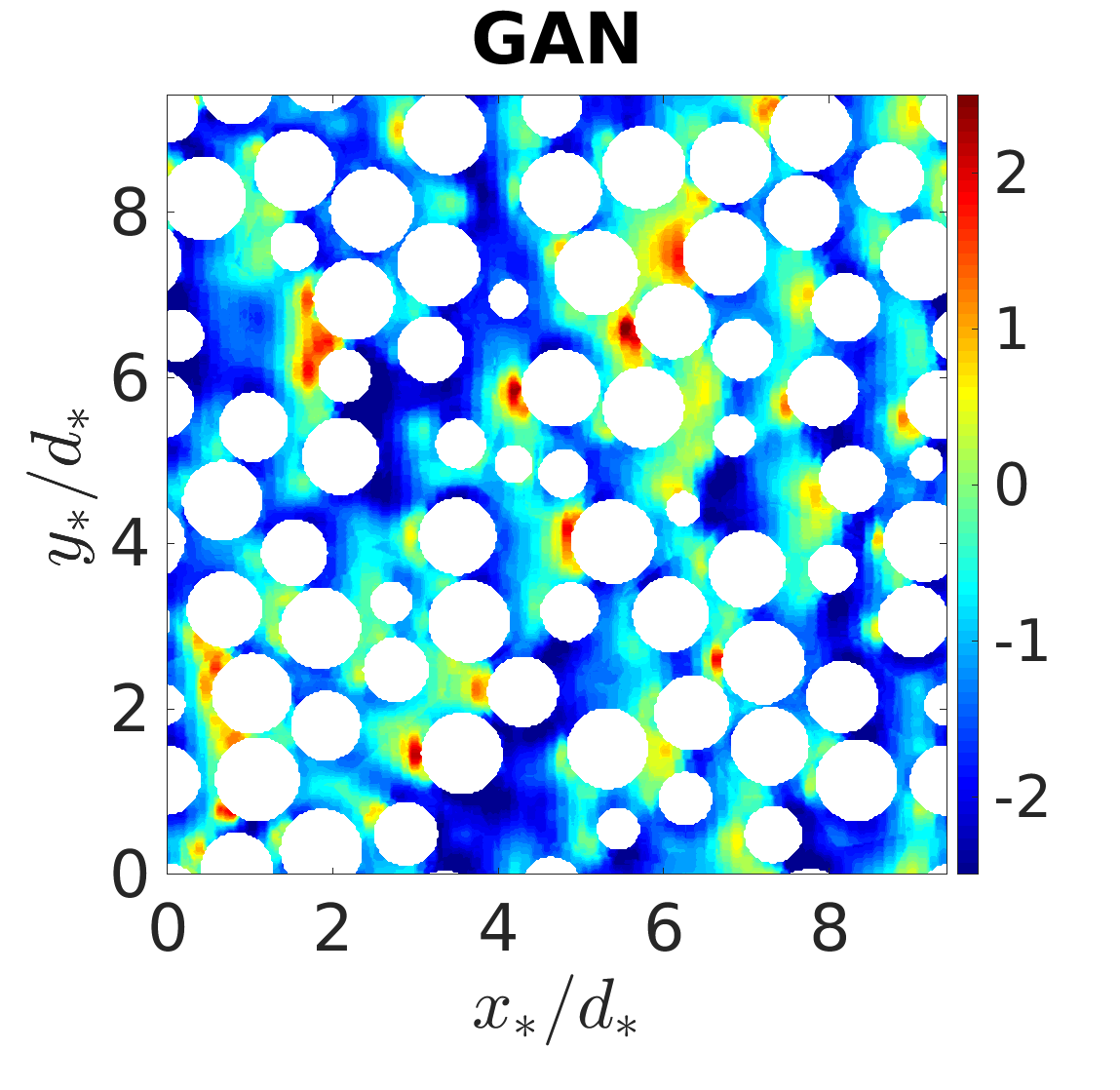}
		\includegraphics[width=0.328\textwidth,keepaspectratio=true]{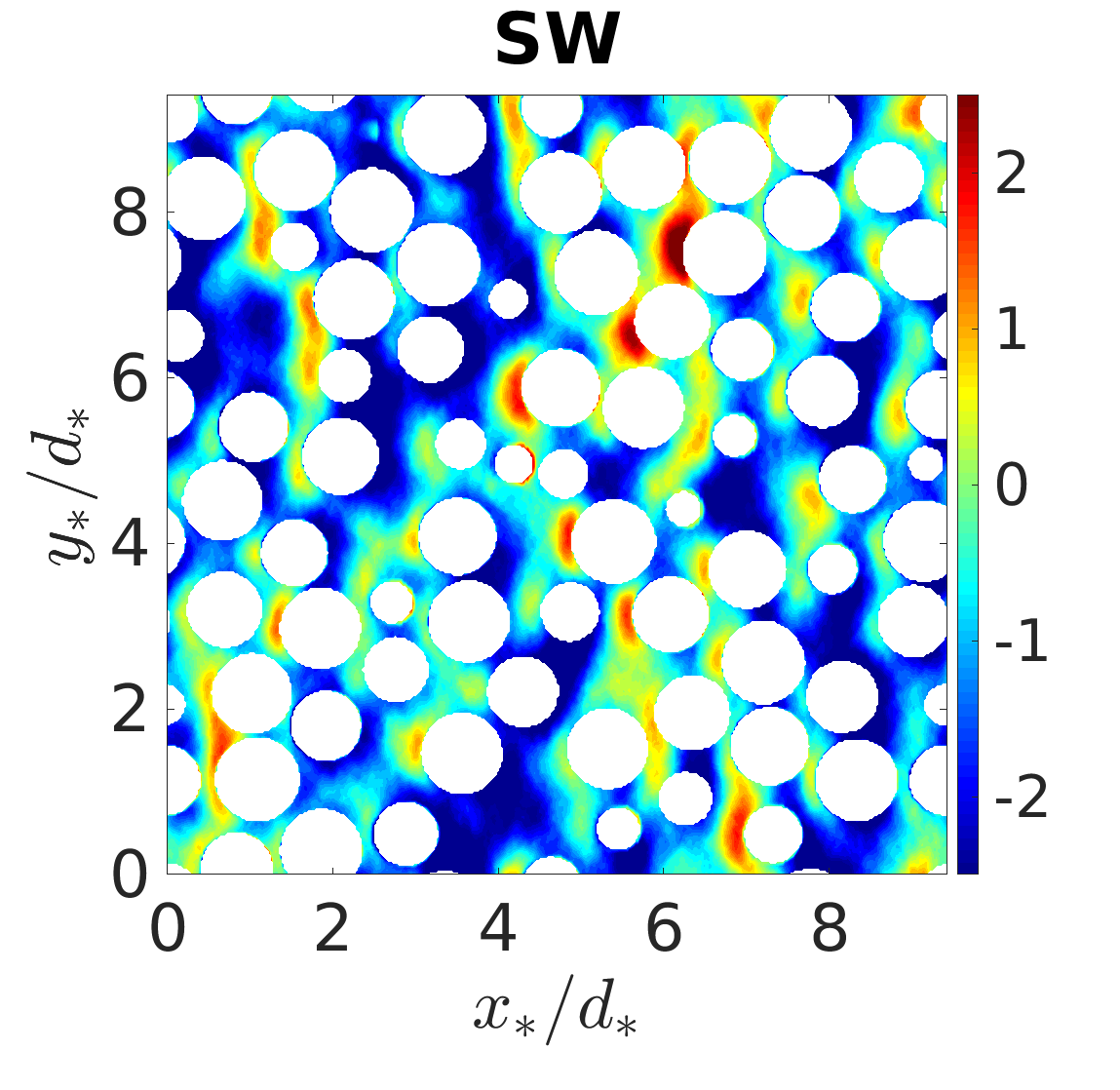}
		\caption{}
	\end{subfigure}
	\caption{Pressure for central $x-y$ plane of a test realization from  (a) Case 1 and (b) Case 8}
	\label{fig:recon_p}
\end{figure}

For illustration purposes the $3 \pi \times 3 \pi \times 3 \pi$ cubic domain's central $x-y$ plane has been presented for a test realization each from Cases 1 and 8. Contours of streamwise velocity, in-plane cross-stream velocity and pressure are presented in Figures~\ref{fig:recon_u}, \ref{fig:recon_v}, \& \ref{fig:recon_p}. The generator used to produce these results was trained using all of the training data, i.e., $N_{sub,tr} = N_{tr}$. In these figures the reconstructed flow field using the GAN methodology is compared with the corresponding PR-DNS solution and also with the flow field predicted using the superposable wake (SW) model \cite{moore-bala2019}. It can be seen from the figures that the reconstructed flow field using GAN methodology is much closer to the PR-DNS solution compared to superposable wake model in both the low and high volume fraction samples. Test performance of the two models for the central $x-y$ slice in both samples have been tabulated in Table~\ref{tab:recon_r2}. The $R^2$ metric presented in the table for $u,\, v,\, p$ is the same as that followed in \citet{moore-bala2019} and it measures the correspondence between the model and the PR-DNS. The superior performance of the GAN prediction over the superposable wake model is clear. In \citet{moore-bala2019} it was pointed out that the superposable wake yields the optimal solution within the pairwise interaction approximation. Thus, the improved performance of the GAN model is due to its ability to account for the $N$-body interaction among the particles. As a result, the improvement over superposable wake is higher in the higher volume fraction case.

\begin{table}
	\begin{center}
			\begin{tabular}{|c|ccc|ccc|}
				\hline
				& & GAN & & & SW & \\
				& $u$ & $v$ & $p$ & $u$ & $v$ & $p$\\ \hline
			Case 1 & 0.9214 & 0.8609 & 0.9262 & 0.7184 & 0.6173 & 0.7275\\	
			Case 8 & 0.8416 & 0.8230 & 0.9043 & 0.5635 & 0.5458 & 0.5965\\\hline	
			\end{tabular}
		\caption{Test performance of the models for the central $x-y$ plane of samples from cases 1 and 8.}
		\label{tab:recon_r2}
	\end{center}
\end{table}
 
\section{Conclusion and Future Scope}
The article describes a methodology to predict flow field in a dispersed multiphase setup that involves randomly distributed monodispersive spheres at different combinations of particle volume fraction and volume-averaged Reynolds number. This methodology is based on a data-driven technique called Generative Adversarial Network (GAN). The networks used in this model essentially comprise of Convolutional Neural Networks. For accurately predicting fluid flow around each particle, the model considers a neighborhood around the particle, termed as the particle's sub-domain. This approach thus requires curation of the Direct Numerical Simulation data, which is required in the first place for training the GAN model. We first obtain normalized perturbations of the flow field within each sub-domain, which are then used as inputs and outputs for different networks. The model is made up of three neural networks namely, Generator, Discriminator, and Attention-CNN. The model's framework can be described as follows. The generator's objective is to produce DNS-like flow field on a coarse-grained mesh in a sub-domain for a given input information of locations of all neighboring particles within the sub-domain and volume-averaged Reynolds number. The training of the generator is pitted against the discriminator, whose objective is to correctly distinguish the actual DNS flow field from the generator produced synthetic flow. The purpose of Attention-CNN is to produce a fine-grained flow in a concentrated region around the particle of interest, known as attention-domain, from the flow output of the generator. Performance of the model (generator, attention-cnn) has been evaluated on test data for the different Reynolds number and volume fraction combinations. The results shows that the present GAN architecture is able to capture the fluid flow around the random distribution of particles to a very good extent. 

The trained models are then used to recreate flow field over the entire domain using voronoi tessellation. Visual and quantitative evaluation of the GAN generated flow fields indicate that this methodology holds great potential. In particular, we observe that the GAN generated synthetic flow fields to be substantially better than those generated using superposable wake approximation presented in \citet{moore-bala2019}. Since the superposable wake model is based on pairwise interaction among the particles it can be concluded that the improved performance of GAN is due to its ability to account for the $N$-body interaction among the particles. Thus, the improvement over pairwise interaction approximation is stronger at higher volume fraction.

Inclusion of symmetries through equivariant networks \cite{wang2020incorporating} and physics based loss terms \cite{raissi2017physics, subramaniam2020turbulence} have indicated in other problems substantial potential for improvement of the generalization capabilities of neural networks. 
Thus, the performance of the GAN model presented in this paper can be further improved by incorporating such augmentation methods. In addition to these generalizing techniques, an interesting investigation that can be carried out is the study of model's performance when it is trained on multiple cases simultaneously. It can provide insight into how the model takes significant variation in particle volume fraction and Reynolds number among samples into consideration.

\section{Acknowledgements}
This work was sponsored by the Office of Naval Research (ONR) as part of the Multidisciplinary University Research Initiatives (MURI) Program, under grant number N00014-16-1-2617. This work was also partly supported and benefited from the U.S. Department of Energy, National Nuclear Security Administration, Advanced Simulation and Computing Program, as a Cooperative Agreement to the University of Florida under the Predictive Science Academic Alliance Program, under Contract No. DE-NA0002378. This work was also partly supported by National Science Foundation under Grant No. 1908299.

\appendix
\section{}\label{appA}
\subsection{Loss Functions}
As described in section \ref{methdlgy} the current ML model contains three neural networks namely, Generator ($\mathcal{G}$), Discriminator ($\mathcal{D}$), and Attention CNN ($\mathcal{A}$). Loss functions used for the training of these networks have been defined below. The loss functions of $\mathcal{G}$ and $\mathcal{D}$ are derived from the objective function given by (\ref{D_OF}).

Generator loss function:
\begin{equation}\label{g_loss}
\mathcal{L}_{\mathcal{G}} = \mathbb{E}_{\scriptsize{\bc} \in P_{c}}\left[-\mathcal{D}(\mathcal{G}(\bc))\right]+\gamma {\|(\bxi-\beta).I_{f}\|}_{1} \, ,
\end{equation}

Discriminator loss function:
\begin{equation}
\mathcal{L}_{\mathcal{D}}  =  \mathbb{E}_{\scriptsize{\bxi} \in P_{\xi}} \left[-\log \mathcal{D}(\bxi)\right]+\mathbb{E}_{\scriptsize{\bc} \in P_{c}}\left[-\log (1-\mathcal{D}(\mathcal{G}(\bc)))\right]+\lambda \mathbb{E}_{\scriptsize{\widetilde{\bxi}}}\left[ {\|(\nabla \mathcal{D})_{\scriptsize{\widetilde{\bxi}}}\|}_{2}^{2} \right]
\end{equation}

Attention CNN loss function:
\begin{equation}\label{cnn_loss}
\mathcal{L}_{\mathcal{A}} = \gamma {\|(\bpsi-\bzeta).I_{f}\|}_{1} \, ,
\end{equation}
where, $(\bxi-\beta)\,.\,I_{f}$ represents element-wise multiplication between $(\bxi-\beta)$ and $I_f$, which are arrays of same size along \textit{x, y, z}, and ${\|\,\cdot\,\|}_{1}$ stands for $L1$-norm. Here $I_f$ is the 3D indicator function defined in \eqref{eq3.3}. In this study, ($\gamma,\lambda$) are chosen to be (200,1000) and the sensitivity of results to a $\pm10\%$ change in these parameters is not large.

\subsection{Architectural and Training Details}
Neural network construction and training was performed using PyTorch 
framework \cite{cite_pytorch}. The notation system given in table~\ref{tab:nn_notation} was followed in describing architectures of networks. The details of the 14 layers of the generator $\mathcal{G}$ are presented in table~\ref{tab:g_arch} following the notation given in table~\ref{tab:nn_notation}. The details of the 6 layers of the discriminator $\mathcal{D}$ are presented in table~\ref{tab:d_arch}. The details of the 3 layers of the attention-cnn $\mathcal{A}$ are presented in table~\ref{tab:cnn_arch}. 
All ML models reported in this work were trained for forty epochs. Asymptotic nature of loss values for the networks was observed in all cases before the end of fortieth epoch. Adam optimizer \cite{adam_opt} was used for optimizing the coefficients/parameters of the three networks. Learning rate scheduler of type ReduceLROnPlateau was used for the generator and attention-cnn. An initial learning rate of $10^{-4}$ was used for $\mathcal{G}$ and $\mathcal{A}$. In $\mathcal{D}$ the initial learning rate was $10^{-5}$. All the convolutional layers were initialized with a normal distribution having a mean of 0 and a standard deviation of 0.2, and the PyTorch default weight initializaton was used for Batch Normalization. A batch size of eight was used for all training processes to ensure sufficient parameter/coefficient update iterations occur by the end of forty epochs. 

\begin{figure}
	\begin{center}
		\begin{subfigure}{\textwidth}
			\includegraphics[width=1.\linewidth]{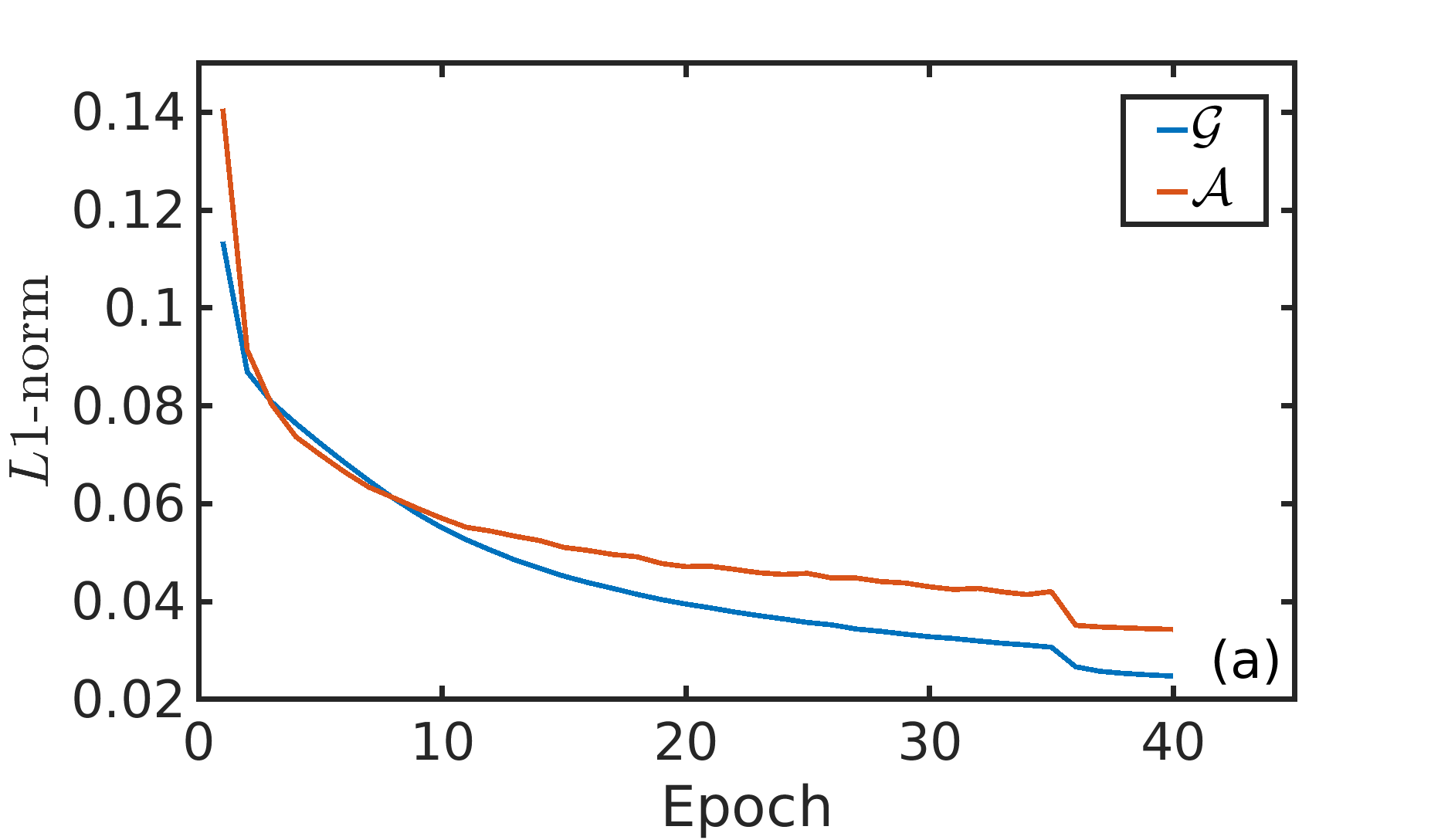}
		\end{subfigure}
		\begin{subfigure}{\textwidth}
			\includegraphics[width=1.\linewidth]{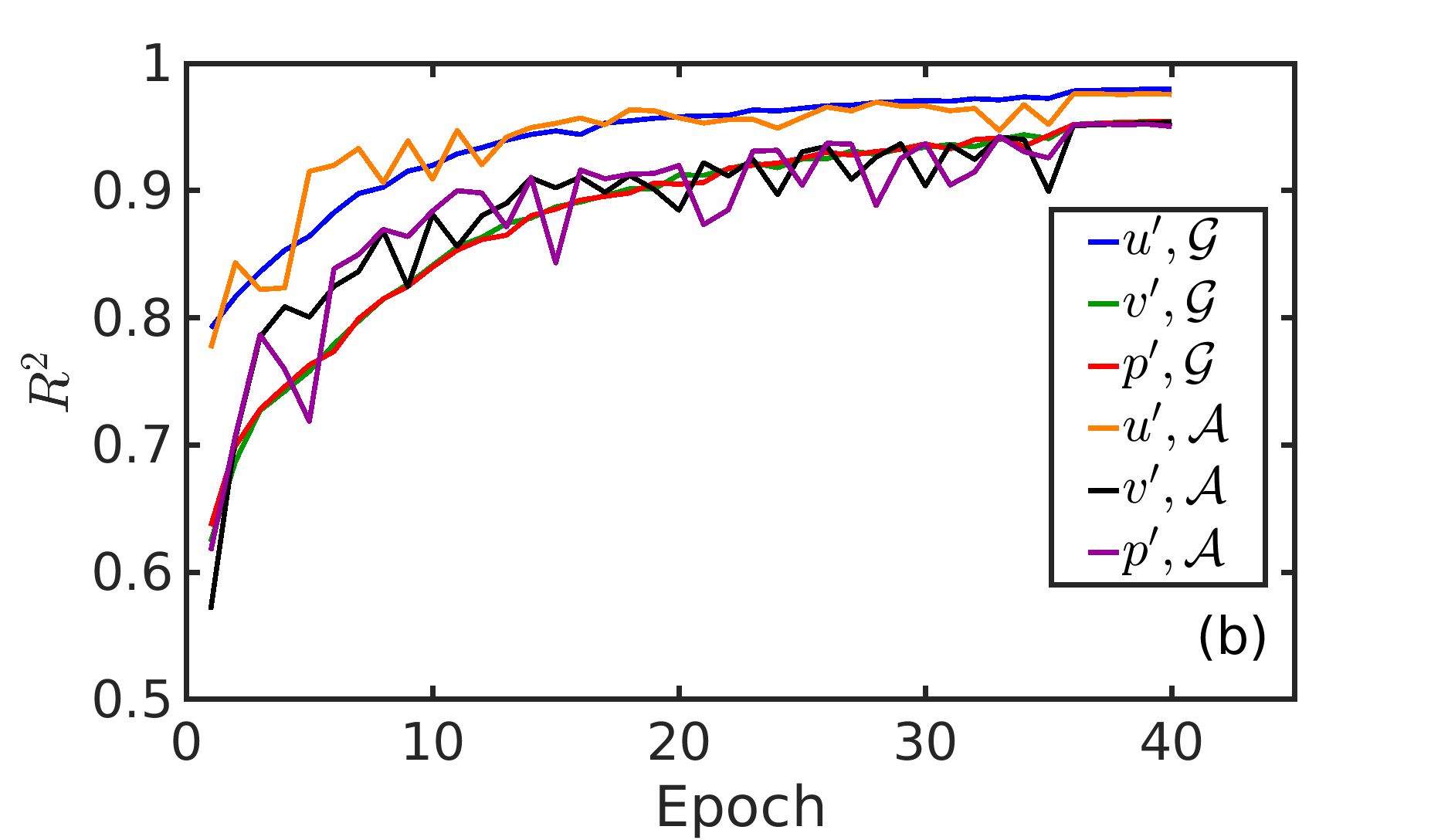}
		\end{subfigure}
	\end{center}
	\caption{Training loss evolution with epochs for a $N_{sub,tr} = N_{tr}/2$ subset from Case $Re = 39.47$, $\phi = 0.11$ using (a) $L1$-norm and (b) $R^2$}
	\label{fig:train_loss}
\end{figure}

Evolution of mean training loss per epoch has been shown in Figure~\ref{fig:train_loss} for the generator and the attention-cnn trained on a $N_{sub,tr} = N_{tr}/2$ dataset belonging to case 1. $L1$-norm presented in subfigure (a) corresponds to ${\|(\bxi-\beta).I_{f}\|}_{1}$ for $\mathcal{G}$ and ${\|(\bpsi-\bzeta).I_{f}\|}_{1}$ for $\mathcal{A}$. Training loss is also shown in terms of $R^{2}$ in subfigure (b) as $R^2$ is the metric used to quantify networks' test performance. A similar training loss behavior exists for the remaining model training runs presented in this work. 

\begin{table}
	\begin{center}
		\begin{tabular}{|l@{\hskip 5mm}l|}
			\hline
			$inpt$ & input to a network\\
			$oupt$ & output of a network\\
			$chls_{in}$ & incoming channels for a layer\\
			$chls_{out}$ & outgoing channels of a layer\\
			$Tsr_{in}$ & input tensor size of a layer\\
			$Tsr_{out}$ & output tensor size of a layer\\
			$fltr$ & \begin{tabular}{@{}l@{}}size of filter\\(same is all three directions)\end{tabular}\\
			$strd$ & stride in a layer\\
			$pdg$ & \begin{tabular}{@{}l@{}}padding at each boundary\\in a layer\end{tabular}\\
			$af$ & activation function\\
			$BN$ & \begin{tabular}{@{}l@{}}batch normalization\\\cite{batch_norm}\end{tabular}\\
			$skp$ & skip connection\\
			$Conv(Tsr_{in},chls_{in},chls_{out},fltr,strd,pdg,Tsr_{out})$ & convolution layer\\
			$Linear(Tsr_{in},chls_{in},chls{out},Tsr_{out})$ & linear layer\\
			$TConv(Tsr_{in},chls_{in},chls_{out},fltr,strd,pdg,Tsr_{out})$ & transpose convolution layer\\
			\hline
		\end{tabular}
	\end{center}
	\caption{Neural network notation system}
	\label{tab:nn_notation}
\end{table}

\begin{table}
	\begin{center}
		\begin{tabular}{|l|}
			\hline
			$inpt$ = $\bc$\\
			$B1$  $\colon$ $Conv(64^3,1,16,3,1,1,64^3)$, $BN$, $af$ = LeakyReLU(0.2)\\
			$B2$  $\colon$ $Conv(64^3,16,32,3,1,1,64^3)$, $BN$,$af$ = LeakyReLU(0.2)\\
			$B3$  $\colon$ $Conv(64^3,32,64,4,2,1,32^3)$, $BN$, $af$ = LeakyReLU(0.2)\\
			$B4$  $\colon$ $Conv(32^3,64,128,4,2,1,16^3)$, $BN$, $af$ = LeakyReLU(0.2)\\
			$B5$  $\colon$ $Conv(16^3,128,256,4,2,1,8^3)$, $BN$, $af$ = LeakyReLU(0.2)\\
			$B6$  $\colon$ $Conv(8^3,256,512,4,2,1,4^3)$, $BN$, $af$ = LeakyReLU(0.2)\\
			$B7$  $\colon$ $TConv(4^3,512,256,4,2,1,8^3)$, $BN$, $af$ = LeakyReLU(0.2), $skp = 5$\\
			$B8$  $\colon$ $TConv(8^3,512,128,4,2,1,16^3)$, $BN$, $af$ = LeakyReLU(0.2), $skp = 4$\\
			$B9$  $\colon$ $TConv(16^3,256,64,4,2,1,32^3)$, $BN$, $af$ = LeakyReLU(0.2), $skp = 3$\\
			$B10$ $\colon$ $TConv(32^3,128,32,4,2,1,64^3)$, $BN$, $af$ = LeakyReLU(0.2), $skp = 2$\\
			$B11$ $\colon$ $TConv(64^3,64,16,3,1,1,64^3)$, $BN$, $af$ = LeakyReLU(0.2), $skp = 1$\\
			$B12$ $\colon$ $TConv(64^3,32,128,3,1,1,64^3)$, $BN$, $af$ = LeakyReLU(0.2)\\
			$B13$ $\colon$ $TConv(64^3,128,64,3,1,1,64^3)$, $BN$, $af$ = LeakyReLU(0.2)\\
			$B14$ $\colon$ $Conv(64^3,64,4,3,1,1,64^3)$\\
			$oupt$ = $\beta$\\
			\hline
		\end{tabular}
	\end{center}
	\caption{$\mathcal{G}$ architecture}
	\label{tab:g_arch}
\end{table}

\begin{table}
	\begin{center}
		\begin{tabular}{|l|}
			\hline
			$inpt$ = $\bc \cup \{ \bxi$  or $\beta \}$,\\
			concatenation of $\bxi$ or $\beta$ and $\bc$ along channels\\
			$B1$ $\colon$ $Conv(64^3,5,16,3,1,1,64^3)$, $af$ = LeakyReLU(0.2)\\
			$B2$ $\colon$ $Conv(64^3,16,32,3,1,1,64^3)$, $af$ = LeakyReLU(0.2)\\
			$B3$ $\colon$ $Conv(64^3,32,64,4,2,1,32^3)$, $af$ = LeakyReLU(0.2)\\
			$B4$ $\colon$ $Conv(32^3,64,128,4,2,1,16^3)$, $af$ = LeakyReLU(0.2)\\
			$B5$ $\colon$ $Conv(16^3,128,256,4,2,1,8^3)$, $af$ = LeakyReLU(0.2)\\
			flatten (256 channels, $8^3$ tensor)\\
			$B6$ $\colon$ $Linear$(flattened tensor,$256^*8^*8^*8$,1, single value), $af$ = sigmoid\\
			$oupt = q$\\
			\hline 
		\end{tabular}
	\end{center}
	\caption{$\mathcal{D}$ architecture}
	\label{tab:d_arch}
\end{table}

\begin{table}
	\begin{center}
		\begin{tabular}{|l|}
			\hline
			$inpt = \beta$\\
			$B1$ $\colon$ $TConv(16^3,4,64,4,2,1,32^3)$, $BN$, $af$ = LeakyReLU(0.2)\\
			$B2$ $\colon$ $TConv(32^3,64,128,4,2,1,64^3)$, $BN$, $af$ = LeakyReLU(0.2)\\
			$B3$ $\colon$ $Conv(64^3,128,4,3,1,1,64^3)$\\
			$oupt = \bzeta$\\
			\hline 
		\end{tabular}
	\end{center}
	\caption{$\mathcal{A}$ architecture}
	\label{tab:cnn_arch}
\end{table}

\subsection{Computational Cost}
Neural network training and testing was carried out using 4 Nvidia GeForce RTX 2080Ti GPUs for all of the results presented in this work. Training time required for a single batch of size 8 is 1.83 seconds per epoch. Testing time taken for a single batch containing 8 samples is 0.573 seconds. 

\bibliography{manuscript.bib}

\end{document}